\documentclass[12pt]{article}
\usepackage{epsf}

\usepackage{epsfig,graphics}
\usepackage {graphicx}
\usepackage {epsfig}
\usepackage{subcaption}
\usepackage {tabularx} 
\usepackage{rotate}	
\usepackage{slashed}
\usepackage{bbm}
\usepackage{color}
\usepackage{tikz}
\usetikzlibrary{decorations.pathmorphing} % Añade esta línea si vas a usar decoraciones como 'snake'
\usepackage{amsmath}
\usepackage{amsfonts}
\usepackage{amssymb}
\usepackage{graphicx}
\usepackage{cite}

\usepackage{fancyhdr}
\usepackage{hyperref}
\usepackage{diagbox}
\newcommand{\bmat}{\left(\begin{array}}
\newcommand{\emat}{\end{array}\right)}

\def\yzero{\smash{\hbox{$y\kern-4pt\raise1pt\hbox{${}^\circ$}$}}}
\def\p{\partial}
\def\a{\alpha}
\def\b{\beta}
\def\g{\gamma}
\def\G{\Gamma}
\def\d{\delta}
\def\beq{\begin{equation}}
\def\eeq{\end{equation}}
\def\beqa{\begin{eqnarray}}
\def\eeqa{\end{eqnarray}}
\def\Om{\Omega}
\def\om{\omega}
\def\th{\theta}
\def\vt{\vartheta}

\def\-{\hphantom{-}}

\def\s2{\frac{1}{\sqrt2}}

\def\oh{\frac{1}{2}}

\def\ch{{\cal H}}
\def\cf{{\cal F}}

\def\Dsl{\,\raise.15ex\hbox{/}\mkern-13.5mu D} %this one can be subscripted
\def\IZ{Z\kern-.4em  Z}

\def\CN {{\cal N}}

\def\CK {{\cal K}}
\def\CR {{\cal R}}
\def\ii           {{\rm i}}
\def\be{\begin{equation}}
\def\ee{\end{equation}}
\def\bea{\begin{eqnarray}}
\def\eea{\end{eqnarray}}
\def\bes{\begin{subequations}}
\def\ees{\end{subequations}}

\def\oh{\frac{1}{2}}

\def\re{\mbox{Re}\, }
\def\im{\mbox{Im}\, }
\def\tr{\mbox{Tr}}

\def\bN{\mathbbm{N}}
\def\Z{\mathbbm{Z}}
\def\R{\mathbbm{R}}

\catcode`\@=11   
\newdimen\@rotdimen
\newbox\@rotbox  

\def\@vspec#1{\special{ps:#1}}%  passes #1 verbatim to the output
\def\@rotstart#1{\@vspec{gsave currentpoint currentpoint translate
   #1 neg exch neg exch translate}}% #1 can be any origin-fixing transformation
\def\@rotfinish{\@vspec{currentpoint grestore moveto}}% gets back in synch 
\def\@rotr#1{\@rotdimen=\ht#1\advance\@rotdimen by\dp#1%
   \hbox to\@rotdimen{\hskip\ht#1\vbox to\wd#1{\@rotstart{90 rotate}%
   \box#1\vss}\hss}\@rotfinish}
\def\@rotl#1{\@rotdimen=\ht#1\advance\@rotdimen by\dp#1%
   \hbox to\@rotdimen{\vbox to\wd#1{\vskip\wd#1\@rotstart{270 rotate}%
   \box#1\vss}\hss}\@rotfinish}%
\def\@rotu#1{\@rotdimen=\ht#1\advance\@rotdimen by\dp#1%
   \hbox to\wd#1{\hskip\wd#1\vbox to\@rotdimen{\vskip\@rotdimen
   \@rotstart{-1 dup scale}\box#1\vss}\hss}\@rotfinish}%
\def\@rotf#1{\hbox to\wd#1{\hskip\wd#1\@rotstart{-1 1 scale}%
   \box#1\hss}\@rotfinish}%
\def\rotate{\@ifnextchar[{\@rotate}{\@rotate[l]}}
\def\@rotate[#1]#2{\setbox\@rotbox=\hbox{#2}\@nameuse{@rot#1}\@rotbox}

\catcode`\@=12
\topmargin
-1.5cm
\textwidth
15.5cm
\textheight
23.5cm
\oddsidemargin
0.7cm
\evensidemargin
0.7cm

\begin{document}

%----------------------------------------------------------------------%
%  numbering equations with section number
%----------------------------------------------------------------------%
\makeatletter
\@addtoreset{equation}{section}
\makeatother
\renewcommand{\theequation}{\thesection.\arabic{equation}}
%----------------------------------------------------------------------%
%  title page
%----------------------------------------------------------------------%
\pagestyle{empty}
%\vspace*{1.0in}
\vspace{-0.2cm}
\rightline{ IFT-UAM/CSIC-24-40}
%\rightline{FTUAM-13-19}
%\rightline{\tt hep-th/xxxxxxx}
\vspace{0.5cm}
\begin{center}

%\vspace{0.5cm}

%\vspace{0.5cm}

\LARGE{Yukawa Couplings at Infinite Distance  \\ and Swampland Towers  in Chiral Theories }
\\[13mm]
\large{Gonzalo F.~Casas,$^\diamondsuit$  Luis E.~Ib\'a\~nez$^{\clubsuit  \diamondsuit}$ and Fernando  Marchesano$^\diamondsuit$}
			\\[12mm]
		\small{
			$^\clubsuit$  Departamento de F\'{\i}sica Te\'orica \\
				Universidad Aut\'onoma de Madrid,
				Cantoblanco, 28049 Madrid, Spain}  \\[5pt]
			$^\diamondsuit$  {Instituto de F\'{\i}sica Te\'orica UAM-CSIC, c/ Nicolas Cabrera 13-15, 28049 Madrid, Spain} 
			\\[8mm]
%({\it Questo e una genialit\'a o e tutto una cassata})
\small{\bf Abstract} \\[6mm]
\end{center}
\begin{center}
\begin{minipage}[h]{15.22cm}

We study limits of vanishing Yukawa couplings of 4d chiral matter fields in Quantum Gravity, using as a laboratory type IIA orientifolds with D6-branes. In these theories chiral fermions arise at brane intersections, where an infinite tower of charged particles dubbed {\em gonions} are localised. We show that in the limit $Y\rightarrow 0$ some of these towers become asymptotically massless, while at the same time the kinetic term of some chiral fields becomes singular and at least two extra dimensions decompactify. For limits parametrised by a large complex structure saxion $u$, Yukawa couplings have a  behaviour of the form 
$Y \, \sim \, {1}/{u^r}$, with $r$ some positive rational number.
Moreover, in this limit some of the gauge couplings associated to the Yukawa vanish. The lightest gonion scales are  of order $m_{\rm gon} \sim  g^s M_{\rm P}$ with $s>1$,  
verifying the magnetic WGC with room to spare and with no need of its tower/sublattice versions. We also show how this  behaviour can be understood in the context of the {\em emergence} of kinetic terms in Quantum Gravity. All these results  may be very relevant for phenomenology, given the fact that some of the Yukawa couplings in the Standard Model are very small.

\end{minipage}
\end{center}
\newpage

%----------------------------------------------------------------------%
%  Resetting of counters
%----------------------------------------------------------------------%
\setcounter{page}{1}
\pagestyle{plain}
\renewcommand{\thefootnote}{\arabic{footnote}}
\setcounter{footnote}{0}
%----------------------------------------------------------------------%
%  Paper begins
%----------------------------------------------------------------------%

%\end{document}

%\end{document}

\tableofcontents

%%%%%%%%%%%%%%%%%%%
%%%%%%%%%%%%%%%%%%%
 	
\section{Introduction}
\label{s:intro}

The Swampland Programme \cite{Vafa:2005ui,Brennan:2017rbf,Palti:2019pca,vanBeest:2021lhn,Grana:2021zvf} attempts to identify model-independent implications of Quantum Gravity (QG) for EFTs coupled to gravity. One of the leitmotifs  is the presence of light towers of infinite number of states when one goes to the boundary of the moduli space \cite{Ooguri:2006in}. This is also related to the fact that, when vacuum expectation values of fields tend to infinity, different types of global symmetries appear, which are inconsistent with QG. The canonical example is the case of a $U(1)$ theory with gauge coupling $g$. The magnetic Weak Gravity Conjecture (WGC) \cite{Arkani-Hamed:2006emk} tell us that as $g\rightarrow 0$, a tower of charged particles with characteristic scale $m_*$ appears such that
	\beq
	m_* \ \lesssim g M_{\rm P}^{D/2-1} \, ,
	\eeq
	with $D$ the number of dimensions and $M_{\rm P}$ the Planck mass. 
	One sees the presence of such a tower as a reflection of the fact that in the strict $g=0$ limit  there would be
	a global $U(1)$ symmetry which would be forbidden by general QG arguments.
	There are many other situations in which towers appear, signalling either a decompactification (via Kaluza--Klein (KK)-like towers) or an emergent critical string \cite{Lee:2019wij}, when one takes infinite distance limits in the moduli space. 
	
	By now many different large moduli limits have been explored and tested against string theory results in a variety of dimensions and with different amounts of supersymmetry, but setups with gauge couplings have mostly been tested in the presence of $\geq 8$ SUSY generators, see \cite{Palti:2019pca,vanBeest:2021lhn} for reviews. Setups with 4d ${\cal N} =1,0$ SUSY and asymptotically vanishing gauge couplings remain relatively unexplored. However, these are the cases of most phenomenological interest, particularly due to the observed chirality of the Standard Model (SM) which requires ${\cal N}\leq 1$.  
	
	In the present paper we aim to improve the state-of-the-art by studying  string compactifications leading to 4d ${\cal N}= 1$ {\it chiral} spectra at infinite distance limits.\footnote{Previous works addressing infinite distance limits in 4d ${\cal N} =1$ chiral constructions include \cite{Lee:2019tst,Lee:2020gvu}.} We are particularly interested in classes of models which contain the ingredients necessary for a realistic model with SM gauge group and fermion generations. Such settings will involve new ingredients like e.g. the existence of perturbative trilinear 
	Yukawa couplings  $Y_{ijk}$ involving three charged chiral multiplets.  One can then pose the following question: Is the limit
	$Y_{ijk} \to 0$ at infinite distance? What happens along such a limit? Do towers appear? 
	If they do, what is their structure?
	The answer to these questions is not immediately obvious, since, unlike the case of the $U(1)$ gauge symmetry
	above, it is not clear what goes wrong (if at all) when $Y_{ijk}\rightarrow 0$. 
 Some previous studies of Yukawa couplings in the context of the Swampland Programme include
 \cite{Palti:2020tsy,Castellano:2022bvr,Castellano:2023qhp,Cribiori:2023gcy}.
	
	 In addition to its theoretical interest, the study of $Y_{ijk}\rightarrow 0$ limits is also interesting phenomenologically, 
	 since most of the Yukawa couplings in the  SM are numerically small. Thus  in order to reproduce such small
	 values this limit is the right one to explore. This is particularly the case of the Yukawa couplings of neutrinos
	 which require values $Y\lesssim 10^{-12}$  in the case of purely Dirac neutrinos, if one wants to match the
	 experimental data. We explore this particularly interesting case in a separate companion paper \cite{Casas:2024clw}, see also the related analysis in \cite{Castellano:2023qhp}.

	To study these questions we use as a laboratory ${\cal N}=1$ Calabi--Yau (CY) orientifold compactifications of type IIA string theory.
	In those theories 
	chiral matter is localised at the intersection of D6-branes, which wrap three-cycles in the CY.  Such 
	type of models have been studied in much detail and specific examples  with the SM gauge group (or some extension) 
	with three chiral families have been constructed (see \cite{Blumenhagen:2005mu,Blumenhagen:2006ci,Marchesano:2007de,Lust:2009kp,Ibanez:2012zz,Marchesano:2022qbx,Marchesano:2024gul}  for reviews).  The properties of these constructions 
	may be considered as representative  of the space of string compactifications with chiral matter, since different dualities connect them with 
	other constructions like type IIB CY orientifolds (via mirror symmetry) or heterotic compactifications with $U(1)$ bundles.
	Thus we believe that our results go beyond this particular class of theories. We also believe that, due to chirality,  many of our results remain true 
	in models with SUSY spontaneously broken. 
	
	Let us summarise some of our results. We find that, indeed, limits in which a Yukawa coupling of a 4d ${\cal N}=1$ EFT goes to
	zero are singular and, in general, lead to the appearance of  different towers of states, depending on the chosen field direction. In 4d $\CN=1$  non-chiral theories, KK-like and string towers are known to arise from the gravitational sector of the compactification \cite{Font:2019cxq,Klaewer:2020lfg,Lanza:2020qmt,Lanza:2021udy,Lanza:2022zyg,Cota:2022yjw}. Moreover, if D-branes are present, KK-towers arising from their worldvolume are also expected. However, as soon as one considers chiral D-brane models there is a new, more exotic class of  towers that appears. They are made of massive particles transforming under bifundamental representations of the D-brane gauge group, and which in our setup are localised at the intersections of  pairs of D6-branes D6$_a \cap$D6$_b$, just where the chiral particles live. They were dubbed {\em gonions} in \cite{Aldazabal:2000cn} and arise from open strings stretched between both branes.	Despite coming from string fluctuations, their mass spectrum	below the string scale $M_s$ is not Hagedorn-like, but is described by a formula of the type $m_n^2\simeq  n  m_{\rm gon}^2$, where $m_{\rm gon}^2\simeq \theta_{ab}M_s^2$ and 
	$\theta_{ab}$ is an intersection angle between the two branes.\footnote{In type IIB mirror setups that feature D-branes with magnetic worldvolume fluxes, these gonion states correspond to Landau levels.}

\setlength{\belowcaptionskip}{-10pt}
\begin{figure}[ht!]
\begin{center}
\includegraphics[width=10cm]{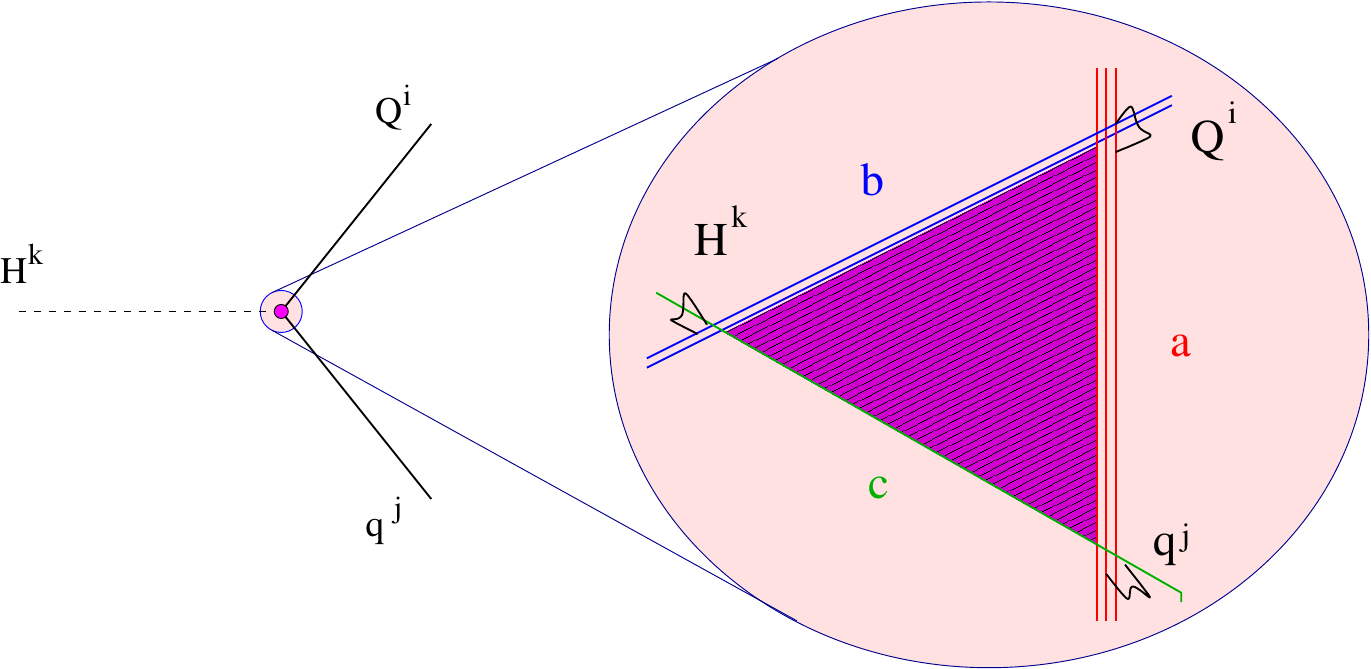}
\caption{Yukawa coupling in type IIA orientifolds with intersecting D6-branes. The holomorphic piece $W_{ijk}$ depends on the shaded region that connects the three D6-brane intersections, while $\Theta_{ijk}$ depends on the local intersection angles. 
\label{fig:yuki}}
\end{center}
\end{figure}	
	
	Perturbative Yukawa couplings come from worldsheet disc diagrams with boundary on three D6$_\a$-branes $\a = a, b, c$ and that connect three of their pairwise intersections \cite{Aldazabal:2000cn,Cremades:2003qj}, see figure \ref{fig:yuki}. Schematically the physical, canonically normalised Yukawa couplings have an expression of the form
	\beq
	Y_{ijk}\, =\, e^{\phi_4/2} {\rm Vol}_X^{1/4}  \Theta_{ijk}^{1/4} W_{ijk} \, .
	\label{yukito}
	\eeq
	Here ${\rm Vol}_X$ is the compactification volume and $W_{ijk}$ the holomorphic piece of the Yukawa coupling generated from world-sheet instanton corrections. Additionally,  ${\phi_4}$ stands for the 4d dilaton and 
	$\Theta_{ijk}$ only depends on the intersection angles (and hence on the complex structure). From the above expression it is not clear which K\"ahler directions will take us to a region of vanishing Yukawas, if any. More interesting are the large complex structure limits at fixed K\"ahler moduli, in which one has $|Y_{ijk}| \sim e^{\phi_4/2}\Theta_{ijk}^{1/4}$.  Taking  ${\rm Vol}_X^{1/4} |W_{ijk}| \sim {\cal O}(1)$, we find that one can write the Yukawa couplings in terms of the gonion masses
	at the three intersections involved in the Yukawa, in the form
	\beq
	|Y_{ijk}| \  \simeq \  e^{-\phi_4} \left( \frac {m_{\rm gon}^i}{M_{\rm P}}\frac {m_{\rm gon}^j}{M_{\rm P}}\frac {m_{\rm gon}^k}{M_{\rm P}}\right)^{1/2} \ ,
 \label{yukigon}
	\eeq
	%)$.
	where $m_{\rm gon}^i$ is the lightest gonion mass at a given intersection. This expression is fairly general in CY orientifolds,\footnote{Slight modifications arise in Yukawas that involve what we dub ${\cal N}=2$ sectors, see section \ref{s:Ylimit}.} since this piece of the Yukawa essentially only depends on local information about the intersection points.
	It shows how, whenever a Yukawa coupling goes to zero in the weak coupling regime $e^{\phi_4} < 1$, some light gonion tower arises. Moreover, one can see that this structure comes from the kinetic terms of the three fields that participate in the Yukawa, which in 4d EFT units read $K_{i\bar{i}}\simeq M_{\rm P}/m_{\rm gon}^i $. As a result, the Yukawas become singular because, in the 4d EFT, at least one of the particles involved in the Yukawa has a singular metric.  One also observes that, when a gonion mass $m_{\rm gon}$ goes to zero, at least two D-brane and bulk dimensions decompactify.

    To compute the Yukawa couplings from the above expression one needs to know the lightest gonion masses at each of the three intersections, or in other words the intersection angles. Such angles are a function of the complex structure saxions of the compactification, but the dictionary between both is in general very involved. We however propose an estimate of $m_{\rm gon}$ in terms of the tensions $T_{\rm D4}$ of 4d string made up from D4-branes wrapping three-cycles of the CY, which couple to space-time filling D6-branes through their Fayet-Ilioupolous (FI) terms. In particular, a D6$_a$-brane  that intersects its orientifold image will have an FI-term of the form $\pi \xi_a = Q_a^K T_{{\rm D4},K}$, with $2Q_a^K \in \Z$, and $K$ running over an appropriate basis of 4d BPS strings. This FI-term sets the mass of the lightest scalars at the intersection, and it is the signed sum of three intersection angles. From here we estimate the lightest gonion scale at such intersection as $m_{\rm gon}^2 \simeq g_a^2 T_{{\rm D4}, K}$, for some appropriate choice of $K$.  With this prescription one can estimate the behaviour of the Yukawa couplings at large complex structure in a variety of settings.  We consider STU-like models of the type described in \cite{Font:2019cxq} as well as the EFT string limits of \cite{Lanza:2020qmt,Lanza:2021udy} or variants thereof that are compatible with intersecting D6-branes at supersymmetric angles. This general analysis can be made explicit in toroidal orientifold models, including an example with a semi-realistic Pati--Salam gauge group.
	 
	 The casuistics are wide, but there are several general lessons. For limits parametrised in terms of a largest complex structure saxion $u$, asymptotically vanishing Yukawa couplings have a general behaviour of the form
	 \beq
	 Y_* \, \sim \, \frac {1}{u^r}\, ,
  \label{Yrelu}
	 \eeq
	 with $r$ a rational positive number. In most of the examples analysed one has multiples of $1/4$, inherited from the exponent in eq.(\ref{yukito}), and more precisely  $r=1/4,1/2,3/4,1$, although other fractional values for $r$ also appear. As a consequence, Yukawas between branes $a$, $b$, $c$ are related to at least one of their gauge couplings $g_*$ as 
	  \beq
	  Y_* \, \sim \,  g_*^{2 r} \, . %, \qquad \text{with} \quad \lambda >0.
   \label{Yrelg}
	  \eeq
  This already tells us what goes wrong when a chiral Yukawa coupling goes to zero: some gauge couplings also go to zero, and global symmetries would arise. 

  It is also interesting that the charged particles in these settings, the towers of gonions, do not satisfy the tower/sublattice 
  WGC\cite{Heidenreich:2016aqi,Montero:2016tif,Andriolo:2018lvp,Heidenreich:2019zkl} by themselves, as all the elements of the tower have the same charge. Moreover, in general they do not saturate the WGC. 
	  From the $U(1)$ WGC one expects $m_{\rm gon}\leq g M_{\rm P}$, but  $m_{\rm gon}$ is oftentimes well below. 
	  An interesting class of limits that we find which illustrate this fact feature only two decompactifying dimensions and 
	  a single dominant gonion tower. We find the relations 
	  \beq
	  m_{\rm gon} \, \sim \,    g_* M_s \, \sim \,  g_*^2 M_{\rm P}  \, , \qquad   Y_*\, \sim \, g_*   \ , \qquad g_* \sim e^{\phi_4}\, .
   \label{favo}
	  \eeq
	 Note the double suppression by $g_*^2$ of the gonion masses with respect to the Planck mass, compared to the standard WGC bound. This implies that, even though there is no lattice of charged states,
  the WGC itself is not in risk upon dimensional reduction.

  The study of the structure of gonion towers reveals further interesting properties with regards to the 
  proposal of {\it Emergence} in Quantum Gravity \cite{Harlow:2015lma,Grimm:2018ohb,Heidenreich:2018kpg}. In that scheme it is proposed that the kinetic terms of particles in the infrared, arise from loops involving towers of states arising for large moduli. This idea has been tested and discussed
  in the contexts of string theory vacua with $\geq 8$ supercharges, see \cite{Harlow:2015lma,Grimm:2018ohb,Heidenreich:2018kpg,Heidenreich:2017sim,Lee:2018urn,Ooguri:2018wrx,Marchesano:2022axe,Castellano:2022bvr,Castellano:2023qhp,Blumenhagen:2023tev,Hattab:2023moj}, and it is interesting to see if it is consistent with 4d chiral theories, as studied in this paper. We show that indeed, in the
  presence of gonion towers saturating the species scale, chiral massless fields at a given intersection get kinetic terms with a behaviour in agreement with the known tree level results. For this result to be obtained  it is crucial the 
  particular behaviour of gonion towers, i.e., the form of the spectrum and crucially the fact that 
  all gonions in a given tower have the same charge. We also check in simple settings how also the gauge kinetic terms 
  are produced by loops from gonion towers.

The rest of the paper is organised as follows. In  section \ref{s:orientifolds} we present a brief overview of the construction of type IIA 4d ${\cal N} = 1$ CY orientifolds with chiral matter. We describe the computation of the  physical Yukawa couplings and matter kinetic terms in toroidal settings and obtain explicit formulae applicable to more general CY examples, by exploiting the local character of both quantities. In section \ref{s:Ylimit} we study the limit $Y_{ijk}\rightarrow 0$ in general CY orientifolds and show how in this limit the Yukawas may be expressed in terms of gonion tower masses, which become light along the limit. We also study how the FI-terms and gonion masses are related to the tension of 4d strings, something that we use to propose an estimate of the gonion scales in terms of compactification moduli. This estimate is then used to argue that in general $Y \to 0$ implies the vanishing of some D6-brane gauge coupling, and to describe the different mass scales and Yukawa scalings in simple limits.  In order to illustrate the emerging picture, in section \ref{s:torito} we analise examples of these limits in explicit toroidal orientifolds, including a Pati--Salam semirealistic model.   In section \ref{s:emergence} we study the structure of the towers in this class of orientifolds with reference to the Species Scale \cite{Dvali:2007hz,Dvali:2007wp,Dvali:2008ec}
in these constructions, as well as its consistency with the Emergence Proposal.  We find that many qualitative features of the towers, kinetic terms, and Yukawa couplings in these models may be understood within this hypothesis, in which it is assumed that kinetic terms of IR particles may be obtained from loop computations involving the towers. Final conclusions	and an outlook are presented in section \ref{s:conclu}. 
 
 Complementary material is presented in the appendices. Appendix \ref{ap:spectrum} describes the gonion spectrum at two D6-brane intersections. Appendix \ref{ap:mirror} describes how to translate type I vacua with $U(N)$ bundles to the language used in the main text. Appendix \ref{ap:Hmom} describes the selection rules for perturbative Yukawas in intersecting D6-brane models.

%%%%%%%%%%%%%%%%%%%
%%%%%%%%%%%%%%%%%%%
		
\section{Yukawa couplings in 4d \texorpdfstring{$\CN=1$}{Lg} orientifolds }
\label{s:orientifolds}
	
In this section we review the general description of 4d $\CN=1$ type IIA  orientifold compactifications with intersecting D6-branes. These constructions feature a massless chiral spectrum that arises from the open string sector, and a set of Yukawa couplings generated by worldsheet instantons. While the explicit computation of these couplings has only been performed in toroidal orbifolds, the localisation properties of D-branes allow us to deduce the expression \eqref{yukito} for a general Calabi--Yau setting. As a byproduct, one can match the kinetic terms for the chiral fields at intersections with the results previously obtained in the literature and, eventually, express them in terms of gonion masses. In section \ref{s:emergence} we interpret this relation between gonion towers and kinetic terms from the viewpoint of the Emergence Proposal.

\subsection{Type IIA orientifolds and intersecting D6-branes}
\label{ss:interD6}

Let us consider type IIA string theory on a background of the form $X_4 \times X_6$, where $X_6$ is a compact Calabi--Yau three-fold with K\"ahler form $J$ and holomorphic three-form $\Omega$, and with volume form $\frac{1}{6}J^3 = \frac{1}{4} \re \Omega \wedge \im \Omega$. To this background we apply the standard orientifold quotient generated by $\Omega_{\rm ws} (-1)^{F_L}{\cal R}$ \cite{Blumenhagen:2005mu,Blumenhagen:2006ci,Marchesano:2007de,Lust:2009kp,Ibanez:2012zz,Marchesano:2022qbx}, where $\Omega_{\rm ws}$ is the worldsheet parity reversal operator, ${F_L}$ is the space-time fermion number for the left-movers and ${\cal R}$ an anti-holomorphic involution of $X_6$ acting as ${\cal R} (J, \Om) = (-J, \bar{\Om})$. The fixed point locus of ${\cal R}$ specifies one or several special Lagrangian three-cycles calibrated by $\re \Om$, where the O6-plane are located, and that we denote by $\Pi_{\rm O6}$. To cancel the induced RR tadpole, one can add background fluxes threading $X_6$ and stacks of D6-branes wrapping its three-cycles.\footnote{In principle, one may also cancel the tadpole with a more general D-brane sector that includes coisotropic D8-branes \cite{Font:2006na}. However, as of today the properties of these objects are not well understood in smooth Calabi--Yau geometries, and for this reason we will not consider them in the following.} Since our main interest is the chiral sector of the resulting 4d EFT, let us assume that background fluxes are absent and that the tadpole is cancelled by the D6-brane content, which satisfies the following equation
\be
\sum_\a N_\a \left( [\Pi_\a] + [\Pi_{\a*}] \right) = 4 [\Pi_{\rm O6}]\, .
\label{tadpole}
\ee
Here $\Pi_\a$ is a three-cycle wrapped by a stack of $N_\a$ D6-branes, and $\Pi_{\a*} = {\cal R} (\Pi_\a)$ its orientifold image, while the brackets denote their  homology classes in $H_3(X_6, \Z)$. To preserve the 4d $\CN=1$ supersymmetry of the background, such three-cycles must satisfy the special Lagrangian conditions $\cf + iJ|_{\Pi_{\a}} = 0$ and $\im \Om|_{\Pi_{\a}} = 0$, where $\cf = B|_{\Pi_{\a}} + \frac{\ell_s^2}{2\pi} F$ is the gauge-invariant D6-brane worldvolume field strength and $\ell_s = 2\pi \sqrt{\a'}$ the fundamental string length. 

Given this setup, the massless spectrum of the 4d EFT can be divided into two: the closed and the open string sectors. The former includes the gravity multiplet and the bulk moduli of the compactification, which can be subdivided into K\"ahler and complex structure moduli. At large compactification volumes the complexified K\"ahler moduli can be defined as $T^a = b^a + i t^a$, where 
\be
J_c \equiv B+ i J = (b^a+ i t^a) \om_a \, ,
\ee
and $[\om_a]$ provide a basis for the cohomology classes $H^2_-(X_6, \R)$ which are $\CR$-odd. Similarly, the complex structure moduli $U^K = \zeta^K + i u^K$ can be defined by the expansion 
\be
\Om_c \equiv C_3 + i e^{-\phi} \re \Om = ( \zeta^K + i u^K) \alpha_K\, ,
\label{Omcexp}
\ee
where $C_3$ is the three-form RR potential, $\phi$ is the 10d dilaton and $[\alpha_K]$ a basis for the $\CR$-even cohomology classes $H^3_+(X_6, \R)$. The kinetic terms of all these bulk fields are encoded in a K\"ahler potential of the form $K \equiv K_K + K_Q$, where \cite{Grimm:2004ua}
\be
K_K   \equiv   -{\rm log} \left({\rm Vol}_{X_6}\right) = -{\rm log} \left(\frac{i}{48} \CK_{abc} (T^a - \bar{T}^a)(T^b - \bar{T}^b)(T^c - \bar{T}^c) \right) \, ,
\label{KK}
\ee
with ${\cal K}_{abc} \equiv \ell_s^{-6} \int_{X_6} \omega_a \wedge \omega_b \wedge \omega_c$ the triple intersection numbers of $X_6$ and 
\begin{equation}
 K_Q \equiv -2 \log \ch = - 2 \log \left( \frac{i}{8\ell_s^6} \int_{X_6} e^{-2\phi}  \Om \wedge \bar{\Om} \right)  = 4\phi_4 \, ,
 \label{KQ}
\end{equation}
with $\phi_4 = \phi - \oh \log {\rm Vol}_{X_6}$ the 4d dilaton. 

These expressions give a good approximation of the kinetic terms in regimes where both worldsheet and D2-brane instanton corrections can be ignored, so that the 4d EFT displays approximate shift symmetries on the fields $b^a$, $\zeta^K$, which are considered axions. Typically, one chooses $\om_a$ and $\alpha_K$ as harmonic forms such that $b^a$ and $\zeta^K$ have unit periodicity. Then, as  discussed in \cite[section 6.4]{Lanza:2021udy}, one can identify $\ell_s^{-2} [\om_a]$ as Poincar\'e duals to $\CR$-even four-cycles of $X_6$ wrapped by NS5-branes, and  $\ell_s^{-3}[\alpha_K]$ as duals to $\CR$-odd three-cycles wrapped by D4-branes. Both objects appear as strings in the 4d EFT, that couple magnetically to the axions. The tension of such strings is a function of the saxions $t^a$ and $u^K$, which are also the fields that describe the infinite-distance geodesics in the large-volume large-complex-structure regime of the compactification. 

Indeed, let us focus on 4d strings coming from D4-branes. Normalising the axion $\zeta^K$ to have unit period implies that $[\Sigma_K^-] = {\rm P.D.} [\ell_s^{-3} \a_K]$ is an $\CR$-odd three-cycle of $X_6$ that, when wrapped by D4-brane, yields a 4d string that sources this axion. If this 4d string is BPS, then its tension in units of the reduced Planck mass $M_{\rm P}$ is given by
\be
\ell_K \equiv - \oh \frac{\p K}{\p u^K} =  - \frac{1}{2\ch}  \int_{\Sigma_K^-} e^{-\phi} \im \Om\, ,
\label{dualsax}
\ee
where we have used results of \cite{Hitchin:2000jd}, that imply that $\ch$ is a function of the $u^K$ only. Following  \cite{Lanza:2021udy} we name the $\ell_K$ as dual saxions, and we choose $[\Sigma_K^-]$ to generate a subcone of 4d BPS strings dubbed {\em EFT strings}, which have the property that their tension only vanishes at  infinite distance boundaries of the moduli space of large complex structures. Near those loci the string is expected to induce a breakdown of the 4d EFT, via an infinite tower of oscillation modes similar to those of a tensionless critical string. 

\setlength{\belowcaptionskip}{0pt}
\begin{table}[htb]
\renewcommand{\arraystretch}{1.25}
\begin{tabular}{ll}
\hline
Non-Abelian gauge group & $\prod_\a SU(N_\a)$\\
Massless $U(1)$s &  $\sum_\a c_\a U(1)_\a$ such that $\sum_\a c_\a ([\Pi_{\a}] - [\Pi_{\a*}]) = 0$  \\
Chiral multiplets & $\sum_{\a<\b}\, I_{\a\b} ({\bf N}_\a, {\bf \bar{N}}_\b) \, + \, I_{\a\b*} ({\bf N}_\a, {\bf \bar{N}}_\b) $\\
& $\frac{1}{2} (I_{\a\a*} -  I_{\a {\rm O6}}) {\bf S}_\a\, + \, \frac{1}{2} (I_{\a\a*} +  I_{\a {\rm O6}}) {\bf A}_\a$\\
\hline
\end{tabular}
\caption{4d EFT chiral spectrum, in terms of the intersection number $I_{\a\b} = [\Pi_\a] \cdot [\Pi_\b]$. ${\bf S}_a$ and ${\bf A}_a$ stand for the symmetric and antisymmetric representations of $U(N_a)$. If $\Pi_\a = \Pi_{\a*}$, the D6-brane instead hosts either an $SO(2N_\a)$ or $USp(N_\a)$ gauge group.}
\label{t:specori}
\end{table}

The massless open string states determine the gauge sector of the 4d EFT. More precisely, the 4d chiral spectrum is specified by the topological data of the three-cycles $\Pi_\a$ wrapped by D6-branes, as summarised in table \ref{t:specori}. To describe the 4d gauge couplings it is useful to consider a basis of three-cycle classes $\{[\Sigma^J_+], [\Sigma_K^-]\}$, where $[\Sigma_K^-]$ are as above and $[\Sigma^J_+]$ is a basis $\CR$-even classes such that $[\Sigma_K^-] \cdot [\Sigma^J_+] = 2 \delta^J_K$. Expanding the three-cycles wrapped by the D6-branes on such a basis
\be
[\Pi_\a] = P_{\a\, J} [\Sigma^J_+] + Q_\a^K  [\Sigma_K^-] \, , \qquad [\Pi_{\a*}] = P_{\a\, J} [\Sigma^J_-] - Q_\a^K  [\Sigma_K^-] \, ,
\ee
one obtains that the gauge kinetic function associated to each stack is
\be
2\pi i f_{\a\a}  =  P_{\a\, K}  U^K \, ,
\label{faa}
\ee
so that the gauge coupling is $g_\a^{-2} = (2\pi)^{-1} P_{\a\, K}  u^K$. The Abelian sector of the gauge group is particularly interesting, as several $U(1)$ factors develop a Stu\"ckelberg mass as a byproduct of a generalised Green--Schwarz mechanism \cite{Aldazabal:2000dg,Ibanez:2001nd}. The piece of the 4d EFT Lagrangian describing such couplings is
\be
\oh M_{\rm P}^2 \int_{X_4} G_{KL} \left( d\zeta^K - 2 Q^K_\a A^\a\right) \wedge * \left( d\zeta^L - 2 Q^L_\b A^\b\right) \, ,
\label{StuLag}
\ee
where $G_{KL} = \frac{1}{2} \p_{u^K} \p_{u^L} K_Q$ and $A^\a$ represents the $U(1)_\a$ gauge boson. It follows that the $U(1)$ St\"uckeberg mass matrix reads \cite{Ghilencea:2002da}
\be
M_{\a\b}^2 = 4 G_{KL} Q^K_\a  Q^L_\b g_\a g_\b \,  M_{\rm P}^2 \, .
\label{masstu}
\ee
In terms of the K\"ahler potential, \eqref{StuLag} is encoded in the replacement $u^K \to u^K - Q^K_\a  V^\a$ in \eqref{KQ}, with $V^\a$ the vector multiplet associated to $U(1)_\a$. From here one can deduce that the chiral fields at D6-brane intersections enter a D-term potential with field-dependent Fayet-Iliopoulos terms \cite{Cremades:2002te}. In particular one finds
\be
V_D = \oh \sum_\a g_\a^2 \left(\xi_\a +  \sum_i q^i_\a K_{i\bar{i}} \Phi_i \bar{\Phi}_{\bar{i}} \right)^2    \, , \qquad \text{with} \quad \pi \xi_\a =   Q^K_\a\ell_K M_{\rm P}^2\, .
\label{VD}
\ee
Here $\Phi_i$ represents the scalar component of the 4d chiral field at two D6-brane intersections, with kinetic terms $K_{i\bar{i}}$ and charge $q^i_\a$ under $U(1)_\a$. The D-term-induced mass for fields at the intersections $\Pi_\a \cap \Pi_\b$ or $\Pi_\a \cap \Pi_{\b*}$ is\footnote{We have redefined the FI terms compared to other references like \cite{Cremades:2002te,Blumenhagen:2006ci}: $[\xi_\a]_{\rm here} = [\xi_\a/g_\a^2]_{\rm there}$.}
\be
q^i_\a g_\a^2 \xi_\a + q^i_\b g_\b^2 \xi_\b \, .
\label{massinter}
\ee

The special Lagrangian condition $-\im (e^{-\pi i \varphi_\a} \Om)|_{\Pi_{\a}} = 0$ with $\varphi_\a \in 2\Z$ implies $\xi_\a =0$, and so when imposed on all D6-branes the masses \eqref{massinter} vanish. As we move in complex structure moduli space $\{u^K\}$, the calibration phase $\varphi_\a$ will vary and we may generate either positive or tachyonic masses for the scalar fields at the intersections, describing walls of marginal stability  \cite{Kachru:1999vj}. Microscopically, one can understand such walls by zooming in near an intersections of two three-cycles $\Pi_\a$ and $\Pi_\b$, where they can be approximated by two special Lagrangian three-planes in flat space, with calibration phases $\varphi_\a$ and $\varphi_\b$. Because of this, their relative orientation can be described by a $U(3)$ rotation $(e^{\pi i\th_{\a\b}^1}, e^{\pi i\th_{\a\b}^2}, e^{\pi i\th_{\a\b}^3})$ such that $\th_{\a\b}^1 +  \th_{\a\b}^2 + \th_{\a\b}^3= \varphi_\b - \varphi_\a \mod \Z$. One can then apply the angle criterion of Nance and Lawlor \cite{Nance1987/88,Lawlor1989} to determine whether or not  it is energetically favourable for these three-cycles to recombine into one. When it is, there is a tachyonic mass term \eqref{massinter} for the field $\Phi_n$ at such intersection, which obtains a vev along the recombination process. The condition 
\be
\th_{\a\b}^1 +  \th_{\a\b}^2 + \th_{\a\b}^3 \in 2\Z\, ,
\label{susyang}
\ee
describes two three-planes related by an $SU(3)$, supersymmetric rotation \cite{Berkooz:1996km}. In this case both special Lagrangians three-cycles $\Pi_\a$ and $\Pi_\b$ have the same calibration phase and are mutually supersymmetric. 

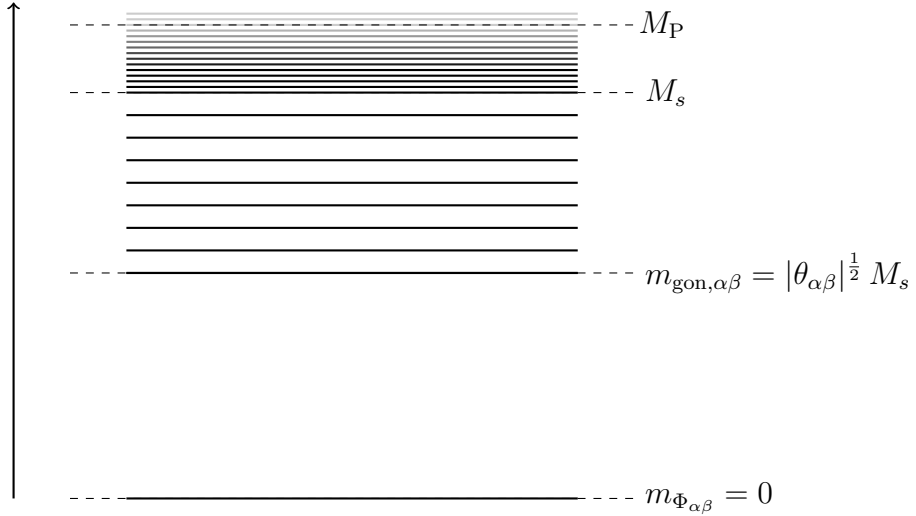
\begin{figure}[h]
    \centering
    \hspace{4.5em}
\begin{tikzpicture}[scale=1.50]
% Dibuja una línea horizontal en el centro
\draw[thick] (0,0) -- (4,0);
\draw[dashed] (-0.5,0) -- (4.5,0) node[right]{$m_{\Phi_{\a\b}}=0$};
\draw[dashed] (-0.5,2) -- (4.5,2)node[right]{$m_{{\rm gon},\a\b}=|\theta_{\a\b}|^{\frac{1}{2}}\,M_s$};
\draw[dashed] (-0.5,3.6) -- (4.5,3.6)node[right]{$M_s$};
\draw[thick] (0,2) -- (4,2);
\draw[thick] (0,2) -- (4,2);
\draw[thick] (0,2.2) -- (4,2.2);
\draw[thick] (0,2.4) -- (4,2.4);
\draw[thick] (0,2.6) -- (4,2.6);
\draw[thick] (0,2.8) -- (4,2.8);
\draw[thick] (0,3) -- (4,3);
\draw[thick] (0,3.2) -- (4,3.2);
\draw[thick] (0,3.4) -- (4,3.4);
\draw[thick] (0,3.6) -- (4,3.6);
% Ms scale
\draw[thick] (0,3.65) -- (4,3.65);
\draw[thick] (0,3.7) -- (4,3.7);
\draw[thick] (0,3.75) -- (4,3.75);
\draw[thick] (0,3.8) -- (4,3.8);
\draw[thick,black!90] (0,3.85) -- (4,3.85);
\draw[thick,black!80] (0,3.9) -- (4,3.9);
\draw[thick,black!70] (0,3.95) -- (4,3.95);
\draw[thick,black!60] (0,4) -- (4,4);
\draw[thick,black!50] (0,4.05) -- (4,4.05);
\draw[thick,black!40] (0,4.10) -- (4,4.10);
\draw[thick,black!30] (0,4.15) -- (4,4.15);
\draw[thick,black!20] (0,4.2) -- (4,4.2);
\draw[thick,black!20] (0,4.25) -- (4,4.25);
\draw[thick,black!20] (0,4.3) -- (4,4.3);
\draw[dashed] (-0.5,4.2) -- (4.5,4.2) node[right]{\hspace{-0.2em}$M_{\rm P}$};

%linea vertical
\draw[->,thick] (-1,0) -- (-1,4.4);
\end{tikzpicture}
\caption{The spectrum of string oscillator states close to an intersection at small angle $\theta_{\a\b}$.
There is a tower of gonions with the same charge as the massless mode $\Phi_{\a\b}$ and
characteristic mass $m_{{\rm gon},\a\b}$. The tower runs up to the string scale $M_s$, at which the full
Hagedorn string spectrum  with integer oscillators is recovered. 
\label{fig: gonionspectrum}}
\end{figure}
Remarkably, the angles $\th_{\a\b}^r$ not only determine the mass of the lightest scalar field at an intersection, but also the spectrum of an infinite tower of massive chiral multiplets that transform in the vector-like representation $({\bf N}_\a, {\bf \bar{N}}_\b) + ({\bf \bar{N}}_\a, {\bf N}_\b)$. If we consider the case $|\th_{\a\b}^r| \ll 1$, $r =1,2,3$, the lightest scalars of such a tower display the following mass spectrum \cite{Berkooz:1996km,Aldazabal:2000dg}
\be
m^2_{\a\b} (\text{scalar})_{\vec{k}, \pm r}  = 2\pi \left[\lambda_{\a\b}^{\vec k} \pm  |\th_{\a\b}^r |  \right] \,  e^{2\phi_4} M_{\rm P}^2 \, ,
\label{towergon}
\ee
which has been depicted in figure \ref{fig: gonionspectrum}, and where
\be
\lambda_{\a\b}^{\vec k} = \sum_r k_r |\th_{\a\b}^r | + \oh \sum_p  |\th_{\a\b}^p| \, , \qquad k_r \in \Z_{>0}\, . 
\ee
Such charged massive states were dubbed {\em gonions} in \cite{Aldazabal:2000cn}, and they form an infinite tower of states localised at D6-brane intersections that becomes particularly relevant in models with very small intersection angles. If the condition \eqref{susyang} is satisfied, then the $\oh$-spin states present the same mass spectrum, combining with the scalars into a tower of 4d $\CN=1$ chiral multiplets. Finally, besides for scalars and fermions, such a tower also exists for $W$-bosons:
\be
m^2_{\a\b} (\text{vector})_{\vec{k}}  = 2\pi \lambda_{\a\b}^{\vec k}  \,  e^{2\phi_4} M_{\rm P}^2 \, .
\label{towerWgon}
\ee
We see that in both cases we find a Kaluza--Klein-like tower of  states charged under $U({\bf N}_{\a})\times U({\bf N}_{\b})$.  As we reach the string scale $M_s = \ell_{s}^{-1}$  additional states appear, such that we recover a Hagedorn-like spectrum typical of a critical string, see figure \ref{fig: gonionspectrum}.
In fact, at this scale we also find states with higher 4d spin, also charged under the gauge group. For more details on the gonion spectrum see Appendix \ref{ap:spectrum} . 

Additional aspects of the 4d EFT are revealed from the dimensional reduction of the D6-brane action \cite{Grimm:2011dx,Kerstan:2011dy}. In particular, a smooth three-cycle $\Pi_\a$ has $b_1(\Pi_\a)$ special Lagrangian deformations and Wilson line moduli, that are perceived as adjoint fields.\footnote{Adjoint fields develop a superpotential generated by worldsheet instantons \cite{Kachru:2000ih,Kachru:2000an}, but the resulting masses are typically negligible at large compactification volumes. Additional superpotential terms are generated for those elements of $b_1(\Pi_\a)$ dual to non-trivial two-cycles of $X_6$ \cite{Marchesano:2014iea}. Such terms incorporate open and closed string moduli in the flux-induced scalar potential, as discussed in \cite{Carta:2016ynn,Herraez:2018vae,Escobar:2018tiu,Marchesano:2019hfb}.} At the 4d EFT level the D6-brane moduli redefine the bulk fields  $\{U^K\}$ \cite{Grimm:2011dx,Kerstan:2011dy,Carta:2016ynn,Herraez:2018vae}, such that the K\"ahler potential is modified and loses its factorised structure. The same feature is expected for the chiral fields localised at D6-brane intersections. Their kinetic terms $K_{i\bar{i}}$ should arise from the bulk K\"ahler potential $K = K_K + K_Q$ once the bulk fields have been properly redefined. Although in this case obtaining such a redefinition is much more involved, below we will propose an Ansatz based on the known results for physical Yukawa couplings, which we now turn to discuss.

\subsection{Yukawas in toroidal and Calabi--Yau orientifolds}
	
Besides the chiral spectrum and FI-terms, key quantities in  any 4d  chiral EFT are the physical Yukawa couplings among chiral fields. In the context of type IIA CY orientifolds with D6-branes, perturbative couplings are generated by worldsheet disc instantons connecting three D6-brane intersections. The result takes the form \cite{Aldazabal:2000cn,Cremades:2003qj}
\be
Y_{ijk} = h_{qu} \sum_{\vec{n}} d_{\vec{n}}\, e^{2\pi i \left[ \int_{D_{\vec{n}}} J_c - \int_{\p D_{\vec{n}}} \nu \right] }\, . % e^{2\pi i \vartheta (D_{\vec{n}})}\, .
\label{yukinter}
\ee
Here $D_{\vec{n}} \subset X_6$ is a holomorphic disc with boundary on the three cycles $\Pi_a$, $\Pi_b$, $\Pi_c$, and going through the intersections $i \in \Pi_a \cap \Pi_b$, $j \in \Pi_b \cap \Pi_c$, $k \in \Pi_c \cap \Pi_a$. The vector $\vec{n}$ indicates different elements $[D_{\vec{n}}]$ of the relative homology class $H_2(X_6; \Pi_a \cup \Pi_b \cup \Pi_c, \Z)$ with such boundary conditions, and $d_{\vec{n}}$ the multiplicity of each class. Finally, $\nu$ is the Wilson line background at each D6-brane and $h_{qu}$ is a contribution from fluctuations around the semiclassical worldsheet instanton solution \cite{Hamidi:1986vh,Dixon:1986qv}. In practice, one can compute this last quantity via CFT correlation functions \cite{Cvetic:2003ch,Abel:2003vv,Lust:2004cx,Bertolini:2005qh} or via dimensional reduction of mirror compactifications of magnetised D9-branes \cite{Cremades:2004wa,DiVecchia:2008tm}. The resulting overall scheme is that

\begin{itemize}

\item[-] The sum over worldsheet instantons depends on the K\"ahler moduli of the compactification and on the D6-brane adjoint moduli. 

\item[-] The prefactor $h_{qu}$ depends on the complex structure moduli, and in particular on the D6-brane intersection angles involved in the Yukawa coupling.

\end{itemize}

These statements should be taken with care, because of the field redefinitions that take place when translating the geometric description of the moduli space of the compactification into 4d supergravity variables. However, one can still get a precise picture by considering the explicit expressions for Yukawa couplings obtained in toroidal compactifications. Indeed, on $\otimes_{r=1}^3 ({\bf T}^2)_r$ naively one obtains a coupling of the form \cite{Cremades:2003qj,Cremades:2004wa}:
\be
Y_{ijk} = e^{\phi_4/2} \prod_{r=1}^3 \left( \im T^r\right)^{1/4} \left[ \Theta^{(r)}\right]^{1/4} e^{\oh H^{(r)}} W_{ijk}^{(r)}\, ,
\label{yukT6}
\ee
where
\be
\Theta^{(r)} = 2\pi \frac{\Gamma(1-|\chi_{ab}^r|)}{\G(|\chi_{ab}^r|)} \frac{\Gamma(1-|\chi_{bc}^r|)}{\G(|\chi_{bc}^r|)}  \frac{\Gamma(1-|\chi_{ca}^r|)}{\G(|\chi_{ca}^r|)}\, ,
\ee
is the piece of $h_{qu}$ that depends on the intersection angles. More precisely, $\chi_{\a\b}^r$ is the internal  twist felt in the $r^{\rm th}$ complex plane by the 4d scalar of the $\a\b$ sector   that is involved in the three-point coupling. For small intersection angles we either have $|\chi_{\a\b}^r| = |\th_{\a\b}^r|$ or $1 - |\th_{\a\b}^r|$, see Appendix \ref{ap:spectrum} for more details. Additionally 
\bea
\label{Whol}
W_{ijk}^{(r)} & = & \vartheta \left[ \begin{array}{c} \delta_{ijk}^r \\ 0 \end{array}  \right] \left( \nu^r, |I_{ab}^r I_{bc}^r I_{ca}^r| T^r \right)\, ,
 \\
H^{(r)} & = &  2\pi i \frac{\nu^r \im \nu^r}{\im T^r} |I_{ab}^r I_{bc}^r I_{ca}^r|^{-1}\, ,
\eea
encode the worldsheet instanton dependence on the K\"ahler and D6-brane moduli, with $\vartheta$ the Jacobi theta function, $I_{\a\b}^r$ the intersection number on $({\bf T}^2)_r$, $\nu^r = I_{ab}^r \nu_c^r + I_{bc}^r \nu_a^r + I_{ca}^r \nu_b^r$ a combination of open string adjoint moduli and $\d_{ijk}^r = \frac{i^r}{I_{ab}^r} + \frac{j^r}{I_{bc}^r} +\frac{k^r}{I_{ca}^r}$ scanning over the different intersections. We refer to \cite{Cremades:2003qj,Cremades:2004wa} for more details on these expressions. From them, and assuming the absence of D6-brane moduli, one can deduce the following structure for the physical Yukawas couplings in a general Calabi--Yau compactification
\be
Y_{ijk} =  e^{\phi_4/2} {\rm Vol}_X^{1/4} \, \Theta_{ijk}^{1/4}\,  W_{ijk} \, ,
\label{yukCY}
\ee
where $\Theta_{ijk}$ only depends on the intersection angles, which now can differ for each triplet $ijk$, and $W_{ijk}$ is a holomorphic function on the complexified K\"ahler and D6-brane position moduli. The intersection angles, and therefore  $\Theta_{ijk}$ are invariant under an overall rescaling of all saxions $u^K \to \lambda u^K, \forall K$, since this amounts to perform a 4d dilaton rescaling  $e^{-\phi_4} \to \lambda e^{-\phi_4}$, and leave the periods of $\Omega/[\int \Omega \wedge \bar{\Omega}]^{1/2}$ fixed. This implies that $\Theta_{ijk}$ is a homogeneous function of degree zero on the saxions $\{u^K\}$, and as a result the physical Yukawas are homogeneous functions of degree $-1/2$ on them.

As we will now discuss, the result  \eqref{yukCY} has a neat interpretation in terms of the K\"ahler metric for the matter fields at the intersection.

\subsection{Kinetic terms of chiral fields}
	 
One interesting byproduct of an expression like \eqref{yukCY} is that it gives valuable information on the kinetic term for the chiral fields at the D6-brane intersections. Indeed, assuming no kinetic mixing between such fields, one should be able to express the physical Yukawa coupling in terms of the usual 4d $\CN=1$ supergravity formula
\be
Y_{ijk} = e^{K/2} \left( K_{i\bar{i}}     K_{j\bar{j}}  K_{k\bar{k}} \right)^{-1/2} \, W_{ijk}\, . \label{eq: Yukawasugra}
\ee
Comparing with \eqref{yukCY} and using the above expressions for $K = K_K + K_Q$ one obtains 
\be
 K_{i\bar{i}}  K_{j\bar{j}}  K_{k\bar{k}} = e^{3(K/2-\phi_4)}  \Theta_{ijk}^{-1/2} \, .
\ee
That is, we recover a simple dependence on the K\"ahler potential, 4d dilaton and a function on the intersection angles. Let us consider the explicit result for $\Theta_{ijk}$ in toroidal compactifications and use that the matter field metrics should only depend on the angles at their own intersection. One arrives at the following result
\be
K_{i\bar{i}} = \frac{e^{K/2-\phi_4}}{\sqrt{2\pi}}   \prod_{r=1}^3 \left( \frac{\G(|\chi_{i}^r|)}{ \Gamma(1-|\chi_{i}^r|)}\right)^{1/2}\, .
\label{metricgen}
\ee
This implies that the K\"ahler metrics are functions of degree $-1$ on the saxions $\{u^K\}$.

To see how this expression connects with previous results in the literature, let us consider the case where $\th_{\a\b}^1, \th_{\a\b}^2 > 0$ and $\th_{\a\b}^3 < 0$. As it follows from the discussion in Appendix \ref{ap:spectrum}, the lightest 4d scalar in this sector corresponds to the twist vector $\chi_{\a\b} = (\th_{\a\b}^1, \th_{\a\b}^2, \th_{\a\b}^3+1, 0) = (|\th_{\a\b}^1|, |\th_{\a\b}^2|, |1-\th_{\a\b}^3|, 0)$. Hence its kinetic term is
\be
K_{i\bar{i}} =  \frac{e^{K/2-\phi_4}}{\sqrt{2\pi}}   \left[ \frac{\G(|\th_{\a\b}^1|)}{ \Gamma(1-|\th_{\a\b}^1|)}  \frac{\G(|\th_{\a\b}^2|)}{ \Gamma(1-|\th_{\a\b}^2|)}  \frac{\G(1-|\th_{\a\b}^3|)}{ \Gamma(|\th_{\a\b}^3|)} \right]^{1/2} \stackrel{|\th^r_{\a\b}| \ll 1}{\simeq} \left[ \frac{e^{K-2\phi_4}|\th_{\a\b}^3|}{2\pi |\th_{\a\b}^1||\th_{\a\b}^2|}\right]^{1/2}\, ,
\label{metricex}
\ee
where in the last equality we have assumed that $|\th_{\a\b}^r| \ll 1$ and $|\theta^3_{\a\b}| = |\theta_{\a\b}^1| + |\theta_{\a\b}^2|$.\footnote{If the last assumption is not true, one should add a factor $e^{-\gamma_E(|\theta_{\a\b}^1| + |\theta_{\a\b}^2|- |\theta^3_{\a\b}|)}$ in the expression for $K_{i\bar{i}}$, where $\g_E$ is the Euler-Mascheroni constant \cite{Lust:2004cx,Bertolini:2005qh}. As a result the rhs of \eqref{metricex} remains valid.} This agrees with several results in the literature \cite{Lust:2004cx,Font:2004cx,Bertolini:2005qh,Cremades:2004wa,Akerblom:2007uc,Billo:2007py,DiVecchia:2008tm,Camara:2009uv}, up to a possible additional dependence on the K\"ahler moduli that is irrelevant for the Yukawas. 

The presence of the 4d dilaton in \eqref{metricgen} will play an  important role in our discussion. To motivate this sort of dependence, let us consider two intersecting D6-branes in a smooth compact Calabi--Yau. As mentioned above, one expects the fields at D6-brane intersections to redefine the 4d supergravity variables that come from compactification moduli. If one considers the following Ansatz
\begin{equation}
    K = - 2\log \left({\rm Vol}_X^{1/2} {\cal H}\right) =-2 \log \left( \hat{\rm V}_X^{1/2}\hat{\cal H} - f( \Phi^i, \bar{\Phi}^{\bar{i}} ) \right) = -2 \log \left( \hat{\rm V}_X^{1/2}\hat{\cal H} - \oh k_{i\bar{i}} \Phi^i \bar{\Phi}^{\bar{i}} + \dots \right) ,
\end{equation}
where $\hat{\rm V}_X^{1/2}$, $\hat{\cal H}$ are the same functions as ${\rm Vol}_X^{1/2}$, ${\cal H}$, but depending on the redefined variables $\hat{T}^a$, $\hat{u}^K$, and in the last step we have performed an expansion in the intersection fields. If we perform the same expansion in the logarithm we find
\begin{equation}
     K = -2 \log \left(\hat{\rm V}_X^{1/2}\hat{\cal H}\right) + \frac{k_{i\bar{i}}}{ \hat{\rm V}_X^{1/2}\hat{\cal H}} \Phi^i \bar{\Phi}^{\bar{i}} + \dots 
\end{equation}
For vanishing vevs of the intersection fields we can identify $\hat{\rm V}_X^{1/2}\hat{\cal H}$ with $e^{-K/2}$, and so our result above translates into
\be
k_{i\bar{i}} = \frac{e^{-\phi_4}}{\sqrt{2\pi}}  \prod_{r=1}^3 \left( \frac{\G(|\chi_{i}^r|)}{ \Gamma(1-|\chi_{i}^r|)}\right)^{1/2}\, .
\label{littlek}
\ee
Notice that, besides the 4d dilaton dependence, the rest of the K\"ahler metrics depend on purely local data, namely the intersection angles. As emphasised before, local intersection data and the associated spectrum of states are independent on the global features of the compactification manifold. As such, we expect the result \eqref{littlek} to be also valid for intersecting D6-branes in smooth Calabi--Yau manifolds. 

In the following sections we will be interested in considering Yukawa couplings and K\"ahler metrics along limits of small intersecting angles. Let us first consider the case where all three intersection angles are small but non-vanishing. If as in \eqref{metricex} we take $\th_{\a\b}^1, \th_{\a\b}^2 > 0$ and $\th_{\a\b}^3 < 0$, then the matter field metric will look like the rhs of this equation. If we now impose the supersymmetry condition $|\th_{\a\b}^3| = |\th_{\a\b}^1| + |\th_{\a\b}^2|$ then either all angles will be of the same order of magnitude or one of them will be much smaller than the other two. In both cases the matter field metrics will go like $(\th_{\a\b}^{\rm min})^{-1/2}$ where $\th_{\a\b}^{\rm min} \equiv {\rm min} \{|\th_{\a\b}^r|\}$. From looking at the spectrum \eqref{towergon} we realise that this determines the mass scale of the lightest gonion tower at the intersection, which we denote as $m_{\rm gon}^i$. More precisely we find
\be
k_{i\bar{i}} \simeq \frac{M_{\rm P}}{m_{\rm gon}^i}\, .
\label{metricgon}
\ee
One can see that this result is valid for other sign choices of the angles, or even when not all of them are small. Indeed, the setup where two intersecting D6-branes have $|\th_{\a\b}^r| \ll 1$ is quite standard in D-brane compactifications, and it is for instance realised by the mirror of two magnetised D9-branes at very large volume. One may also consider the mirror of a magnetised D9-brane and a D5-brane \cite{Cremades:2004wa}, or that of two intersecting D7-branes with worldvolume fluxes \cite{Font:2004cx,Aparicio:2011jx}. In both cases in the limit of dilute fluxes one of the angles goes to zero, say $\th_{\a\b}^1 \to 0$, while the remaining two remain finite and become equal in magnitude $|\th_{\a\b}^2| - |\th_{\a\b}^3| \to 0$. Plugged into \eqref{metricgen} and taking the limit, one again recovers \eqref{metricgon}. Finally, when no angle is small, $m_{\rm gon}^{\rm min}$ is of the order of the string scale, and so \eqref{metricgon} is still a reasonable estimate of \eqref{littlek}. In section \ref{s:emergence} we will provide a derivation of this remarkably simple result from the viewpoint of emergence.

There are however exceptions to this rule, like when two D6-branes $\a$ and $\b$ do not intersect transversely, and one of the angles is identically zero. For instance, one may have two D6-branes intersecting over a one-cycle, as in the Pati--Salam example of section \ref{s:torito}. In that case there are only two gonion towers below the string scale, and the kinetic terms for the fields at the intersection are not determined by such gonion masses. While one can still recover the correct kinetic terms as a limit of the expression \eqref{metricgen}, it does not make sense to relate it with a (non-existent) gonion tower. Instead, it is more appropriate to express such a kinetic terms as
\be
K_{i\bar{i}} =   e^{K/2} h_{i\bar{i}} \prod_{r=1}^2 \left( \frac{\G(|\chi_{i}^r|)}{ \Gamma(1-|\chi_{i}^r|)}\right)^{1/2} \simeq   e^{K/2} h_{i\bar{i}}  \, ,
\label{metricexep}
\ee
where $h_{i\bar{i}}$ is a homogeneous function of degree one on the saxions $\{u^K\}$, and in the rhs we have used supersymmetry. In explicit examples one can see that $h_{i\bar{i}}$ satisfies
\be
h_{i\bar{i}} \lesssim g^{-1}_\a  g^{-1}_\b \, .
\label{hiiineq}
\ee
Further exceptions should arise from setups where the would-be lightest gonion tower is removed by some effect, like an orbifold projection or worldsheet instanton induced masses. In the mirror type IIB frame, \eqref{metricexep} is expected to apply for magnetised D7-branes wrapping the same four-cycle, like in \cite{Conlon:2008qi}, or for intersecting D7-branes with no chiral matter at their intersection \cite{Font:2004cx}. Note that in flat space \eqref{metricexep} occurs when two D-branes preserve a common $\CN=2$ supersymmetry. Abusing language, we  will dub all these exceptions as $\CN=2$ sectors, even when they host chiral matter. Notice that $K_{i\bar{i}}$ are always homogeneous functions of degree $-1$. 

Finally, notice that \eqref{littlek} is nothing but the K\"ahler metric for a chiral field, measured in units of the string scale $M_s$. Whenever $m_{\rm gon}^i/M_{\rm P} \to 0$, this metric blows up, and also does so in units of the lower, 4d EFT scale $\Lambda_{\rm EFT}$.  This indicates that one should decouple kinematically this chiral field from the 4d EFT, and that we are approaching a singular limit of the EFT. As we will argue in the next section, this sort of behaviour occurs whenever a Yukawa coupling \eqref{yukCY} vanishes asymptotically for a large complex structure limits. The vanishing Yukawa is accompanied by a vanishing D-brane gauge coupling, under which chiral matter at some intersection is charged. This triggers the appearance of a gonion tower at that intersection, preventing the EFT to reach such a singular, infinite distance limit.

%%%%%%%%%%%%%%%%%%%
%%%%%%%%%%%%%%%%%%%
	 
\section{The \texorpdfstring{$Y \to 0$}{Lg} limit and infinite distance}
\label{s:Ylimit}

In this section we explore the limits $Y \to 0$ for Yukawas of chiral matter fields. Along large complex structure limits, one finds simple expression for the Yukawa asymptotic behaviour in terms of gonion masses. In particular, if the Yukawa connects D6-brane intersections with three gonion towers each, one recovers the expression \eqref{yukigon}, which shows that a limit in which $Y \to 0$ is accompanied by a massless gonion tower. If the Yukawa involves an intersection with only two gonion towers, dubbed $\CN=2$ sector in section \ref{s:orientifolds}, the expression is slightly different. However, we argue that in all cases at least one of the gauge couplings related to the Yukawa vanishes asymptotically, which gives a rationale of why a gonion scale tends to zero. We do so by estimating the lightest gonion masses in terms of the gauge couplings of the D6-branes and the tensions of the 4d strings that they couple to via their FI-terms. We use this same estimate to discuss the different mass scales in simple complex structure limits.

\subsection{Small Yukawas and gonion masses}
\label{ss:smallYuk}	 
	 
Given the expression \eqref{yukCY} for the Yukawa couplings in Planck units, one may analyse those limits in which the different Yukawa couplings go to zero value. For this, let us distinguish their dependence on three different set of moduli:

\begin{itemize}

\item[-] The Yukawas depend on D-brane moduli in the adjoint representation of the gauge group, through the holomorphic piece $W_{ijk}$. As it is clear for the toroidal example \eqref{Whol} some Yukawas vanish at special vevs of the complexified D6-brane position moduli $\nu_r$ \cite{Cremades:2003qj}, at finite distance points in moduli space. The presence of such moduli is however unwanted in realistic constructions, so as above we will consider Yukawas arising from D-branes with no adjoint moduli. This can be achieved by considering D6-branes that wrap three-cycles $\Pi_\a$ with $b_1(\Pi_\a) =0$, or whose adjoint moduli are projected out via some orbifold action, as in \cite{Blumenhagen:2005tn}.

\item[-] The squared Yukawas depend on the K\"ahler moduli as:
\be
|Y_{ijk}|^2 = A\, {\rm Vol}_X^{1/2} |W_{ijk}|^2\, ,
\label{Yukkahler}
\ee
where $A>0$ is a function of the complex structure saxions $\{u^K\}$, that include the 4d dilaton and that we treat as an ${\cal O}(1)$ constant factor.  To reach the limit $|Y| \to 0$ one may consider either taking ${\rm Vol}_X \to 0$ or to grow some K\"ahler moduli such that $W_{ijk} \to 0$ exponentially fast. The first case is oftentimes obstructed when quantum corrections are taking into account. In the second case some entries of $Y_{ijk}$ may indeed vanish, but for this to be physically relevant the rank of the Yukawa tensor should decrease as well. Since such a rank has been shown to be a topological quantity in mirror setups \cite{Cecotti:2009zf,Cecotti:2010bp}, it is not clear that these are actual vanishing Yukawa limits. Moreover, even if \eqref{Yukkahler} vanished this would however not guarantee the vanishing of the actual Yukawa, because corrections coming from D-brane instantons \cite{Abel:2006yk,Blumenhagen:2007zk,Cvetic:2008hi,Ibanez:2008my,Blumenhagen:2009qh,Marchesano:2009rz,Anastasopoulos:2010hu} would not necessarily vanish along these limits. Finally, these large volume limits take us to a region of 10d strong coupling, and so to analyse then properly one should reconsider it in the context of M-theory. For these reasons we will not consider them in the following. 

\item[-] The squared Yukawas depend on the complex structure saxions as:
\be
|Y_{ijk}|^2 = B\, e^{\phi_4} \Theta_{ijk}^{1/2} \, ,
\label{Yukcpx}
\ee
where $B > 0$ is function of the K\"ahler moduli, that we treat as an ${\cal O}(1)$ numerical factor. If the Yukawa involves an $\CN=2$ sector, then one of the angles within $\Theta_{ijk}$ must be set to zero via the appropriate limit. In any case, $|Y_{ijk}|^2$ is always a homogeneous function of degree $-1$ on the saxions $\{u^K\}$, so taking the trajectory $u^K \to \lambda u^K, \forall K$ with $\lambda \to \infty$ makes $|Y_{ijk}|^2$ scale like $\lambda^{-1}$. At the same time, due to \eqref{faa} all D6-brane gauge couplings scale like $g_\a \sim \lambda^{-1/2} \sim  |Y|$. Therefore, asymptotically one recovers a large global symmetry, in agreement with the fact that this is an infinite distance limit. From general Swampland arguments, one expects the appearance of asymptotically massless  towers of particles charged under these gauge symmetries, namely gonion towers.  

Indeed, because this trajectory amounts to the weak coupling limit $e^{\phi_4} \to e^{\phi_4}/\lambda$ and to leave $\Theta_{ijk}$ constant, all gonion masses  scale like $m_{\rm gon} \sim M_s  \sim \lambda^{-1} M_{\rm P}$.\footnote{Notice that \eqref{yukCY} can be rewritten as $|Y_{ijk}|^2 = g_s \Theta_{ijk}^{1/2} |W_{ijk}|^2$, where $\Theta_{ijk}$ and $W_{ijk}$ only depend on the CY geometric moduli, and that limit amounts to take $g_s \to 0$.} Hence the gonion towers essentially act like that of a fundamental string, and we obtain the relation
\be
m_{\rm gon} \sim |Y|^2 M_{\rm P}\, .
\label{YWGCdil}
\ee
Clearly, $|Y|, g \to 0$ occurs at the same time as $m_{\rm gon}/M_{\rm P} \to 0$. From the results of \cite{Font:2019cxq,Lanza:2021udy}, one expects the presence of additional towers of Kaluza--Klein and winding modes with the same asymptotic behaviour as $m_{\rm gon}$ (see section \ref{ss:scales} for more details) together with bulk fundamental string excitations.  Thus, we find a tower spectrum that implements an emergent string limit, with the novelty that some string excitations are charged under the gauge group hosted by D6-branes. 

\end{itemize}

Let us now consider more involved trajectories in the space of saxions $\{u^K\}$, in which not all of them but only a subset is very large.\footnote{Considering small $u^K$ takes us away from our  regime of control. In that case D2-brane instanton effects become relevant and the corresponding axionic shift symmetries of \eqref{KQ} are no longer present.}  In such regions $e^{\phi_4} \ll 1$, and so naively the Yukawa couplings \eqref{yukCY} are very suppressed unless this is compensated by $\Theta_{ijk}$. In this sense it is useful to distinguish different cases:

\begin{itemize}

\item[{\it i)}] The Yukawa does not involve any $\CN=2$ sector. Then, using the same reasoning to derive \eqref{metricgon} we conclude that in these regions we have 
\be
|Y_{ijk}|^2 \simeq B\, e^{-2\phi_4}  \frac{m_{\rm gon}^i}{M_{\rm P}}  \frac{m_{\rm gon}^j}{M_{\rm P}}  \frac{m_{\rm gon}^k}{M_{\rm P}}  \, , 
\label{YWGCcpx}
\ee
where $m_{\rm gon}^i$ is the mass scale of the lightest gonion tower at the $i^{\rm th}$ intersection in $\Pi_a \cap \Pi_b$, etc. Just like \eqref{yukCY}, we expect this relation to hold in general Calabi--Yau compactifications, as long as we are in large complex structure regions. 

Since $e^{-\phi_4} \gg 1$ by assumption, any trajectory along which $|Y| \to 0$ implies that at least one gonion tower becomes massless in Planck units. Reciprocally,  because $e^{-\phi_4} m_{\rm gon}/M_{\rm P} <1$, if one gonion tower goes to zero in Planck units all the Yukawas in which such an intersection is involved will also vanish asymptotically. In section \ref{ss:Y0g0} we argue that then at least one gauge coupling under which this gonion tower is charged must also vanish asymptotically, signalling that we are approaching an infinite distance limit. 

In general, large complex structure limits of infinite distance imply that $e^{\phi_4} \to 0$ and all Yukawas of the form \eqref{YWGCcpx} vanish. To see their relation with the lightest gonion mass, let us consider a trajectory where the scalings satisfy
\be
\left( \frac{m_{\rm gon}^{i}}{M_{\rm P}}\right)^{1/w_i} \sim  \left( \frac{m_{\rm gon}^{j}}{M_{\rm P}}\right)^{1/w_j} \sim \left( \frac{m_{\rm gon}^{k}}{M_{\rm P}}\right)^{1/w_k} \sim  e^{\phi_4}\, , \quad w_i, w_j, w_k \geq 1\, .
\ee
Defining $m_{\rm gon}^{ijk} \equiv {\rm min} (m_{\rm gon}^i, m_{\rm gon}^j, m_{\rm gon}^k)$ and $w_{ijk} \equiv {\rm max} (w_i, w_j, w_k)$, one obtains
\be
m_{\rm gon}^{ijk} \sim |Y_{ijk}|^{\frac{2w_{ijk}}{w_i+w_j+w_k-2}} M_{\rm P}\, ,
\label{case2}
\ee
Notice that the maximum exponent in the rhs is $|Y|^2$, as in the case \eqref{YWGCdil}.

\item[{\it ii)}] One (possibly chiral) $\CN=2$ sector is involved in the Yukawa, say the intersections in $ i\subset \Pi_a \cap \Pi_b$. Then the Yukawa asymptotic behaviour is  captured by
\be
|Y_{ijk}|^2 \simeq B\, e^{-2\phi_4} h_{i\bar{i}}^{-1} \frac{m_{\rm gon}^j}{M_{\rm P}}  \frac{m_{\rm gon}^k}{M_{\rm P}} = B\, h_{i\bar{i}}^{-1}  \frac{m_{\rm gon}^j}{M_{s}}  \frac{m_{\rm gon}^k}{M_{s}} \, .
\label{YWGCN2}
\ee
In this case it is not true that the Yukawa coupling must necessarily vanish whenever $e^{\phi_4} \to 0$. Some of the large complex structure infinite distance trajectories will not modify neither $h_{i\bar{i}}$  nor the intersection angles involved in the Yukawa, and as such \eqref{YWGCN2} will remain invariant along the limit.  

A trajectory for which $|Y| \to 0$ implies that either $m_{\rm gon}/M_{\rm P} \to 0$ or $h_{i\bar{i}} \to \infty$. The first case is similar to the one above, while in the second one we also expect a vanishing gauge coupling due to \eqref{hiiineq}. Moreover, in the examples of section \ref{s:torito} we will identify $h_{i\bar{i}}^{-1} \simeq m_{\rm w}/M_{\rm P}$, where $m_{\rm w}$ is a winding scale for bifundamental matter (see ft. \ref{footnote:winding}). This suggests that there is always a gauge coupling and a tower of states charged under it that asymptotes to zero mass whenever $|Y| \to 0$. 

\item[{\it iii)}] If more than one $\CN=2$ sector is involved in the Yukawa, one may proceed as before by performing further replacements $m_{\rm gon}^j/M_{\rm P} \to h_{j\bar{j}}$. However, in concrete setups the corresponding Yukawas do not seem to be related to a chiral sector of the compactification, unless one generates chirality via an orbifold projection.\footnote{For instance, it was shown in \cite{Beasley:2008dc} that for D7-branes wrapping Hirzebruch or del Pezzo surfaces, Yukawas involving more than one $\CN=2$ sector vanish identically.} For this reason we will not consider this case in the following. 

\end{itemize}

To sum up, the expressions \eqref{YWGCcpx} and \eqref{YWGCN2} encode the possible relations between the lightest gonion towers and the Yukawa couplings along different limits. However, they do not explain how these two quantities depend on the compactification moduli, nor which other mass scales are relevant in this setup. One can gain some insight on both of these aspects by relating gonion masses to 4d string tensions, as we now discuss.

\subsection{Gonions, strings and monopoles}
\label{ss:monopole}

One difficulty in evaluating expressions for the Yukawa couplings like \eqref{yukCY} or \eqref{YWGCcpx} is that they depend on a very different set of data. On the one hand, the 4d dilaton ${\phi_4}$ is a function of the complex structure moduli vevs, which are global compactification data. On the other hand, the intersection angles and the gonion masses depend on local data at each D6-brane intersection. While gonion masses clearly depend on the complex structure fields, the dictionary between global and local data is very involved as soon as we consider non-trivial Calabi--Yau metrics. In spite of this, in the following we would like to propose an estimate for the lightest gonion mass at the set of intersections $\Pi_a \cap \Pi_b$, based on the properties of 4d strings and monopoles in this class of compactifications. This estimate will be later on used to discuss the spectrum of energy scales that appear at regions of small Yukawa couplings. 

\begin{figure}[ht!]
\begin{center}
\includegraphics[width=14cm]{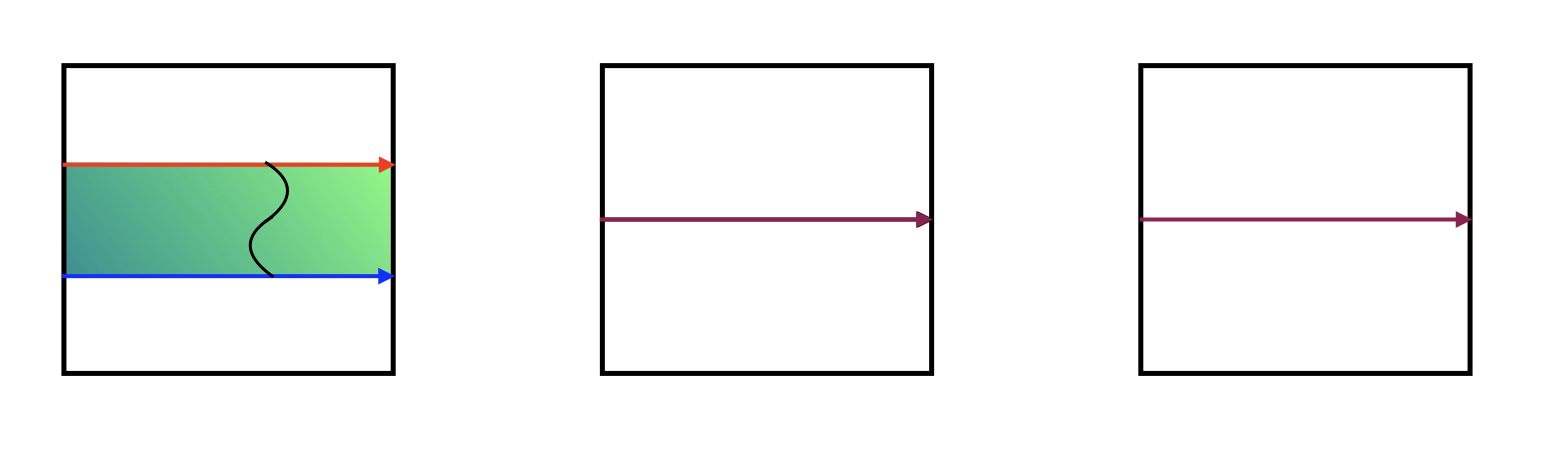}
\caption{Two D6-branes after the adjoint Higgssing $U(2) \to U(1) \times U(1)$. Electrically charged particles arise from open strings stretched between both D6-branes. Magnetic charges arise from D4-branes on four-chains connecting both three-cycles. 
\label{fig:parallel}}
\end{center}
\end{figure}

Let us consider a type IIA orientifold compactification with intersecting D6-branes, as described in section \ref{ss:interD6}. To each three-cycle $\Pi_\a$ wrapped by a D6-brane one can associate an Abelian gauge factor $U(1)_\a$, and some linear combinations of these $U(1)$'s remain massless after the St\"uckelberg mechanism has been implemented. As captured in table  \ref{t:specori}, the condition for a massless $U(1)$ is that the linear combination $c_\a [\Pi_\a]$ minus its orientifold image is a trivial homology class of \cite{Camara:2011jg}. This implies that such a combination of three-cycles can be connected by a four-chain $\Sigma_4$ wrapped by a D4-brane, which appears as a 4d monopole of this $U(1)$ factor \cite{Marchesano:2014bia}. The simplest case in which such a monopole appears is when one considers a stack of two D6-branes wrapping a three-cycle $\Pi$ with a position modulus $\nu$, which one  can use to perform an adjoint Higgssing $U(2) \to U(1)_a \times U(1)_b$ by giving a relative vev $\nu_a - \nu_b = \zeta \ell_s$ to the three-cycles. After the Higgssing, particles with electric and magnetic charges under the massless combination  $U(1) = \oh [U(1)_a - U(1)_b]$ appear, and for $\zeta \ll 1$ they satisfy
\be
m_{\rm mono} \simeq g^{-2} m_{\rm ele}\, ,
\label{rel}
\ee
as figure \ref{fig:parallel} illustrates for the case of flat space. 
\setlength{\belowcaptionskip}{-15pt}

\begin{figure}[ht!]
\begin{center}
\includegraphics[width=14cm]{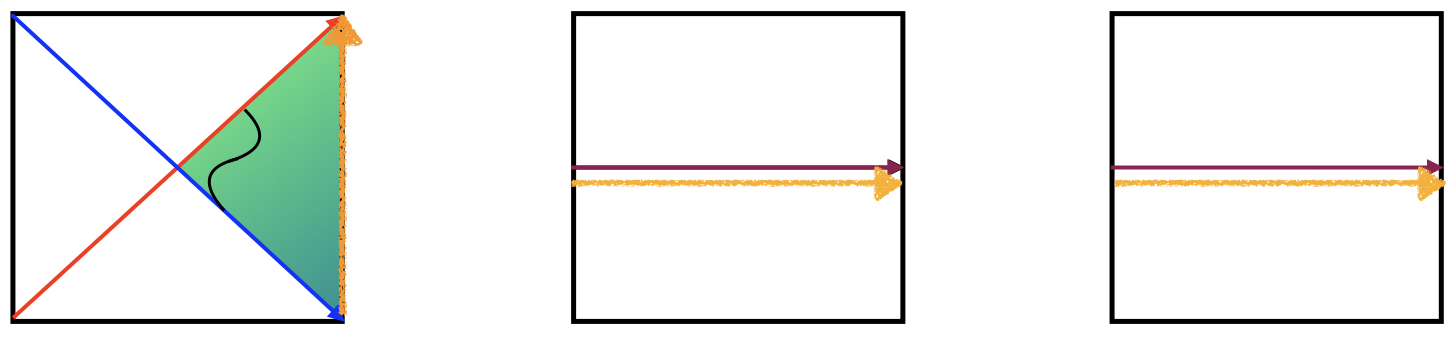}
\caption{Configuration of D6-branes intersecting at one angle $\theta$. Particles with bifundamental electric charge still arise from open strings stretched between both branes, including the gonion tower. However, now monopoles (D4-branes on the shaded green area) need to have 4d strings ending on them (D4-branes on the orange three-cycle).
\label{fig:massive}}
\end{center}
\end{figure}

We now consider the case of a $U(1)$ with a non-vanishing St\"uckelberg coupling. Now $[\Pi_a] \equiv \sum c_\a [\Pi_\a]$ will be a non-trivial three-cycle class in $H_3(X_6, \Z)$ (even after subtracting its orientifold image) and so there is no four-chain connecting the constituent three-cycles. To have the analogue of the previous setup we must first consider a D4-brane wrapping a three-cycle on the class $[\Pi_a] - [\Pi_{a*}]$, and then a second D4-brane on a four-chain $\Sigma_4$  that connects this D4 with the D6-branes wrapping $\Pi_a$ and $\Pi_{a*}$. This new configuration fits well with our field theory intuition regarding monopoles of $U(1)$'s with a 
$B \wedge F$ coupling \cite{Banks:2010zn}. The 4d monopole, which in our setup is represented by the D4-brane in $\Sigma_4$, is not a gauge invariant operator by itself, and to cure this  it must have a 4d string ending on it, represented by the D4-brane (or D4-branes) wrapping  $[\Pi_{\rm D4}] = [\Pi_a] - [\Pi_{a*}]$. See figure \ref{fig:massive} for a specific flat-space representation of this setup. 
\setlength{\belowcaptionskip}{-15pt}

One may now compare the mass of electrically and magnetically charged objects. In a 4d $\CN=1$ setting the mass of the monopole is not protected by supersymmetry,\footnote{See \cite{Casas:2023wlo} for BPS versions of these configurations, in $\CN=2$ settings.} 
 but the tension of the 4d string is, and we have the field theory relation between this tension and the mass of the open strings at the D6-brane intersection. For simplicity, let us consider a single D6-brane on $\Pi_a$ and its orientifold image on $\Pi_{a*}$. In a configuration like in figure \ref{fig:massive}, we have ${\rm Vol} (\Pi_{\rm D4}) = 2 \tan (\theta/2) {\rm Vol} (\Pi_{a})$, so one obtains the relation
\be
T_{\rm string} = 2\pi |\xi_a| \simeq g_a^{-2} |m_{\rm ele}^{2}|\, ,
\label{TFI}
\ee
where $\xi_a$ is the FI-term of $U(1)_a$. Here we assume a regime such that $\xi_a \ell_s^2 \ll 1$, which corresponds to a small intersection angle $\theta \ll 1$ and to the field theory approximation. Besides being the (tachyonic) mass of the lightest state, using \eqref{towergon} one can see that $m_{\rm ele}$ also corresponds to the scale of the gonion tower at the D6-brane intersection. We thus see that, in this case, the gonion scale reduces to $g_a T_{\rm string}^{1/2}$, which is a simple function of the complex structure saxions.

This configuration, however, is too simple to represent realistic D6-branes models. In particular, the 4d string that couples to the D6-branes corresponds to an EFT string, in the sense of \cite{Lanza:2020qmt,Lanza:2021udy,Lanza:2022zyg}. As mentioned above, this is a particular class of 4d BPS strings, whose tension only vanishes at infinite distance boundaries in moduli space, precisely where an approximate axionic shift symmetry becomes exact. As a result, the corresponding pair of D6-branes displays a non-vanishing FI-term at any point in moduli space, and $\Pi_a$ and $\Pi_{a*}$ are always non-mutually BPS, as indeed happens in the figure, where there is a single non-trivial angle and $m_{\rm ele}^{2}$ describes a tachyonic mode. 

In the intersecting D6-brane models of section \ref{ss:interD6}, the analogue of relation \eqref{TFI} is given by \eqref{VD} and \eqref{massinter}. To each D6-brane wrapping a three-cycle $\Pi_\a$ we associate the homology class  $[\Pi_\a] - [\Pi_{\a^*}] = 2 Q_\a^K [\Sigma_K^-]$, which corresponds to a 4d string charge. The tension $T_\a = 2\pi \xi_\a$ related to this string charge depends on the complex structure saxions $\{u^K\}$, and the functions $g_\a^2 T_\a$ determine the mass of the lightest open strings at the different D6-brane intersections through \eqref{massinter}. The dictionary with gonion towers is however no longer as simple as in figure \ref{fig:massive}, because now $g_\a^2 T_\a$ corresponds to a (signed) sum of EFT string tensions and of gonion scales, and cancellations can occur. 

Indeed, in the setup of section \ref{ss:interD6} we have chosen the three-cycles $[\Sigma_K^-]$ to generate the cone of EFT strings. The corresponding 4d strings are dubbed elementary EFT strings, and their tension is given by $T_K = \ell_K M_{\rm P}^2$. The 4d string tension associated to $\Pi_\a$ reads $T_\a  =  2Q^K_\a T_K$, and it cannot correspond to an EFT string tension because by assumption it vanishes at some locus in the interior of the moduli space.  Necessarily some $Q^K_\a$ must be negative, and so the D-flatness condition $T_\a = 2\pi \xi_\a =0$ for $\Pi_\a$ translates into a relation between elementary EFT string tensions. Additionally, the condition $g_\a^2 T_\a = g_\b^2 T_\b$ implies the supersymmetry relation \eqref{susyang} between the sum of intersection angles at $\Pi_\a \cap \Pi_\b$, while  $g_\a^2 T_\a = - g_\b^2 T_\b$ does the same for $\Pi_\a \cap \Pi_{\b^*}$. This suggests that the gonion masses at these intersection must depend on the complex structure saxions through the functions $g_\a^2 T_{K_\a}$, $g_\b^2 T_{K_\b}$, where ${K_\a}$ runs over the values such that $Q_\a^{K_\a} \neq 0$. In particular, if all these functions have a certain scaling along an infinite distance limit, we would expect such gonion masses to scale in the same way. 

In the following we would like to translate this intuition into a proposal for estimating the lightest gonion scale coming from  the intersections $\Pi_\a \cap \Pi_\b$, at asymptotic regions in moduli space. Let us first consider the case where $\Pi_\a \cap \Pi_\b$ corresponds to an $\CN=2$ sector, so there are only two intersection angles $\th_{\a\b}^1$ and $\th_{\a\b}^2$. For the three-cycle $\Pi_\a$ we split the subset of indices $K_\a \in K$ as $\{ K_\a\} = \{K_\a^+, K_\a^-\}$ such that  $Q^{K_\a^+}_\a > 0$ and $ Q^{K_\a^-}_\a <0$, and we do the same for $\Pi_\b$. We now define 
\be
m_{\a\b, +}^2 = g_\a^2 T_\a^+ -  g_\b^2 T_\b^-\, , \qquad m_{\a\b, -}^2 = -g_\a^2 T_\a^- +  g_\b^2 T_\b^+\, ,
\label{maxM}
\ee
where $T_\a^+ = \sum_{K_\a^+} Q^{K_\a^+}_\a T_{K_\a^+}$ and $T_\a^- = \sum_{K_\a^-} Q^{K_\a^-}_\a T_{K_\a^-}$, and similarly for $T_\b^\pm$.\footnote{An alternative definition would involve the decomposition $[\Pi_\a] -[\Pi_{\a^*}] =  2 \tilde{Q}^K_\a    [\tilde{\Sigma}_K^-]$, where $[\tilde{\Sigma}_K^-]$ is a basis of odd three-cycles hosting 4d BPS strings, in the particular region under consideration. From here one can define analogous quantities to $T_\a^\pm$, $m_{\a\b, \pm}^2$ and $m_{\a\b, {\rm min}}^2$. Because the cone of 4d BPS strings is larger than that of 4d EFT strings (see \cite{Lanza:2021udy} for examples), the former involves smaller tensions that vanish at finite-distance boundaries of the moduli space. It would be interesting to see whether this would give a more accurate gonion mass estimate.} Notice that $m_{\a\b, \pm}^2 > 0$ at any point in moduli space, and that at the supersymmetric locus we have that $m_{\a\b, +}^2 = m_{\a\b, -}^2 $, which implies $|\th_{\a\b}^1| = |\th_{\a\b}^2|$. As in \eqref{TFI}, if the intersection angles are small it is natural to identify $m_{\a\b, \pm}$ with the gonion scale at this intersection. 

The case of three intersection angles is more involved. Let us for instance consider the case where $|\th_{\a\b}^r| \ll 1$ and $|\theta^3_{\a\b}| = |\theta_{\a\b}^1| + |\theta_{\a\b}^2|$. Then one can identify $m_{\a\b, \pm}$ with the largest gonion scale of the three and, if they are all of the same order, it provides a good estimate of the lightest one as well. If, on the contrary, there is a hierarchy between min$(|\theta_{\a\b}^1|, |\theta_{\a\b}^2|)$ and $|\theta^3_{\a\b}|$ then  $m_{\a\b, \pm}$ only serves as an upper bound for the lightest gonion scale $m_{\rm gon}^{\rm min}$. One expects that such a scale is approximated by one of the summands that appear in either $m_{\a\b, +}^2$ or $m_{\a\b, -}^2$, although a priori it is not clear which one. However, one may set a lower bound on $m_{\rm gon}^{\rm min}$ by defining
\be
m^2_{\a\b, {\rm min}} = {\rm min}_K\, \left( \left| g_\a^2 Q_\a^{K} - g_\b^2 Q_\b^K \right| T_K  \right) , 
\label{minM}
\ee
with no summation over indices involved. In fact, because gonions are located at D6-brane intersections, one can propose a more precise prescription, in which the indices $K$ that appear in \eqref{minM} contribute to the intersection number $I_{\a\b} = 2 (Q_\a^K P_{\b\, K} - Q_\b^K P_{\a\, K})$. In practice, this amounts to require that either $P_{\a\, K}$ or $P_{\b\, K}$ is non-vanishing for a given value of $K$, or in 4d EFT terms that
\be
\nabla g_\a^{-2} \cdot \nabla T_K \neq 0\, , \qquad \text{and/or} \qquad \nabla g_\b^{-2} \cdot \nabla T_K \neq 0\, .
\label{refi}
\ee

To sum up, when at least one intersection angle is small, we estimate that the scale for the lightest gonion tower at the intersections $\Pi_\a \cap \Pi_\b$ lies in the range
\be
m_{\a\b, {\rm min}} \lesssim m_{\rm gon}^{\rm min} \lesssim m_{\a\b, \pm}\, .
\label{estimate}
\ee
Notice that both $m_{\a\b, {\rm min}}/M_{\rm P}$ and $m_{\a\b, \pm}/M_{\rm P}$ are homogeneous functions of degree $-1$ on the complex structure saxions $\{u^K\}$, just like $m_{\rm gon}^{\rm min}/M_{\rm P}$ should be.

In specific setups, one may give an even more precise prescription for this estimate. Let us for instance consider a region in which a single field $u^I \in \{ u^K\}_K$ is much larger than the rest. Then by the results of \cite{Lanza:2020qmt,Lanza:2021udy,Lanza:2022zyg} we have that $T_I \sim M_{\rm P}^2 /u_I$ in this region. If this EFT string appears in the decomposition $T_\a = Q_\a^K T_I$ and $g_\a^{2}$ depends on $u^I$, then we have that the definition \eqref{minM} gives $m_{\a\b, {\rm min}}/M_{\rm P} \sim 1/u^I$, which is the smallest possible value given the degree of this function. A typical setup where this occurs is in (the mirror of) type I compactifications with magnetised D9-branes, with $u^I = e^{-\phi} {\rm Vol}_X$ and $Q_\a^I = \frac{1}{6} \int_X F_\a \wedge F_\a \wedge F_\a$  \cite{Blumenhagen:2000wh,Cremades:2004wa,Blumenhagen:2005pm}. In this case it is known that the role of $T_I$ in $T_\a$ is to account for the $\alpha'$ corrections to the FI-term $\xi_\a$ \cite{Rabadan:2001mt}, and as such its presence is irrelevant for estimating the gonion masses. Therefore, in this kind of regimes $T_I$ should be excluded when computing the lower bound for the gonion scale, and the rhs of \eqref{minM} should be modified as ${\rm min}_{K} \left( \dots\right) \to {\rm min}_{K\neq I} \left( \dots\right)$.

\subsection{How \texorpdfstring{$Y \to 0$}{Lg} implies \texorpdfstring{$g \to 0$}{Lg}}
\label{ss:Y0g0}

In the following we would like to argue that, in the setup at hand, a large complex structure limit in which $Y_{ijk} \to 0$ implies the vanishing of a gauge coupling under which the chiral fields $i$, $j$ or $k$ are charged. Our reasoning will rely on the gonion mass estimate \eqref{estimate} and the analysis made in \cite{Lanza:2021udy} for type II orientifolds limits of the kind that we are considering in this work. 

We begin by considering the asymptotic expressions for the Yukawa couplings \eqref{YWGCcpx} and \eqref{YWGCN2}. We have that $Y \to 0$ implies either $h_{i\bar{i}} \to \infty$ or $m_{\rm gon}/M_{\rm P} \to 0$. As mentioned above, in the first case a gauge coupling should vanish due to the inequality \eqref{hiiineq}, which we will assume to hold for any $\CN=2$ sector. One may then turn to the second case, and in particular assume an asymptotically massless gonion tower that transforms in the bifundamental representation of $U(1)_\a \times U(1)_\b$. Then, using the estimate \eqref{estimate} one finds two non-exclusive options: 
\begin{itemize}

\item[a)] At least one of the two gauge couplings $g_\a^2$ or $g_\b^2$ goes to zero.

\item[b)] At least one of the string tensions $T_K$ in  \eqref{minM} goes to zero in Planck units.

\end{itemize}
In case a) we are done. To analyse case b) we will assume that a) does not happen, which implies $m_{\rm gon} \gtrsim T_{\rm D4}^{1/2}$. In the following we will argue that this is not a reasonable scale for a gonion tower with a vanishing Yukawa. 

Since by construction $T_{\rm D4}$ is the tension of a 4d EFT string, the framework considered in \cite{Lanza:2021udy} is particularly useful to evaluate this scale. In there, one considers one or several field directions $u^I \subset \{u^K\}$, growing at the same rate 
\be
u^I \sim \lambda \to \infty\, , 
\label{limituI}
\ee
while the remaining ones stay constant. These were dubbed EFT string limits in \cite{Lanza:2020qmt,Lanza:2021udy,Lanza:2022zyg} and they will be reviewed in more detail in the next subsection. In fact, the following analysis also applies to a more general class of limits, where those saxions that do not grow linearly with $\lambda$ are not constant, but instead bounded from above. We dub such limits as {\em quasi-EFT string} limits, see section \ref{s:torito} for an explicit example. 

The homogeneity of ${\cal H}$ in the complex structure saxions $\{u^K\}$ implies the following equality between them an their duals \eqref{dualsax}
\be
u^K \ell_K = 2\, .
\label{noscale}
\ee
As a result, when we grow a field direction as in \eqref{limituI} necessarily $\ell_I \sim 1/ \lambda$, and so the corresponding EFT string tension scales as $T_{{\rm D4}, I} = \ell_I M_{\rm P}^2 \sim M_{\rm P}^2/\lambda$. The reverse is not true, because there are many $\ell_K$ that vanish asymptotically when $u^I \to \infty$. In particular, in \cite{Lanza:2021udy} the number $p$ of dual saxions $\ell_K$ that asymptote to zero were called the degeneracy of the limit, $p=1$ corresponding to non-degenerate limits in which only $\ell_I \sim 1/\lambda$ and the rest of the dual saxions were asymptotically constant.  

By assumption, both gauge couplings $g_\a$ and $g_\b$ are independent of the growing direction \eqref{limituI}. If $u^K_{\a, \b}$ are the saxions on which these couplings depend, then due to the condition \eqref{refi}, the lightest gonion mass is determined by the duals $\ell_K^{\a,\b}$ of these saxions. As a result, in order to achieve a limit in which $m_{\rm gon}/M_{\rm P} \to 0$ with constant gauge couplings, we necessarily need a degenerate limit, because $\ell_I \notin  \ell_K^{\a,\b}$. We will assume that the tensions of the EFT strings degenerate with $T_{{\rm D4},I}$ also scale like $T_{\rm D4} \sim M_{\rm P}^2/\lambda$, although for milder scalings our reasoning works as well.

As mentioned before, along large complex structure limits the 4d dilaton always vanishes asymptotically. Hence one may use its asymptotic behaviour to classify 
the different limits. This is related to the asymptotic behaviour of the K\"ahler potential, which in terms of $\lambda$ reads
\be
K_Q  = - n \log \lambda + \dots 
\label{asymKQsig}
\ee
Here $n$ is typically an integer, which in the language of \cite{Grimm:2018ohb} corresponds to the singularity type. For us, due to the relation $K_Q = 4\phi_4$, it describes the asymptotic behaviour of the fundamental string scale:
\be
M_s \sim M_{\rm P} \lambda^{-n/4}\, . 
\label{asymMs}
\ee
In the following we analyse the gonion scale for  different values of $n$:
\begin{itemize}

\item[-] $n=1$

Here we have that $T_{{\rm D4}, I}^{1/2} \sim M_s/\lambda^{1/4} \ll M_s$, and it is the EFT string that sets the species scale along the limit. The orientifold limits analysed in \cite{Lanza:2021udy} suggest that this is an emergent string limit, and so by the Emergent String Conjecture there should be a unique emergent string, which tranlates into a non-degenerate limit. As mentioned above, this contradicts our assumption that  $m_{\rm gon}/M_{\rm P} \to 0$.

\item[-] $n=2$

In this case $T_I \sim M_s^2$. If the gauge couplings are constant and the limit is degenerate, we can have all gonions towers at the scale $M_s$. While this is a plausible scenario, it implies that the whole D-brane sector related to the Yukawa remains invariant along the limit, when measuring things in string units. In other words, the D-brane subsector $a$, $b$, $c$ are isolated from the rest of the compactification, where the infinite distance/gravity-decoupling trajectory happens. As a result, one should recover a 4d EFT with non-vanishing gauge couplings and Yukawas that decouples from gravity along the limit, as in the setups of \cite{Conlon:2006tj} and \cite{Marchesano:2023thx}. Then, applying an argument similar to that in \cite{Conlon:2006tj}, it would not make sense that the Yukawa is described by \eqref{YWGCcpx}, because one would then have $Y \sim e^{\phi_4} \to 0$ instead of a constant physical Yukawa. The only reasonable option seems that the Yukawa involves an $\CN=2$ sector, and so the expression to consider is \eqref{YWGCN2}. Then $Y \to 0$ implies that we should either have $h_{i\bar{i}} \to \infty$, which was already excluded, or  $m_{\rm gon}/M_{s} \to 0$, which contradicts our assumptions.

\item[-] $n>2$

In this case we find that $T_{{\rm D4},I} \sim \lambda^{n/2-1} M_s^2$, and so $M_s \ll T_I^{1/2}$. In this case applying \eqref{estimate} with constant gauge couplings would give us $m_{\rm gon} \sim \lambda^{\frac{n-2}{4}} M_s$, which does not make sense because $m_{\rm gon} \leq M_s$.

\end{itemize}

Finally, as we discuss in section \ref{s:emergence}, that a light gonion mass implies the vanishing of a gauge coupling can be motivated from the viewpoint of the Emergence Proposal. Indeed, in the presence of a gonion tower, the gauge couplings of the gauge bosons
that couple to them receive large contributions at one loop from the graphs in figures \ref{emergauge} and \ref{emerfermion}. The associated gauge kinetic functions
grow to infinity as the gonion masses go to zero.
In any event, it would be interesting to see if some of the working assumptions that we have used could be circumvented in some setup, and in particular if they are modified when one considers more involved limits for the growing fields, like the growth sectors of \cite{Grimm:2018cpv}. In the following we will aim to get more intuition on these and other limits of light gonion towers, by analysing the scale spectrum that emerges along them.

\subsection{Mass scales in simple limits}
\label{ss:scales}

To gain a better physical picture of the limits under discussion, in the following we discuss the relevant mass scales in a couple of simple setups. First, we consider an STU-like model that arises from freezing complex structure moduli with the vanishing FI-term condition \eqref{susyang}, which we analyse along the lines of \cite{Font:2019cxq}. Second, we comment on the mass scales that appear in the type IIA EFT string limits of \cite{Lanza:2021udy}. 

\subsubsection*{STU-like model}

To build a simple STU-like model one may consider a type IIA orientifold compactification with several complex structure moduli, and then impose that the FI-terms $\xi_\a$ for some D6-branes wrapping three-cycles $\Pi_\a$ vanish.\footnote{This requirement is imposed by hand, in order to maintain a non-trivial chiral spectrum along the trajectories that we take.} One then obtains a slice of the saxionic moduli space of much lower dimension, which we will assume to be parametrised by only two moduli: $s\equiv u^0$ and $u$.

To engineer such a setup and at the same time have some control over the compactification scale, we consider the type IIA dual of the type I and $SO(32)$-heterotic Calabi--Yau compactifications considered in \cite{Blumenhagen:2005pm,Blumenhagen:2005zg}, which  in Appendix \ref{ap:mirror} we translate to the language of section \ref{s:orientifolds}. In particular we consider a compactification with only two complex structure moduli, so that we have
\be
e^{-\phi} \re \Omega = s \a_0 + u^1 \a_1 + u^2 \a_2\, , 
\ee
and a K\"ahler potential that reads
\be
K_Q = - \log s - \log \kappa(u^1,u^2) \, ,
\label{KQSTU}
\ee
where
\be
\kappa(u^1,u^2) \equiv \frac{1}{6} \left[ \kappa_{111} (u^1)^3 + 3\kappa_{112} (u^1)^2 u^2 + 3\kappa_{122} u^1 (u^2)^2 + \kappa_{222} (u^2)^3 \right]\, ,
\ee
with $\kappa_{ijk} \geq 0$. The dual saxions then read 
\be
\ell_0 = \frac{1}{2s}\, , \qquad \ell_1 = \frac{\p_{u^1}\kappa(u^1,u^2)}{2\kappa(u^1,u^2)}\, , \qquad \ell_2 = \frac{\p_{u^2}\kappa(u^1,u^2)}{2\kappa(u^1,u^2)}\, . 
\label{dualsSTU}
\ee

We now introduce a single D6-brane wrapping a three-cycle $\Pi_\a$ with homology class 
\be
[\Pi_\a] =  \oh [A] + \oh f_\a^i [D_i] - \left( \oh \kappa_{ijk} f_\a^j f_\a^k + \frac{1}{24} c_{2\, i} \right)[C^i] + \left(\frac{1}{6}  \kappa_{ijk} f_\a^i f_\a^j f_\a^k + \frac{1}{24} c_{2\, i} f_\a^i  \right)[B]\, ,
\label{homoSTU}
\ee
where $[B] = {\rm P.D.} [\a_0]$, $[D_i] = 2{\rm P.D.} [\a_i]$, $i=1,2$, and $[A]\cdot[B]= 2$, $[C^i] \cdot [D_j] = 2\delta^i_j$. Here $f_\a^i \in \Z$ specify the D6-brane wrapping numbers and are mirror to the magnetic worldvolume flux quanta of a D9-brane, while $c_{2\, i}$ are fixed by the topology of the mirror manifold $Y$. The orientifold image of this D6-brane wraps a three-cycle whose homology class $[\Pi_{\a^*}]$ is obtained by performing the sign flip $f_\a^i \to - f_\a^i$ in \eqref{homoSTU}. The gauge coupling of this D6-brane is given by
\be
\frac{2\pi}{g_\a^2} = s - \left( \oh \kappa_{ijk} f_\a^j f_\a^k + \frac{1}{24} c_{2\, i} \right) u^i\, ,
\label{gaugecSTU}
\ee
while the condition for a vanishing Fayet-Iliopoulos term is
\be
\ell_i f^i_\a = \frac{1}{2s} \left(\frac{1}{6}  \kappa_{ijk} f_\a^i f_\a^j f_\a^k + \frac{1}{24} c_{2\, i} f^i_\a \right) \, .
\label{FISTU}
\ee
Clearly, imposing this condition removes one of the three saxion moduli $\{s, u^1, u^2\}$ of the model. In the regime of control of the mirror type I setup we have the hierarchy $s \gg u^i \gg f^i_\a$, which we will also assume in our type IIA setting. From there one can see that neglecting the rhs of \eqref{FISTU} is a good approximation of the vanishing FI-term condition. From there one sees that $f^1_\a$ and $f^2_\a$ must be of opposite sign and that, unless there is a hierarchy between them or the $\kappa_{ijk}$ we have that $u^1 \simeq u^2 \simeq u$. Moreover, given a solution $(\bar{u}^1, \bar{u}^2)$ of the approximate D-flatness equation, the rescaling $\bar{u}^i \mapsto \lambda \bar{u}^i$ provides a new one. When we restore the rhs of \eqref{FISTU} this scaling symmetry is no longer there, but given the hierarchy $s \gg u^i \gg f^i_\a$ the difference is going to be subleading. Hence, for our purposes, we can assume that after imposing \eqref{FISTU} we have two saxions $s$ and $u$, whose only constraint is $s \gg u$. 

\setlength{\belowcaptionskip}{0pt}
\setlength{\arrayrulewidth}{0.2mm}
\renewcommand{\arraystretch}{1.3}
\begin{table}[h]
\begin{center}
\begin{tabular}{|c|c|}
\hline
Scale & value in $M_{\rm P}^2$ units  \\
\hline\hline
$M_s^2$ & $1/\sqrt{su^3}$ \\
\hline
$T_{\rm D4}$ & $1/s$,  $1/u$\\
\hline
$m_{\rm gon}^2$ & $1/su$ \\
\hline
$m_{\rm KK}^2$ or $m_{\rm w}^2$ & $1/su$ \\
\hline
$m_{\rm Stu}^2$ & $1/s u^2$ \\
\hline
\end{tabular}
\end{center}
\caption{Scales of the STU-like model in Planck units. \label{tab:scales}}
\end{table}

The different mass scales of this model have been collected in table \ref{tab:scales}, in terms of $s$ and $u$. The string scale $M_s = e^{\phi_4} M_{\rm P}$ can be obtained from $K_Q = 4\phi_4$ and \eqref{KQSTU}. The tension of 4d strings made up of D4-branes wrapping odd three-cycles can be computed from $T_K = \ell_K M_{\rm P}^2$ and  \eqref{dualsSTU}. The gonion mass scale $m_{\rm gon}$ refers to the intersections of $\Pi_\a$ with its orientifold image $\Pi_{\a^*}$, and can be estimated via \eqref{minM}. Since $s \gg u$ by assumption, one must not include $K=0$ in \eqref{minM}, and so the two bounding scales $M_{\a\a^*}$ in \eqref{estimate} coincide, with $g_{\a}^{2} = g_{\a^*}^{2} \simeq 1/s$ and $|T_\a^\pm| \sim T_u \sim M_{\rm P}^2/u $. Alternatively, one may compute the gonion scale by going to the mirror type I frame. There, in the approximation of slowly varying field strengths  the gonion masses can be estimated by $f_a^i/A_i$, where $A_i$ is some area of the mirror manifold $Y$. Therefore, they coincide with a mirror Kaluza--Klein scale $m_{\rm KK}(Y)$, which given the isotropy of areas induced by (the mirror of) the FI-term condition \eqref{FISTU} can be estimated as $m_{\rm KK}^2(Y) \sim {\rm vol}_Y^{-1/3} = \sqrt{u/s} M_s^2 = M_{\rm P}^2/su$, where we have applied the mirror map to express the result in type IIA variables. In the type IIA frame $m_{\rm KK}(Y)$ translates into a KK or winding scale, also displayed in table \ref{tab:scales}. Finally, the St\"uckelberg mass can be easily estimated from \eqref{masstu}, and using that $\xi_\a =0$:
\be
M_{\a\a}^2 \sim g_\a^2  Q_\a^K Q_\a^L \frac{\p_K \p_L (s\kappa) }{s\kappa} \, M_{\rm P}^2 \sim \frac{M_{\rm P}^2}{su^2} \, .
\label{StuckSTU}
\ee
 and it is the lowest of all scales. 

Additionally, one may study the dependence of Yukawa couplings on complex structure saxions. With only two D6-branes there are no Yukawas couplings, but one may add an additional D6-brane of the form \eqref{homoSTU} (mirror to a magnetised D9-brane) or a D6-brane on the class $[C^i]$, mirror to a D5-brane. In both cases there are no $\CN=2$ sectors, and so one may estimate the Yukawa coupling either using \eqref{YWGCcpx} or by performing a computation along the lines of \cite{Cremades:2004wa,DiVecchia:2008tm} in the mirror frame. One then obtains 
\be
|Y|^2 \sim 1/s \, . 
\label{YukSTU}
\ee
Notice that, because the gauge couplings of D9-branes are given by \eqref{gaugecSTU}, one has the relation $|Y| \sim g_\a$, which corresponds to $r=1/2$ in \eqref{Yrelu} and \eqref{Yrelg}.

\begin{figure}[ht!]
\begin{center}
\includegraphics[width=10cm]{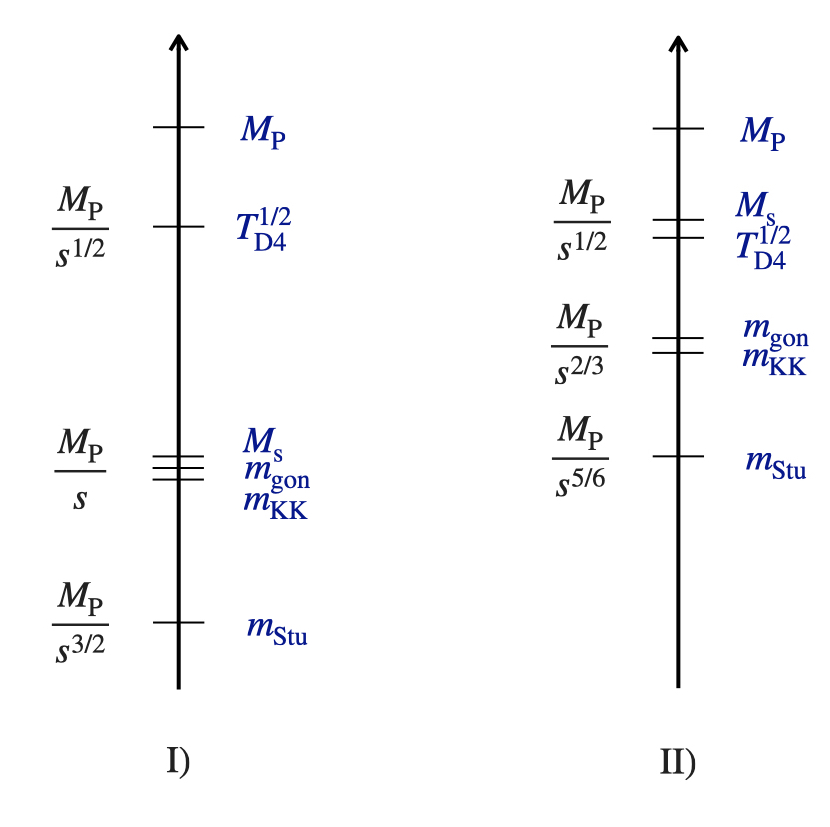}
\caption{Scales in both limits of the STU-like model. Here $m_{\rm KK}$ stands for either a Kaluza--Klein or a winding scale. 
\label{fig:scales}}
\end{center}
\end{figure}

Let us now consider a couple of scaling limits which preserve the hierarchy $s \gg u$:
\be
\text{I)}\quad s \sim u \sim \lambda\, , \qquad  \text{II)}\quad s \sim u^3 \sim \lambda \, .
\ee
The first limit corresponds to scale the 4d dilaton as $e^{\phi_4} \to e^{\phi_4}/\lambda$ without changing the periods of $\Omega$, and the second to the scaling of the mirror volume ${\rm vol}_Y \to \lambda \, {\rm vol}_Y$ without rescaling the 4d dilaton. We see that in the first limit the scaling relations are:
\be
\text{I)}\quad  \frac{m_{\rm Stu}}{M_{\rm P}} \sim \left( \frac{m_{\rm KK}}{M_{\rm P}}\right)^{3/2}\hspace{-.4cm} \sim  \left(\frac{m_{\rm gon}}{M_{\rm P}}\right)^{3/2} \hspace{-.4cm} \sim  \left(\frac{M_s}{M_{\rm P}}\right)^{3/2}  \hspace{-.4cm} \sim  \left(\frac{T_{\rm D4}^{1/2}}{M_{\rm P}}\right)^{3}  \hspace{-.4cm} , \quad m_{\rm gon} \sim |Y|^2 M_{\rm P}\, ,
\label{limitI}
\ee
while in the second one we have instead 
\be
\text{II)}\  \frac{m_{\rm Stu}}{M_{\rm P}} \sim \left( \frac{m_{\rm KK}}{M_{\rm P}}\right)^{5/4}  \hspace{-.4cm} \sim  \left(\frac{m_{\rm gon}}{M_{\rm P}}\right)^{5/4} \hspace{-.4cm}  \sim  \left(\frac{T_{\rm D4}^{1/2}}{M_{\rm P}}\right)^{5/3}  \hspace{-.4cm} \sim \left(\frac{M_s}{M_{\rm P}}\right)^{5/3}  \hspace{-.4cm} , \quad m_{\rm gon} \sim |Y|^{4/3} M_{\rm P}\, .
\label{limitII}
\ee
In both cases we have taken $T_{\rm D4} \sim M_{\rm P}^2 /s$ since it corresponds to the lightest 4d string scale. Notice that the Yukawa relation in \eqref{limitI} coincides with \eqref{YWGCdil}, as expected, while the one in \eqref{limitII} coincides with \eqref{case2} when one set $w_i = w_i = w_k = 4/3$. The generated hierarchy of scales in both kinds of limits has been depicted in figure \ref{fig:scales}. Notice that, even when they scale similarly, $m_{\rm KK} \leq m_{\rm gon}$ because $|f_\a^1|^{1/2} \geq 1$ contributes to the gonion mass, and $m_{\rm gon} < M_s$ because of the hierarchy $s \gg u$. Notice also that in both cases the St\"uckelberg-induced mass is parametrically below the Kaluza--Klein scale.\footnote{A similar hierarchy between KK and St\"uckelberg scales was observed in \cite{Lee:2019tst}.} Above this scale we should still have a 4d EFT in which the massive $U(1)$'s become  dynamical gauge fields. Despite the 4d chiral anomalies associated to some of them, we still have a consistent EFT thanks to the Green--Schwarz mechanism. 

\setlength{\belowcaptionskip}{-5pt}

\subsubsection*{EFT string limits}
 
Another interesting set of limits are the EFT string limits of \cite{Lanza:2020qmt,Lanza:2021udy,Lanza:2022zyg}. Applied to the  context of complex structure saxions of type IIA orientifolds, they arise from the backreaction of 4d string coming from certain D4-branes, and they essentially correspond to take the trajectory $u^K = u^K_0 + q^K \lambda$, with $q^K \in \bN$ and $\lambda \to \infty$, in the moduli space of complex structure saxions, see \cite[section 6.4]{Lanza:2021udy} for more details. Since the K\"ahler potential takes the form \eqref{KQ} where $\ch$ is a homogeneous polynomial of degree two on the saxions $\{ u^K\}$, the value of any dual saxion $\ell_K$ for which $q^K \neq 0$ scales as $\lambda^{-1}$ along the limit. In particular, the  tension of the 4d string sourcing this limit is given by $T_{\rm D4} = q^K \ell_K M_{\rm P}^2$, and along the limit it decreases as $T_{\rm D4} \sim M_{\rm P}^2/\lambda$. A remarkable feature of these limits is that they seem to satisfy the Integral Weight Conjecture, which states that along EFT limits the scale $m_*$ of lightest tower of states scales as \cite{Lanza:2021udy,Lanza:2022zyg}
\be
\frac{m_*^2}{M_{\rm P}^2} \sim \left( \frac{T_{\rm D4}}{M_{\rm P}^2}\right)^w ,
\label{IWC}
\ee
where $w=1,2,3$ is called the scaling weight. In \cite{Lanza:2021udy} several instances of complex structure limits in type IIA orientifolds and of mirror K\"ahler limits in type I and type IIB orientifolds were analysed, finding that only the weights $w=1$ and $w=2$ were realised, with $m_*$ corresponding to either a KK or a winding scale. In the following we will analyse these two possibilities. 

An important restriction to keep in mind is that, given a pair of D6-branes wrapping $\Pi_\a$ and $\Pi_\b$, we are interested in those limits such that the FI-terms $\xi_\a$ and $\xi_\b$ remain zero along them, or else remain small enough such that one can attain a supersymmetric locus by adjusting the non-growing saxions of the compactification. When applied to EFT string limits, this favours linear combinations of dual saxions $Q_\a^K \ell_K$ and $Q_\b^K \ell_K$ that are essentially independent of $\lambda$. The simplest way in which this can happen is if all the dual saxions in such linear combination scale the same along the limit, or in other words if the tensions of all 4d strings that couple to $\Pi_\a$ and $\Pi_\b$ scale similarly along $\lambda \to \infty$. In the following we will assume that this is the case. Notice that, as a result, the upper and lower bounds in \eqref{estimate} coincide, and we have a much sharper estimate for the gonion masses. 

Let us first consider $w=1$ limits, where we have:
\be
\frac{m_{\rm KK/w}}{M_{\rm P}} \sim \frac{T_{\rm D4}^{1/2}}{M_{\rm P}} \sim 1/\sqrt{\lambda}\, .
\label{scalingw1}
\ee
Based on the cases analysed in \cite{Lanza:2021udy}, one expects these limits to be non-degenerate, in the sense that along the limit $T_{\rm D4}$ is the smallest 4d string tension, and that there is no other EFT string with similar or faster scaling. This non-degeneracy is also motivated by the Emergent String Conjecture \cite{Lee:2019wij}, since $w=1$ limits are expected to correspond to emergent string limits. One also expects them to be elementary \cite{Marchesano:2023thx}, meaning that $q^K$ has a single non-vanishing entry and there is only one dual saxion that scales as fast as $1/\lambda$, say $\ell_I$. Then, because there is always at least two dual saxions involved in the FI-terms, our working assumption to keep $\pi \xi_{\a/\b} =  Q^K_{\a/\b} \ell_K$ vanishing along the limit implies that $Q_{\a/\b}^I =0$ and so the dual saxions in the sums $Q_{\a/\b}^K \ell_K$ scale slower than $1/\lambda$. Based on the examples of \cite{Lanza:2021udy}, we in particular expect them to be asymptotically constant. This implies that the string tensions in \eqref{maxM} and \eqref{minM} scale like $M_{\rm P}^2$, and so the gonion masses are determined by the scaling of gauge couplings along the limit.

 In the following we would like to argue that either $g_\a^2 \sim 1/\lambda$ or $Q_\a^K =0$, $\forall K$, so that $\xi_\a \equiv 0$ identically, and the same for $g_\b^2$, $\xi_\b$. The reasoning again stems from the Emergent String Conjecture, which suggests that the D4-brane sourcing the limit should be dual to a fundamental heterotic string on a smooth Calabi--Yau compactification, and the growing saxion dual to the heterotic 4d dilaton. In this frame, the D6-branes wrapping $\Pi_\a$ and $\Pi_\b$ should become $U(n)$ bundles, as in \cite{Blumenhagen:2005pm,Blumenhagen:2005zg}, or alternatively one of them may become a space-time filling NS5-brane.\footnote{Due to the assumption of chirality, it cannot be that both $\Pi_\a$ and $\Pi_\b$ are dual to NS5-branes.} In the former case the 4d dilaton dependence of the heterotic bulk gauge group gives $g \sim 1/\sqrt{\lambda}$, and in the latter $g$ is independent of the 4d dilaton but there is no FI-term. When plugged into \eqref{maxM} or \eqref{minM} all this results into
\be
\frac{m_{\rm gon}^{\rm min}}{M_{\rm P}} \sim 1/\sqrt{\lambda}\, ,
\label{mgonw1}
\ee
which coincides with \eqref{scalingw1}. More precisely, one would expect that $m_{\rm KK/w} \leq m_{\rm gon}^{\rm min} \leq T_{\rm D4}^{1/2}$, as it happens in the STU-like model. 

Let us now turn to $w=2$ limits, where 
\be
\frac{m_{\rm KK/w}}{M_{\rm P}} \sim \frac{T_{\rm D4}}{M_{\rm P}^2} \sim 1/\lambda\, .
\label{scalingw2}
\ee
Typically, these limits are partially degenerate (see e.g. \cite[Appendix B]{Marchesano:2023thx}). As a result one may consider two simple options to build FI-terms independent of $\lambda$. Either they are made of 4d EFT strings whose tension scale {\it i)}  like $T_{\rm D4} \sim M_{\rm P}^2/\lambda$ or {\it ii)}  slower. In the second case we assume that their tension is asymptotically constant in $M_{\rm P}^2$ units. 

In case {\it i)} we have two possibilities: either $g_\a^2 \sim 1/\lambda$ or it is independent of this EFT string limit. In the first case the contribution $g_\a^2T_\a^{+}$ in \eqref{maxM} to the squared gonion mass goes like $M_{\rm P}^2/\lambda^2$, and in the second it goes like $M_{\rm P}^2/\lambda$. In case {\it ii)} we expect things to work like for $w=1$ limits and the gauge coupling to scale like $g_\a^2 \sim 1/\lambda$, since otherwise we would obtain a region in which $m_{\rm gon}^{\rm min} \gg M_s$, which would be inconsistent. Therefore in this second case the contribution $g_\a^2T_\a^{+}$ to the squared gonion mass also goes like $M_{\rm P}^2/\lambda$. Unless the FI-term for the D6-brane wrapping $\Pi_\b$ is trivial, the same analysis applies to the contribution $-g_\b^2 T_\b^-$ to the squared gonion mass, and the final scaling will be the larger of the two. We thus recover two generic scalings:
\be
m_{\rm gon}^{\rm min} \sim M_{\rm P}/\lambda\, , \qquad \text{or} \qquad m_{\rm gon}^{\rm min} \sim M_{\rm P}/1\sqrt{\lambda}\, .
\ee
In the first case the gonion tower scales like the lightest KK or winding modes, and in the second it scales like the lightest 4d EFT string. Note that this result is again compatible with the ordering $m_{\rm KK/w} \leq m_{\rm gon}^{\rm min} \leq T_{\rm D4}^{1/2}$ of the STU-like model.

\setlength{\belowcaptionskip}{-15pt}
\begin{figure}[ht!]
\begin{center}
\includegraphics[width=15cm]{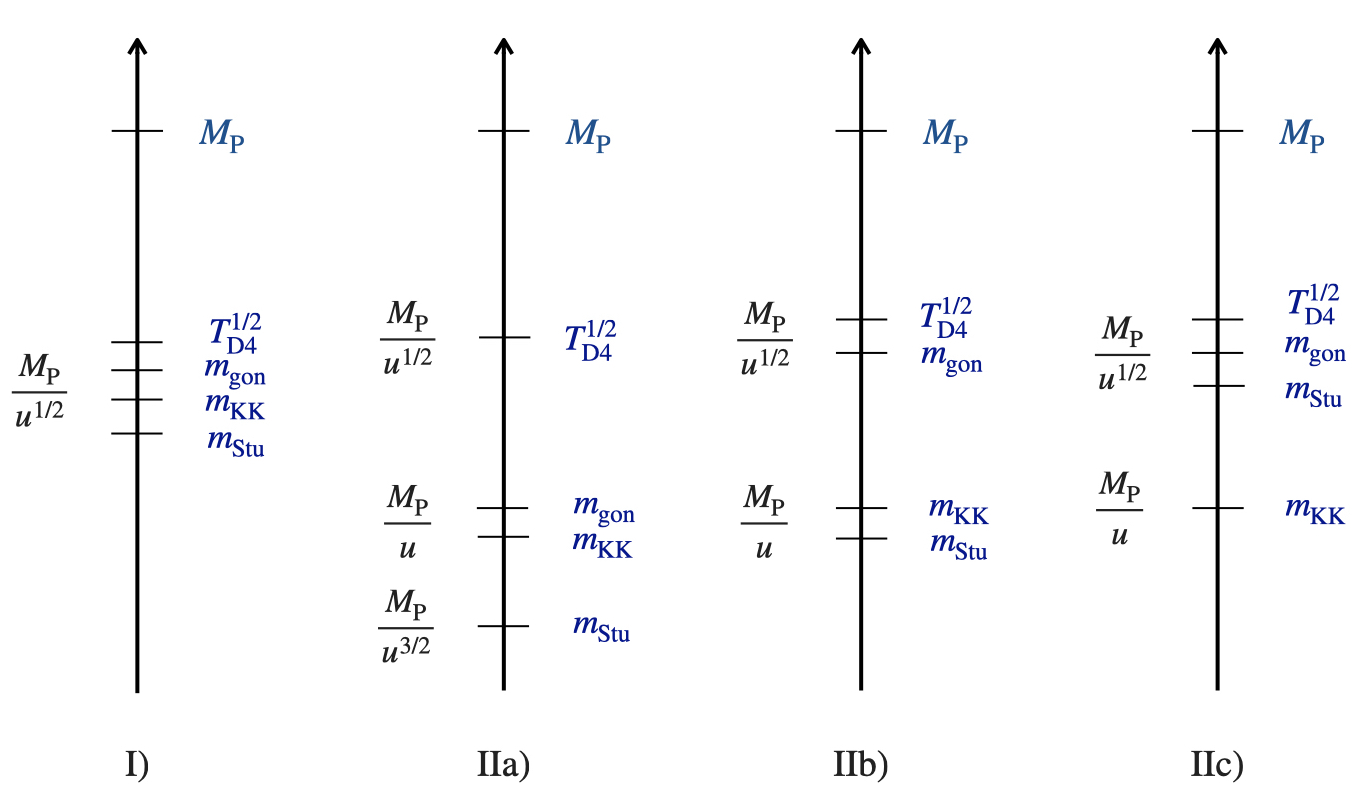}
\caption{Generic asymptotic scales in different EFT string limits. I) represents $w=1$ limits and II$\a$) $w=2$ limits of type ($\a$), with $\a=$ a,~b,~c in \eqref{mgonw2}. To compare with figure \ref{fig:scales} we have identified the EFT string flow parameter $\lambda$ with the growing field direction $u$, and $m_{\rm KK}$ with the lowest Kaluza--Klein or winding scale. The string scale $M_s$ is missing as it cannot be inferred from the general considerations in the main text.   
\label{fig:scalesEFT}}
\end{center}
\end{figure}

Let us discuss the scaling of the St\"uckelberg mass terms \eqref{masstu} in all these cases. Given the structure \eqref{KQ} and that the FI-terms $2\pi \xi_\a = Q_\a^K \ell_K M_{\rm P}$  vanish, we have
\be
M_{\a\b}^2 = g_\a g_\b Q_\a^K Q_\b^L \frac{\p_K \p_L \ch}{\ch} \, M_{\rm P}^2 \, .
\label{Stuckapp}
\ee
For $w=1$ limits, the non-degeneracy and our assumption that $Q^I_{\a/\b} =0$ imply that $M_{\a\b}^2 \sim g_\a g_\b M_{\rm P}^2$. Combined with our previous claim on $g_\a$ for this case, we obtain that  $m_{\rm Stu} \sim M_{\rm P}/\lambda^{1/2}$. For $w=2$ limits, we assume that $\ch^{-1} \p_K \p_L \ch$ has the same scaling as $\ell_K \ell_L$ in terms of $\lambda$, which translates into the estimate $M_{\a\b}^2 \sim g_\a g_\b T_\a^{+} T_\b^{+} M_{\rm P}^{-2}$. Applied to the three setups considered for this case one finds three different behaviours:
%\vspace{-.5cm}
\begin{subequations}
\label{mgonw2}  
\bea
\label{mgonw2a} 
\text{(a)} &  & \frac{m_{\rm gon}^{\rm min}}{M_{\rm P}}  \sim \left(\frac{m_{\rm Stu}}{M_{\rm P}} \right)^{2/3} \sim 1/\lambda \, , \\ 
\label{mgonw2b} 
\text{(b)} &  & \frac{m_{\rm gon}^{\rm min}}{M_{\rm P}} \sim \left(\frac{m_{\rm Stu}}{M_{\rm P}} \right)^{1/2} \sim 1/\sqrt{\lambda} \, , \\ 
\label{mgonw2c} 
\text{(c)} &  & \frac{m_{\rm gon}^{\rm min}}{M_{\rm P}} \sim \frac{m_{\rm Stu}}{M_{\rm P}} \sim 1/\sqrt{\lambda}\, .
\eea
\end{subequations}
All these scenarios are illustrated in figure \ref{fig:scalesEFT}.

\subsubsection*{Back to the Yukawas}

Let us now discuss what the above scenarios based on EFT string limits imply for the Yukawa couplings. Although from this generic viewpoint we have to rely on previous assumptions and cannot be exhaustive, the results will give as an idea on what to expect for the asymptotic behaviour of Yukawas in the type II orientifold limits under discussion. In the next section we will confirm our expectations with explicit constructions. 

We first consider Yukawas along limits with scaling weight $w=1$ and no $\CN=2$ subsector, so that the asymptotic Yukawa behaviour is dictated by \eqref{YWGCcpx}. To evaluate this expression, one needs the behaviour of the 4d dilaton along the limit, which as discussed previously should be of the form \eqref{asymMs}. More precisely, since this should be an emergent string limit, a typical value that one obtains in orientifold setups is $n=1$, from where one obtains a scaling of the form $|Y|^2 \sim 1/\lambda$. Notice that this coincides with \eqref{YukSTU} if we identify $\lambda$ with the field direction $s$ in the STU-like model, which is consistent with the fact that such a field is S-dual to the heterotic 4d dilaton.

Let us now turn to $w=2$ limits. Here the different gonion scalings \eqref{mgonw2} can combine into a single Yukawa, as long as they all correspond to the same asymptotic behaviour of $M_s$. Since we are no longer in an emergent string limit, we will assume that $M_s$ plays the role of the species scale and therefore lies in the range $m_{\rm gon} \leq M_s \leq T^{1/2}_{\rm D4}$. Then, a Yukawa made of three intersections with gonion scaling \eqref{mgonw2a} is compatible with the values $n = 2, 3, 4$, from where one obtains a Yukawa scaling of the form $|Y|^2 \sim 1/\lambda^{3-n/2}$. For $n=2$ one can actually consider any combination of the three scenarios in \eqref{mgonw2}, obtaining the scalings $|Y|^2 \sim 1/\lambda^{q}$ with $q = 2, 3/2, 1, 1/2$. 

To sum up, for Yukawas that satisfy \eqref{YWGCcpx} the above scenarios based on EFT string limits give us the asymptotic behaviour
\be
|Y| \sim \frac{1}{u^r}\, , \qquad \text{with} \quad r = \frac{1}{4}, \frac{1}{2}, \frac{3}{4}, 1\, , 
\label{asymYu}
\ee
where we have again identified $\lambda$ with the field direction $u$. Remarkably, a behaviour of the type $Y\sim 1/u^r$ resembles the recent results obtained in the context of heterotic compactifications in smooth Calabi--Yau manifolds \cite{Butbaia:2024tje,Constantin:2024yxh}, where numeric techniques were employed to compute physical Yukawas. Thus e.g. a behaviour like $Y\sim 0.025/u^{1/2}$ could match 
 the slope of the Yukawas in figure 7 of \cite{Constantin:2024yxh}, if one identifies $u=\psi$. While the resemblance is not obvious in terms of the duality relating both setups, it would be interesting to further explore this possible connection. 

Finally, one could consider Yukawas with $\CN=2$ sectors, whose asymptotic behaviour is dictated by \eqref{YWGCN2}. While in general one expects a milder scaling as compared to \eqref{YWGCcpx}, for  these cases the limits of vanishing Yukawas are more involved to discuss in general, unless one makes some assumption on the factors $h_{i\bar{i}}$. For this reason we will address their computation in explicit examples, which we now turn to discuss. As we will see, in simple limits these Yukawas are also described by the rule \eqref{asymYu}.

%%%%%%%%%%%%%%%%%%%
%%%%%%%%%%%%%%%%%%%
	
\section{Explicit examples}
\label{s:torito}

The scenarios obtained at the end of the last section depend on a number of assumptions and simplifications which may be more or less generic depending on the specific setup. Moreover, from figures \ref{fig:scales} and \ref{fig:scalesEFT} there is no precise way to determine neither the string scale nor the density of states in the KK and gonion towers. In this section we will recover most of these scenarios via an explicit class of intersecting D6-brane models, namely type IIA compactified on toroidal orientifolds, and more precisely on the orbifold first analysed in \cite{Cvetic:2001tj,Cvetic:2001nr}. These constructions have the advantage that both the Kaluza--Klein, winding and gonion scales can be computed explicitly, as well as the density of states for each tower. The same applies to other toroidal orbifold backgrounds, see e.g. \cite{Blumenhagen:2005mu,Blumenhagen:2006ci,Marchesano:2007de} for reviews, to which our discussion can be easily generalised.

To describe these models we follow the conventions of \cite[section 6.4]{Lanza:2021udy}, where the scales of the closed string sector were analysed for different EFT string limits. We start from the orbifold background $X_6 = (({\bf T}^2)_1 \times ({\bf T}^2)_2 \times ({\bf T}^2)_3)/\Gamma$, with $\Gamma = \mathbb{Z}_2 \times \mathbb{Z}_2$ has the same action and choice of discrete torsion as in \cite{Cvetic:2001nr}. Each $({\bf T}^2)_i$ is parametrised by a complex coordinate $z_i = y_{2i-1} + i \tau_i y_{2i}$, where the $y_j$ have unit period and the complex structure is $\tau_i = \ii \frac{R_{2i}}{R_{2i-1}}$, with $R_i$ the compactification radii in units of $\ell_s$. In terms of these coordinates the orientifold anti-holomorphic involution reads ${\cal R}:(z_1,z_2,z_3)\mapsto (\bar z_1,\bar z_2,\bar z_3)$, and the holomorphic $(3,0)$-form reads
\be
\begin{aligned}
\Omega&=\ell_s^3 R_1R_3R_5\, d z_1\wedge d z_2\wedge d z_3\\
&=\ell_s^3 (R_1d y_1+\ii R_2d y_2)\wedge (R_3d y_3+\ii R_4d y_4)\wedge(R_5d y_5+\ii R_6d y_6)\, .
\end{aligned}
\ee
A harmonic basis for even three-form classes $[\a_K] = \ell_s^3 {\rm P.D.}[ \Sigma_K^-]$ is given by 
\be
\begin{aligned}
 \a_0 =  4d y_1\wedge d y_3\wedge d y_5\, \ell_s^{3}\, , &\quad  \a_1=-4 d y_1\wedge d y_4\wedge d y_6\, \ell_s^{3}\, ,\\
 \a_2  =- 4 d y_2\wedge d y_3\wedge d y_6\, \ell_s^{3} \, , &\quad  \a_3 =-4 d y_2\wedge d y_4\wedge d y_5\, \ell_s^{3}\, ,
\end{aligned}
\label{even3form}
\ee
while one for odd three-form classes $[\b^J] = \ell_s^3 {\rm P.D.} [{\Sigma}_+^J]$ is
\be
\begin{aligned}
 \b^0 =-2 d y_2\wedge d y_4\wedge d y_6\, \ell_s^{3} \, , & \quad  \b^1 =2 d y_2\wedge d y_3\wedge d y_5\, \ell_s^{3}\, ,\\
\b^2 =2 d y_1\wedge d y_4\wedge d y_5\, \ell_s^{3}\, , &\quad \b^3 =2 d y_1\wedge d y_3\wedge d y_6\, \ell_s^{3}\, .
\end{aligned}
\label{odd3form}
\ee
Notice that their Poincar\'e duals satisfy $ \Sigma_K^-  \cdot \Sigma^L_+ = 2\delta_K^L$ when one takes into account the orbifold quotient. The relative factor of 2 in between these two basis takes into account that D2-branes wrapping the even three-cycles ${\Sigma}_+^L$ dual to \eqref{odd3form} yield $O(1)$ instantons. Plugging the $\a_K$ into \eqref{Omcexp} one identifies  $\zeta^K = \frac{1}{\ell_s^3} \int_{\Sigma_+^K}  C_3$ and
\be
\begin{aligned}
s \equiv u^{(0)}=\frac{1}{4} e^{-\phi}R_1R_3R_5\,, & \quad u^{(1)}=\frac{1}{4} e^{-\phi}R_1R_4R_6\, ,\\
u^{(2)}=\frac{1}{4}e^{-\phi}R_2R_3R_6\,, & \quad u^{(3)}=\frac{1}{4}e^{-\phi}R_2R_4R_5\, .
\end{aligned}
\ee
The saxions $u^K$ measure the internal volume of $O(1)$ instantons wrapping calibrated three-cycles, while $2u^K$ measure the volume of $U(1)$ instantons. Similarly, $2\zeta^K$ would be the axion with unit periodicity in the unorientifolded theory, while $\re U^K =  \zeta^K$ is the actual unit periodicity axion upon orientifolding. In terms of these fields we have
\be
\ch = \frac{\ii}{32\ell_s^6} \int_{{\bf T}^6}e^{-2\phi}  \Omega \wedge \bar{\Omega}  = 4 \sqrt{u^{(0)}u^{(1)}u^{(2)}u^{(3)}}\, ,
\ee
and so the K\"ahler potential reads $K_Q = - \sum^3_{K=0}\log 2 \im U^K$. The dual saxions are 
\be
{\ell}_K = \frac{1}{2u^K}\, .
\ee

One may describe several scales of the compactification in terms of these bulk moduli. For instance, in the regime $\tau_i \ll 1$, the lightest KK and winding scales are 
\be
m_{{\rm KK}, i} = \frac{2M_s}{R_{2i-1}} = \frac{M_{\rm P}}{A_i^{1/2} 2\sqrt{s u^{(i)}}}\, , \qquad m_{{\rm w}, i} = \oh R_{2i} M_s = \frac{A_i^{1/2} M_{\rm P}}{2  \sqrt{s u^{(i)}}}\, ,
\label{KKwindT6}
 \ee
where $A_i = \frac{1}{4}R_{2i}R_{2i-1}$, and we have used that the fundamental string scale is given by
\be
M_s^2 = \frac{M_{\rm P}^2}{4\sqrt{su^{(1)}u^{(2)}u^{(3)}}}\, .
\label{MF1}
\ee
To build our models we consider fractional factorisable three-cycles classes of the form
\bea
[\Pi_\a] &= &2 [(n_\a^1,m_\a^1) (n_\a^2, m_\a^2) (n_\a^3,m_\a^3)]\, , \qquad n_\a^i, m_\a^i \in \mathbb{Z} 
\label{f3cycle} \\ \nonumber
& = & n_\a^1 n_\a^2 n_\a^3 \Sigma^0_+ - n_\a^1 m_\a^2 m_\a^3 \Sigma^1_+ - m_\a^1 n_\a^2 m_\a^3 \Sigma^2_+ - m_\a^1 m_\a^2 n_\a^3 \Sigma^3_+ 
\\ \nonumber
& + & \oh \left(m_\a^1 m_\a^2 m_\a^3 \Sigma_0^- - m_\a^1 n_\a^2 n_\a^3 \Sigma_1^- - n_\a^1 m_\a^2 n_\a^3 \Sigma_2^- - n_\a^1 n_\a^2 m_\a^3 \Sigma_3^- \right)\, ,
\eea
where we follow the notation of \cite{Cvetic:2001nr}. That is, a three-cycle $(n^1,m^1) (n^2, m^2) (n^3,m^3)$ corresponds, in the covering space ${\bf T}^6$, to the class $[\Pi] = \left( n^1 a_1 + m^1 a_2 \right) \times \left( n^2 a_3 + m^2 a_4 \right) \times \left( n^3 a_5 + m^3 a_6 \right)$, where $a_i$ is the one-cycle class of ${\bf T}^6$ along $y_i$. These fractional three-cycles have a reduced moduli space, and for simplicity we will assume them to cross the same fixed point of the orbifold action, say the origin $z_i=0$. Moreover, if they are not invariant under the orientifold projection one needs to add the orientifold image, which lies at
\be
[\Pi_{\a^*}] = 2 [(n_\a^1,-m_\a^1) (n_\a^2, -m_\a^2) (n_\a^3,-m_\a^3)]\, .
\ee

\subsubsection*{A toy model}

Let us first consider a set of three D6-brane stacks $a$, $b$, $c$, wrapping the three-cycles 
\begin{align}
\nonumber \Pi_a & = 2 (k,1)(k,1)(k,-1)\, , \\ 
\label{toymodel}
\Pi_b & = 2(0,1)(1,0)(0,-1) \, ,\\  \nonumber \Pi_c & = 2(0,1)(0,-1)(1,0)\, \, 
\end{align}
with $k \in \bN$. From here one obtains a gauge group $U(1)_A \times {\rm USp}(2)_b \times {\rm USp}(2)_c$, up to additional D6-branes needed to cancel tadpoles. The D6-branes $b$ and $c$ are invariant under both  the orbifold and orientifold action, and as a result each of them hosts a USp$(2)$ gauge group and has a vanishing FI-term. The D6-brane $a$ is not invariant under ${\cal R}$ and has an orientifold image $\Pi_{a^*}$ and an FI-term of the form
\be
\frac{\pi \xi_a}{M_{\rm P}^2} = -\frac{1}{4s} - \frac{k^2}{4u^{(1)}} - \frac{k^2}{4u^{(2)}} + \frac{k^2}{4u^{(3)}} \, ,
\label{xiatoy}
\ee
where we have identified $s \equiv u^{(0)}$. Using the dictionary 
\be
 2s = e^{-\phi_4}\frac{1}{\sqrt{\tau_1\tau_2\tau_3}}\, , \quad 2u^{(i)}=e^{-\phi_4}\sqrt{\frac{\tau_j\tau_k}{\tau_i} }\, ,
 \label{instvolumen}
\ee
one can see that the condition $\xi_a =0$ is equivalent to impose
\be
\theta_1 + \theta_2 = \theta_3 \, , \qquad \text{where} \quad \pi \theta_i = \tan^{-1} \frac{\tau_i}{k}\,  . 
\ee
where we have used the identity $\tan (\vartheta + \varphi) = \frac{\tan \vartheta + \tan \varphi}{1 - \tan \vartheta  \tan \varphi}$. 
One also finds the following gauge couplings for the different D6-branes
\be
\frac{2\pi}{g_a^2} = 2(k^3 s + k u^{(1)} + k u^{(2)} - k u^{(3)})\, , \qquad \frac{2\pi}{g_b^2} = 2u^{(2)}\, ,  \qquad \frac{2\pi}{g_c^2} = 2u^{(3)}\, , 
\label{gacouplings}
\ee
from where one deduces the following D-term induced (squared) mass \eqref{massinter}
\be
q_a g_a^2 \xi_a = q_a \tan \left[ \pi (\th_3 - \th_2 - \th_1) \right]  M_s^2\, ,
\ee
with $q_a$ the charge of the open string state under $U(1)_a$. Finally, out of the three D6-branes above, the St\"uckelberg mass is only generated for the $U(1)_a$ boson and reads
\be
M_{aa}^2 = \oh g_a^2 \left(\frac{1}{s^2} + \frac{k^4}{(u^{(1)})^2} + \frac{k^4}{(u^{(2)})^2} + \frac{k^4}{(u^{(3)})^2} \right) M_{\rm P}^2 \, .
\ee

What is particularly interesting for us is that we have precise expression for scales of the gonion towers. For instance in the $aa^*$ sector we have that $m_{{\rm gon}, i}^2 = 2 \th_i M_s^2$. In general these scales are an involved function of the complex structure saxions, but it simplifies in certain limits, like for instance $\tau_i/k \ll 1$. In this regime, which realises the STU-like hierarchy $s \gg u^{(i)}$, we have
\be
\frac{m_{{\rm gon}, i}}{M_{\rm P}} \simeq \frac{1}{2\sqrt{k s u^{(i)}}}\, .
\label{mgontoy}
\ee
This expression describes all those gonion scales that are small in string units, while the rest are of the order of $M_s$. We have summarised in table \ref{tab:goniontoy} the  different gonion scales for each open string sector. 
\setlength{\belowcaptionskip}{0pt}
\setlength{\arrayrulewidth}{0.2mm}
\renewcommand{\arraystretch}{1.3}
\begin{table}[h]
\begin{center}
\begin{tabular}{|c|c|c|c|}
\hline
Sector & $m_{{\rm gon}, 1}/M_{\rm P}$ & $m_{{\rm gon}, 2}/M_{\rm P}$ & $m_{{\rm gon}, 3}/M_{\rm P}$ \\
\hline\hline
$aa^*$ & $1/ \sqrt{ksu^{(1)}}$ &  $1/\sqrt{ksu^{(2)}}$ &  $1/\sqrt{ksu^{(3)}}$ \\
\hline
$ab$ &   $e^{\phi_4}$  & $1/ \sqrt{ksu^{(2)}}$ &  $e^{\phi_4}$ \\
\hline
$bc$ &  $e^{\phi_4}$ &   $e^{\phi_4}$ & $e^{\phi_4}$\\
\hline
$ca$ &  $e^{\phi_4}$ & $e^{\phi_4}$ &  $1/\sqrt{ksu^{(3)}}$ \\
\hline
\end{tabular}
\end{center}
\caption{Gonion scales for different sectors of the toy model, up to numeric factors. \label{tab:goniontoy}}
\end{table}

Notice that, by increasing the value of $k$ and keeping the K\"ahler moduli fixed, one may achieve the ordering $m_{{\rm gon}, i} < m_{{\rm KK}, i} <  m_{{\rm w}, i}$, cf. \eqref{KKwindT6}. In this case, the gonion scale $m_{{\rm gon}, i}$ is below any tower from the bulk, even if it cannot  be made parametrically lighter because it shares the same dependence on complex structure saxions with the lightest of them.\footnote{Even then, the scale at which the gonion tower displays its Hagedorn spectrum, namely $M_s$, is always above either $m_{{\rm KK}, i}$ or  $m_{{\rm w}, i}$.}  It is however interesting to notice that, even in this case, there is still a Kaluza--Klein tower which is lighter than any gonion tower. Indeed, on each ${\bf T}^2$ we have the KK and winding modes of each D6-brane, which for the D6-brane $a$ read
\be
m_{{\rm KKD6}_a, i} = \frac{2M_s}{R_{2i-1}\sqrt{k^2 +\tau_i^2}} \simeq \frac{M_{\rm P}}{A_i^{1/2} 2k\sqrt{s u^{(i)}}}\, , \quad m_{{\rm wD6}_a, i} = \frac{R_{2i}}{2\sqrt{k^2 +\tau_i^2}} M_s \simeq  \frac{A_i^{1/2} M_{\rm P}}{2  k\sqrt{s u^{(i)}}}\, .
\label{KKwindD6}
 \ee
In particular we have that $m_{{\rm KKD6}_a, i} < m_{{\rm gon}, i}$ for any value of $k$. Again, the dependence on the complex structure saxions is similar for all these towers, so taking limits on the moduli space of complex structures they cannot be separated parametrically from each other. Hence, when we take different infinite distance limits we find several towers of different nature that become light at the same time. 

Indeed, let us for instance consider the limit $\tau_i =$ const. and $e^{\phi_4} \sim \lambda^{-1} \to 0$ which realises the STU-like limit I), or the limit $e^{\phi_4} \sim \tau_i^{3/2} \sim \lambda^{-3/2} \to 0$ that implements the STU-like limit II). In both cases it is easy to check that the content of table 
 \ref{tab:scales} is realised with $u \simeq u^{(i)}$. In the second case, the lightest towers scale as $m_{\rm tower}^{\rm min} \sim u^{-2} M_{\rm P}$, as in  figure \ref{fig:scales}. More precisely we find the following towers with this scaling: 
 \begin{itemize}
 
  \item[-] Three bulk KK  $m_{{\rm KK}, i}$ and three bulk winding   $m_{{\rm w}, i}$ towers, $i=1,2,3$.
 
 \item[-] In the $aa$ sector, three towers of KK modes $m_{{\rm KKD6}_a, i}$, and three towers of winding modes $m_{{\rm wD6}_a, i}$, $i=1,2,3$. 
 
 \item[-] In the $bb$ sector, one tower of KK modes $m_{{\rm KKD6}_b, 2}$ and one tower of winding modes $m_{{\rm wD6}_c, 2}$.

 \item[-] In the $cc$ sector, one tower of KK modes $m_{{\rm KKD6}_c, 3}$ and one tower of winding modes $m_{{\rm wD6}_c, 3}$.

\item[-] In the $aa^*$ sector, three towers of gonion modes $m_{{\rm gon}, i}$ $i=1,2,3$.

\item[-]  In the $ab$ sector, one tower of gonion modes $m_{{\rm gon},2}$.
 
\item[-]  In the $ca$ sector, one tower of gonion modes $m_{{\rm gon},3}$.

 \end{itemize}

The towers on each item combine multiplicatively among themselves into a lattice, while those in different items only combine additively. In this sense, the towers that have higher density dominate the light spectrum. Notice that due to the spectrum \eqref{towergon} one gonion tower has comparable density as the combination of one Kaluza--Klein and one winding tower. Therefore, in this toy model the towers that dominate the lightest spectrum correspond to the bulk KK and winding, $aa$ sector KK and winding, and $aa^*$ gonion towers. In section \ref{s:emergence} we will further analyse this and other light tower spectra from the viewpoint of the species scale and the Emergence Proposal.

One may also consider a different regime for this toy model, in which only a single complex structure modulus is small compared to the other two. Let us for instance consider the case where $e^{2\phi_4} \sim \tau_2/k  \sim \lambda^{-1} \ll 1$ while $\tau_1/k, \tau_3/k \sim {\cal O}(1)$. In terms of complex structure saxions, this corresponds to the limit
\be
s , u^{(2)} \sim \lambda\, , \qquad u^{(1)}, u^{(3)} \ \text{bounded}\, .
\label{quasiEFTIIa}
\ee
The saxions $u^{(1)}, u^{(3)}$ cannot be constant along the limit, because one needs to adjust them as $s$ and $u^{(2)}$ grow, in order to satisfy the supersymmetry condition $\xi_a =0$, cf. \eqref{xiatoy}. However, they will be functions of $\lambda$ bounded from above.\footnote{For instance, one could set $u^{(1)}=u_0$ and $\frac{k^2}{u^{(3)}} = \frac{1+k^2}{\lambda}+ \frac{k^2}{u_0}$, i.e. $u^{(3)}$ bounded from above.} In this sense, this limit is an example of a quasi-EFT string limit, in the sense defined in section \ref{ss:Y0g0}. For this case in particular, because we have two growing saxions, we would expect the limit to be quasi-EFT with $w=2$. Indeed, we will see that its tower spectrum resembles to great extent the scenario IIa) in figure \ref{fig:scalesEFT}. First, the scales corresponding to 4d strings coming from D4-branes and the fundamental string scale read
\be
T_{{\rm D4}, 0}^{1/2}, T_{{\rm D4}, 2}^{1/2} \sim  \lambda^{-1/2}  M_{\rm P}\, , \qquad M_s \sim  \lambda^{-1/2}  M_{\rm P}\, .
\ee
Second, we find the lightest set of towers at the scale $m_{\rm tower}^{\rm min} \sim \lambda^{-1} M_{\rm P}$. These are:
 \begin{itemize}
 
 \item[-] One tower of bulk KK modes $m_{{\rm KK}, 2}$ and one of bulk winding modes $m_{{\rm w}, 2}$.
 
 \item[-] In the $aa$ sector, one KK mode $m_{{\rm KKD6}_a, 2}$, and one winding mode tower $m_{{\rm wD6}_a, 2}$. 
 
\item[-] In the $aa^*$ sector, one tower of gonion modes $m_{{\rm gon}, 2}$.

\item[-]  In the $ab$ sector, one tower of gonion modes $m_{{\rm gon},2}$.

\end{itemize}
This tower spectrum is much simpler than in the previous limit, and now each item contains a similar density of states. We therefore find agreement with the tower spectrum of the scenario IIA in figure \ref{fig:scalesEFT} if we identify $\lambda \sim u$, and we include the string scale at $M_s = u^{-1/2} M_{\rm P}$. The estimate for the St\"uckelberg scale, however, reads in case $m_{\rm Stu} \sim \lambda^{-1/2} M_{\rm P}$, which is in disagreement with the strict EFT string limit scenario IIa) in the figure. It thus seems that the estimate for this scale is not particularly robust when generalising an EFT string limit to a quasi-EFT string one. Finally, one could also consider a third limit in which the role of $u^{(1)}$ and $u^{(2)}$ is interchanged in \eqref{quasiEFTIIa}. The description above would work the same up to redefinitions, except for the light gonion tower in the $ab$ sector, which would not be present. 

This relatively small difference turns out to be quite relevant when we look at the Yukawa couplings between massless chiral fields. First, the selection rules of Appendix \ref{ap:Hmom} imply that there is a single non-vanishing Yukawa in the massless sector, namely the one involving the sectors $ab$, $bc$ and $ca$. The kinetic terms for each of these sectors are:
\bea
K_{ca} &= &\frac{e^{K/2-\phi_4}}{\sqrt{2\pi}}\left[\frac{\Gamma(\frac{1}{2}-\theta_1)}{\Gamma(\frac{1}{2}+\theta_1)}\frac{\Gamma(\frac{1}{2}-\theta_2)}{\Gamma(\frac{1}{2}+\theta_2)}\frac{\Gamma(\theta_3)}{\Gamma(1-\theta_3)}\right]^{1/2}\label{eq: Kca}\, , \\
K_{ab} &= & \frac{e^{K/2-\phi_4}}{\sqrt{2\pi}} \left[\frac{\Gamma(\frac{1}{2}+\theta_1)}{\Gamma(\frac{1}{2}-\theta_1)}\frac{\Gamma(\theta_2)}{\Gamma(1-\theta_2)}\frac{\Gamma(\frac{1}{2}-\theta_3)}{\Gamma(\frac{1}{2}+\theta_3)}\right]^{1/2}\label{eq: Kab} \, , \\
K_{bc}& =&e^{K/2-\phi_4}\sqrt{\tau_1/2\pi} = 2 e^{K/2} \sqrt{u^{(2)}u^{(3)}/2\pi}\, .
\label{eq: Kbc}
\eea
The metric for the first two sectors is obtained by direct application of \eqref{metricgen}, while the third one corresponds to an $\CN=2$ sector and it is obtained as a limit of the general expression \cite{Font:2004cx}. Notice that from \eqref{eq: Kbc} we obtain $h_{bc} = g_b^{-1} g_c^{-1}$ in \eqref{metricexep}, which saturates the inequality \eqref{hiiineq}. When plugged into the supergravity formula \eqref{eq: Yukawasugra} one obtains the following Yukawa coupling
\be
Y_{abc}=B\, \sqrt{2\pi g_bg_c} \left[\frac{\Gamma(1-\theta_2)}{\Gamma(\theta_2)}\frac{\Gamma(1-\theta_3)}{\Gamma(\theta_3)}\frac{\Gamma(\frac{1}{2}+\theta_2)}{\Gamma(\frac{1}{2}-\theta_2)}\frac{\Gamma(\frac{1}{2}+\theta_3)}{\Gamma(\frac{1}{2}-\theta_3)}\right]^{1/4}\, ,
\label{Yukawatoy}
\ee
where $B$ is a function of the K\"ahler moduli that we treat as a constant. 

Let us now see how the Yukawa behaves under the different limits considered above. For the limit $e^{\phi_4} \sim \tau_i^{3/2} \sim \lambda^{-1/2} \to 0$ that implements the STU-like limit II), we find 
\be
Y_{abc} \sim \left(\frac{ \theta_2 \theta_3}{u^{(2)} u^{(3)}} \right)^{1/4}  \sim \frac{e^{-\phi_4}}{ \sqrt{su^{(2)} u^{(3)}}} \sim \lambda^{-1/3}\, .
\ee
This result can also be recovered from \eqref{YWGCN2}, with $h_{bc} = g_b^{-1} g_c^{-1}$ . The difference between this scaling and the generic one in \eqref{YukSTU}, which would correspond to $Y \sim \lambda^{-1/2}$ is due to the presence of an $\CN =2$ sector in the Yukawa, which typically softens the suppression. 

If instead we consider the quasi-EFT string limit \eqref{quasiEFTIIa} we obtain
\be
Y_{abc} \sim \left(\frac{ \theta_2 }{u^{(2)}} \right)^{1/4}  \sim \frac{e^{-\phi_4/2}}{ \sqrt{s^{1/2}u^{(2)}}} \sim \lambda^{-1/2}\, .
\ee
 Finally, if one considers the limit which interchanges the roles of $u^{(2)}$ and $u^{(1)}$ in \eqref{quasiEFTIIa}, one obtains that the Yukawa coupling remains constant, due to cancellations involving the $\CN=2$ sector. We have summarised these results in table \ref{tab:gYlimitstoy}, together with the asymptotic behaviour for each gauge coupling involved in the Yukawa. The pattern that we find is $Y_{abc} \sim g_bg_c$, and we also see that eqs.\eqref{Yrelu} and \eqref{Yrelg} are satisfied, with $g_*$ the largest of the three gauge couplings and with $r = \frac{1}{3}, \oh$. Finally, notice that the limit \eqref{quasiEFTIIa} realises the scenario \eqref{favo} highlighted in the introduction.

\setlength{\arrayrulewidth}{0.2mm}
\renewcommand{\arraystretch}{1.25}
\begin{table}[h]
\begin{center}
\begin{tabular}{|c|c||c|c|c|c|c|}
\hline
Limit & Scenario & $g_a$ & $g_b$ &  $g_c$ & $Y_{abc}$ \\
\hline\hline
$s \sim (u^{(i)})^3 \sim \lambda $ & STU-like II) &$\lambda^{-1/2}$ & $\lambda^{-1/6}$ & $\lambda^{-1/6}$ & $\lambda^{-1/3}$ \\
\hline
\eqref{quasiEFTIIa} & quasi-EFT IIa)  & $\lambda^{-1/2}$  & $\lambda^{-1/2}$ & const. & $\lambda^{-1/2}$ \\
\hline
\eqref{quasiEFTIIa}$^\prime$ & quasi-EFT IIa)  & $\lambda^{-1/2}$ &  const. & const.  & const. \\
\hline
\end{tabular}
\end{center}
\caption{Different limits discussed in the text, together with the asymptotic behaviour of the gauge and Yukawa couplings. The limit \eqref{quasiEFTIIa}$^\prime$ corresponds to \eqref{quasiEFTIIa} with the behaviour for $u^{(1)}$ and $u^{(2)}$ interchanged.  \label{tab:gYlimitstoy}}
\end{table}

\subsubsection*{A Pati--Salam model}

Let us now consider a more realistic example, which realises a Left-Right model first considered in \cite{Cremades:2002qm,Cremades:2003qj} or its Pati--Salam extension $SU(4)_a \times SU(2)_b \times SU(2)_c$. The embedding of such model into the $\Z_2 \times \Z_2$ orientifold background that we are considering was realised in \cite{Marchesano:2004yq,Marchesano:2004xz}, including the additional D6-branes needed to cancel all tadpoles. For simplicity, we will not include the latter into this discussion. 

The three stacks of D6-branes relevant for our discussion are
\begin{align}
\nonumber \Pi_a & = 8 (1,0)(k,1)(k,-1)\, , \\  \label{PSmodel} \Pi_b & = 2(0,1)(1,0)(0,-1) \, ,\\  \nonumber \Pi_c & = 2(0,1)(0,-1)(1,0)\, \, ,
\end{align}
where $k \in \bN$ determines the number of chiral families. Notice that the D6-branes $b$ and $c$ are similar to the ones in the previous toy model, while the D6-brane $a$ is different. In particular, it does not intersect its orientifold image, and as a result the lightest gonion in the $aa^*$ sector can be made of the order of the fundamental string scale $M_s$ by separating $a$ from $a^*$ in $({\bf T}^2)_1$. The gauge coupling associated to $U(1)_a \subset U(4)_a$ is
\be
\frac{2\pi}{g_a^2} = 2(k^2 s + u^{(1)})\, , 
\ee
while its FI-term reads
\be
\frac{\pi \xi_a}{M_{\rm P}^2} = \frac{k}{4}\left(  - \frac{1}{u^{(2)}} + \frac{1}{u^{(3)}}\right) \, ,
\label{xiaguay}
\ee
such that the supersymmetry condition $\xi_a = 0$ is equivalent to $u^{(2)} = u^{(3)}$ and to $\theta_2 = \theta_3$, with $\pi \theta_j = \tan^{-1} (\tau_j/k)$, $j=2,3$. More precisely, we have that
\be
q_a g_a^2 \xi_a = q_a \tan \left[ \pi (\th_3 - \th_2) \right]  M_s^2\, ,
\ee
with $q_a$ the charge of the open string state under $U(1)_a$. The St\"uckelberg mass term is
\be
M_{aa}^2 = \oh g_a^2 \left(\frac{k^2}{(u^{(2)})^2} + \frac{k^2}{(u^{(3)})^2} \right) M_{\rm P}^2 \, .
\ee
\setlength{\arrayrulewidth}{0.2mm}
\renewcommand{\arraystretch}{1.3}
\begin{table}[h]
\begin{center}
\begin{tabular}{|c|c|c|c|}
\hline
Sector & $m_{{\rm gon}, 1}/M_{\rm P}$ & $m_{{\rm gon}, 2}/M_{\rm P}$ & $m_{{\rm gon}, 3}/M_{\rm P}$ \\
\hline\hline
$aa^*$ & $e^{\phi_4}$ &  $e^{\phi_4} \ \text{or}\ 1/\sqrt{ksu^{(2)}}$ &  $e^{\phi_4} \ \text{or}\ 1/\sqrt{ksu^{(3)}}$ \\
\hline
$ab$ &   $e^{\phi_4}$  & $1/ \sqrt{ksu^{(2)}}$ &  $e^{\phi_4}$ \\
\hline
$bc$ &  $e^{\phi_4}$ &   $e^{\phi_4}$ & $e^{\phi_4}$\\
\hline
$ca$ &  $e^{\phi_4}$ & $e^{\phi_4}$ &  $1/\sqrt{ksu^{(3)}}$ \\
\hline
\end{tabular}
\end{center}
\caption{Gonion scales for different sectors of the PS model, up to numeric factors. \label{tab:gonionguay}}
\end{table}

As in the toy model, one may now consider different infinite distance limits within the regime $\tau_i \ll 1$. The gonion scales are quite similar to those of the toy model, and have been summarised in table \ref{tab:gonionguay}. The main difference is that now all towers in the $aa^*$ sector can be made of the order of the fundamental string scale by separating the $\Pi_a$ from $\Pi_{a^*}$ in the first two-torus. %In the following we will assume that this is the case. 

Another important difference is that $\tau_1$ does not influence neither the supersymmetry condition $\xi_a =0$ nor the gonion tower scales. As a result one may conceive different strict EFT string limits that are compatible with supersymmetry. A few of them are:
\begin{align}
\label{guayI}
s \sim \lambda\, ,  \quad u^{(i)} =\, \text{const.} & \implies  e^{\phi_4} \sim \lambda^{-1/4}\, , \\
\label{guayIIa}
s, u^{(2)} = u^{(3)} \sim \lambda \, ,   \quad u^{(1)} =\, \text{const.} & \implies  e^{\phi_4} \sim \lambda^{-3/4}\, , \\
\label{guayIIb}
u^{(2)} = u^{(3)} \sim \lambda \, ,   \quad s, u^{(1)} =\, \text{const.} & \implies  e^{\phi_4} \sim \lambda^{-1/2}\, , \\
\label{guayIIc}
s, u^{(1)}\sim \lambda \, ,   \quad u^{(2)} = u^{(3)}  =\, \text{const.} & \implies  e^{\phi_4} \sim \lambda^{-1/2}\, ,
\end{align}
with $\lambda \to \infty$. In addition we will consider the limit
\be
s \sim \lambda \, , \quad u^{(2)} = u^{(3)} \sim \lambda^{1/2} \, , \quad u^{(1)} = \, \text{const.} \, \implies  e^{\phi_4} \sim \lambda^{-1/2}\, .
\label{guaynonEFT}
\ee

\setlength{\arrayrulewidth}{0.2mm}
\renewcommand{\arraystretch}{1.25}
\begin{table}[h]
\begin{center}
\begin{tabular}{|c|c||c|c|c|c||c|c|c|c|}
\hline
Limit & Scenario & $\# T_{\rm D4}^{1/2}$ & $\# m_{{\rm KK}}^{\rm bulk}$ &  $\# m_{{\rm KK}}^{\rm brane}$ & $\# m_{{\rm gon}}$ & $g_a$ & $g_b$ &  $g_c$ & $Y_{abc}$ \\
\hline\hline
 \eqref{guayI} & EFT I) & 1 & 6 & 6 & 2 or 4&  $\lambda^{-\oh}$ & const. & const. & $\lambda^{-1/4}$ \\
\hline
 \eqref{guayIIa} & EFT IIa) & - & 6 & 4 & 2 or 4 &  $\lambda^{-1/2}$ & $\lambda^{-1/2}$ & $\lambda^{-1/2}$ & $\lambda^{-3/4}$ \\
\hline
 \eqref{guayIIb} & EFT IIb) & - & 2 & 2 & -  &  const. &  $\lambda^{-1/2}$ & $\lambda^{-1/2}$  & $\lambda^{-1/2}$ \\
\hline
 \eqref{guayIIc} & EFT IIc) & - & 2 & 2 & - &  $\lambda^{-1/2}$ &  const. & const.  & const.  \\
\hline
 \eqref{guaynonEFT} & non-EFT & - & 4& 4&2 or 4 & $\lambda^{-1/2}$ &  $\lambda^{-1/4}$ & $\lambda^{-1/4}$ & $\lambda^{-1/2}$  \\
\hline
\end{tabular}
\end{center}
\caption{Different limits discussed for the Pati--Salam model, together with their corresponding scenario in figure \ref{fig:scalesEFT}, the dimension of the lattice for each kind of leading tower, and the asymptotic behaviour of the gauge and Yukawa couplings. $\# m_{\rm KK}$ stands for both KK and winding towers.  \label{tab:limits}}
\end{table}

For each of these limits one can perform a similar analysis to the one made in the toy model, regarding the spectrum of light towers and the Yukawa couplings. We have summarised the results regarding the tower spectra in table \ref{tab:limits}. One finds that each of the limits \eqref{guayI}-\eqref{guayIIc} falls into a different scenario of figure \ref{fig:scalesEFT}, and that the behaviour for the scales on each scenario is exactly reproduced in this toroidal setting, upon the identification $\lambda = u$. What remains to specify for each of these limits that is particularly relevant for the low energy physics, is the degeneracy of each of the leading towers, which increases when towers add multiplicatively. In table \ref{tab:limits} we have encoded this information by specifying the dimension of the lattice of states for each leading tower, when present in a given limit. Within the same limits, the degeneracy of the largest gonion tower (which is twice compared to that of a KK tower) changes depending on whether or not we separate the branes $a$ and $a^*$ in the first two-torus. The number of leading gonion towers that are replicas of chiral fields involved in the non-vanishing Yukawa $Y_{abc}$ is however not affected by this choice. 

Notice that the limits \eqref{guayIIa} and \eqref{guayIIb} are special, in the sense that $\tau_1 \sim \lambda^{1/2}$ and we are taken away from the regime $\tau_i \ll 1$, $\forall i$. This implies that the lightest KK and winding scales coming from the first torus are not captured by \eqref{KKwindT6}.\footnote{Instead, we have that the lightest KK mode is $m_{{\rm KK}, 1} = \frac{M_s}{R_{2}} \simeq \frac{M_{\rm P}}{2 A_1^{1/2} \sqrt{u^{(2)} u^{(3)}}}=\frac{M_{\rm P}}{2\pi A_1^{1/2}h_{i\bar{i}}}$. \label{footnote:winding}} 
The limit \eqref{guaynonEFT} does not fall into the category of EFT string limits, and the scaling for each of its towers goes as $m_{\rm tower}^{\rm min}\sim \lambda^{-3/4}$. 

The selection rules for Yukawa couplings described in Appendix \ref{ap:Hmom} imply that the only non-vanishing Yukawa formed by zero mode couplings is the one from the sector $abc$, as in the previous example. The kinetic terms for each sector have the form:
\bea
K_{ca} &= & \frac{e^{K/2-\phi_4}}{\sqrt{2\pi}}\left[\frac{\Gamma(\frac{1}{2}-\theta_2)}{\Gamma(\frac{1}{2}+\theta_2)}\frac{\Gamma(\theta_3)}{\Gamma(1-\theta_3)}\right]^{1/2}\label{eq: KcaPati}\, , \\
K_{ab} &= & \frac{e^{K/2-\phi_4}}{\sqrt{2\pi}}\left[\frac{\Gamma(\theta_2)}{\Gamma(1-\theta_2)}\frac{\Gamma(\frac{1}{2}-\theta_3)}{\Gamma(\frac{1}{2}+\theta_3)}\right]^{1/2}\label{eq: KabPati} \, , \\
K_{bc}& =&e^{K/2-\phi_4}\sqrt{\tau_1/2\pi} =2 e^{K/2} \sqrt{u^{(2)}u^{(3)}/2\pi} = e^{K/2} g_b^{-1} g_c^{-1}\,\label{eq: kbcPati} .
\eea
The main novelty with respect to the metrics of the toy model is that the kinetic terms of the sectors $ab$ and $ca$ are insensitive to the geometry of the first torus, since the corresponding angles satisfy $\theta_{ab}^1=\theta_{ca}^1=\oh$. This difference disappears when one considers the physical Yukawa coupling:
\be
Y_{abc}=B\, \sqrt{2\pi g_bg_c} \left[\frac{\Gamma(1-\theta_2)}{\Gamma(\theta_2)}\frac{\Gamma(1-\theta_3)}{\Gamma(\theta_3)}\frac{\Gamma(\frac{1}{2}+\theta_2)}{\Gamma(\frac{1}{2}-\theta_2)}\frac{\Gamma(\frac{1}{2}+\theta_3)}{\Gamma(\frac{1}{2}-\theta_3)}\right]^{1/4}\, ,
\label{Yukawapati}
\ee
which yields an identical expression to \eqref{Yukawatoy}. The behaviour of the Yukawa couplings concerning the limits considered above is also displayed in table \ref{tab:limits}. Notice that one again satisfies the pattern \eqref{Yrelu} and \eqref{Yrelg} with $r=1/4,1/2,3/4,1$.

%%%%%%%%%%%%%%%%%%%
%%%%%%%%%%%%%%%%%%%

\section{Structure of towers: species scale and emergence}
\label{s:emergence}

	We have seen in the previous sections different examples of asymptotic moduli directions in which towers of particles,
	both of KK and gonion type appear.  In the limits in which Yukawa couplings go to zero, towers of gonions appear.
	Table \ref{tab: Pg and pkk} together with figures \ref{fig:scales} and \ref{fig:scalesEFT} show the structure of scales  of some examples, sometimes quite intricate. 
	At this point  it is worth summarising the general structure of towers. One can show that the  different options may be
	neatly classified in terms of the density parameters of the towers, as we now describe. 

\begin{table}[h]
    \centering
\resizebox{1\textwidth}{!}{
\renewcommand{\arraystretch}{1.3}
\begin{tabular}{|c|c|c|c|c|}\hline
            $p_{\rm KK} / p_{\rm gon}$  & $p_{\rm gon}=0$& $p_{\rm gon}=2$ & $p_{\rm gon}=4$ & $p_{\rm gon}=6$ \\ \hline
             $p=2$& EFT IIb), IIc) (PS) & quasi-EFT IIa) (TM) & {X} & {X}\\ \hline
             $p=4$& $\mathcal{N}=2$ sectors & non-EFT (PS) & non-EFT (PS)$^{\star}$ & {X}\\ \hline
             $p=6$& $\mathcal{N}=2$ sectors & EFT IIa) (PS) & EFT IIa) (PS)$^{\star}$ & STU-like II) (TM)\\\hline
\end{tabular}
}
\caption{For the explicit examples considered in section \ref{s:torito}, several limits are displayed in terms of the density parameter $p_t$. PS = Pati--Salam model \eqref{PSmodel} . TM = toy model \eqref{toymodel}. The $^{\star}$ in PS means that the sector $aa^{\star}$ intersects. } 
    \label{tab: Pg and pkk}
\end{table}

	\subsection{Tower densities and the Species Scale}
	
	The density parameters 
	$p_t$ are defined in terms of a tower spectrum as
	\beq
	m_n^{(t)} \ = \ n_t^{1/p_t} m_0^{(t)} \, ,
	\label{tower}
	\eeq 
	with $n_t \in \mathbb{Z}$ and $m_0^{(t)}$ the characteristic scale of the tower. The sum in $n_t$ goes up to the fundamental Quantum Gravity scale of the theory which we will call $\Lambda_{QG}$, so that
	\beq
	\Lambda_{QG} \ =\ N_t^{1/p_t} m_0^{(t)} \ ,
	\eeq
	with $N_t$ the total number of states in the tower which fit below the scale $\Lambda_{QG}$. Thus, e.g., the case of the
	decompactification of a single dimension corresponds to $p_t=1$ with $m_0=m_{\rm KK}$.  The case of $p $ decompactified dimensions
	with the same scale $m_0=m_{\rm KK}$, corresponds to $N_t=N^{p}$, with $N$ the number of states in each of the $p$.
	We will consider in what follows the cases  with $p=2,4,6$  decompactified large dimensions, appropriate to handle 
	chiral theories with $\CN=1$ supersymmetry.
	In particular we are considering theories with $p$ large dimensions and $(6-p)$ dimensions of order of the fundamental scale
	(and hence not associated towers).
	In the case of gonions one has $p_{\rm gon}=2,4,6$ depending on whether,  in the studied local brane intersection, 
	there are gonion towers in one, two, or three complex planes.\footnote{It is interesting to remark that the gonion towers are the first examples of genuine particle towers with $p_t>1$. The other examples with $p_t>1$ 
 so far in the literature correspond to a combination of $p_t$ KK towers, which may be described effectively
 as a single tower with $p_t>1$.}.

	  The different options with $p,p_{\rm gon}=2,4,6$ are 
	shown in table \ref{tab: Pg and pkk}. Note that necessarily $p_{\rm gon}\leq p$, since gonion towers are only generated when the
	corresponding dimensions are large. To indicate this, the entries with $p_{\rm gon} > p$ are marked with an X in the table.  For the other choices of $p,p_{\rm gon}$ we have found explicit examples in section \ref{s:torito}, except for $p_{\rm gon}=0$ and $p>2$, which corresponds to configurations with only $\mathcal{N}=2$ sectors.
	
	In theories of QG in which a large number of species are present, a prominent role is played by the species scale $\Lambda_{QG}$ \cite{Dvali:2007hz,Dvali:2007wp,Dvali:2008ec}.
    It may be defined in different ways, presumably equivalent, and corresponds to the
	scale $\Lambda_{\rm QG}$  at which quantum gravity effects become strong and can no longer be ignored. This scale is well 
	below the Planck scale in the presence of  a large number $N$  of species and in $D$ dimensions is of order
	\beq
	\Lambda_{QG}\ \lesssim  \frac {M_{\rm P}^{(D)}} {N^{1/(D-2)} } \ ,
	\label{species}
	\eeq
	where $M_{\rm P}^{(D)}$ is the Planck scale in $D$ dimensions. When we have  a tower whose total number of states is $N_t=N$,
	one says that this tower {\it saturates} the species bound.  When a tower saturates the species scale then,  combining
	eqs.(\ref{tower}) and (\ref{species}) one obtains (in Planck units)
	\beq
	\Lambda_{QG} \ \simeq \  \left(m_0^{(t)}\right)^{p_t/(D-2+p_t)}
	\eeq
	In our case, we have three types of towers: 1) bulk KK+winding  (multiplicity $p$), 2) KK+winding towers on the
 branes, 3) gonion (multiplicity $p_{\rm gon}$) and 3) 4d strings (either fundamental or dual). In all the limits considered 
	in section \ref{s:Ylimit} that are not emergent string limits, and in particular in all the limits of section \ref{s:torito}, the Species Scale is the string scale, $\Lambda_{\rm QG}\simeq M_s$ and the geometric KK towers saturate the species scale. 
	Thus when we have $p=6$ the above formula yields $M_s\simeq m_{\rm KK}^{3/4}$ which is consistent with the example  in figure \ref{fig:scales}-II) in which	six dimensions are decompactified. When $p=2$, as in  figure  \ref{fig:scalesEFT}-IIa) with two dimensions large, one rather has $M_s\simeq m_{\rm KK}^{1/2}$.
	
	Gonion towers may or may not saturate the species bound. When $p_{\rm gon}=p$ there is saturation but there is not for $p_{\rm gon}<p$. This depends on whether
	the total number of gonion states $N_g$ equals the KK total $N=(M_s/M_{\rm P})^2$. In general  the number of gonions 
	along the $i^{\rm th}$  complex dimension is $\sim (M_s/m_{\rm gon}^{i})^2$ so that 
	\beq
	N_g\ \simeq \  \prod_{i=1}^{p_{\rm gon}/2} \left( \frac {M_s}{m_{\rm gon}^i}\right)^2 \, .
	\eeq
	If the gonion tower saturates the species scale (which only happens for $p_{\rm gon}=p=2,4,6$), one has $N_g=N=(M_{\rm P}^2/M_s^2)$ 
	and one can write (in Planck units)
	\beq
	M_s^{(p_{\rm gon}+2)} \ \simeq \ \prod_{i=1}^{p_{\rm gon}/2}  (m_{\rm gon}^i)^2 \ .
	\label{stringsc}
	\eeq 
	Threfore, one has $m_{\rm gon}\simeq M_s ^{1+2/p}/M_{\rm P}^{2/p} = e^{2\phi_4/p} M_s$. Note that this agrees with the results described in the examples of section \ref{s:torito} and figure \ref{fig:scalesEFT}-IIa) above.
	Note also that in the cases where the gonion tower does not saturate the bound (\ref{stringsc}) does not apply.

    Let us consider for example the case of the toy model \eqref{toymodel} and the gonion towers living at the intersection $aa^*$.  Let us in particular consider $u\simeq u^{(i)}$ and the limit $s\sim u^3\sim \lambda$.
    We have gonions at the three complex planes and six decompactifying dimensions, so that $p_{\rm gon}=p=6$. The gonion masses are all of order $m^i_{\rm gon}\simeq M_{\rm P}/u^2$.  From table \ref{tab:gonionguay} one can compute the total number of gonion states $N_{aa^*}$ in the sector $aa^*$
    \beq
    N_{aa^*} \ \simeq \prod_i\left(\frac {M_s}{m^i_{\rm gon}}\right)^2 \ \simeq \ k^3 s^2 e^{2\phi_4} \ \simeq \ k^3 s\, .
    \eeq
  One can compute the total number of species from the geometric KK states using e.g.
  the species scale expression $N=M_{\rm P}^2/\Lambda_{QG}^2\simeq u^3\sim s$. Hence $N_g=N$ and the gonion tower at $aa^*$ 
  also saturates the species scale. Differently, at the intersection $ab$ there is only a tower in  the second complex plane so that $N_{ab}\simeq (M_s/m_{{\rm gon},2})^2\simeq ku \simeq kN^{1/3}$ and the tower does not saturate  the species scale.

	\subsection{Emergence of kinetic terms }

	We have shown in previous sections how in the limit in which Yukawa couplings vanish, towers of charged states (gonions) 
	become light, and have also discussed how in that limit the metric of the chiral fields localised at brane intersections
	becomes singular in string units. It has been recently argued that all kinetic terms of massless fields in QG may vanish in the UV
	and that they are generated in the IR by summing over loop contributions over towers of states. This is the so called
	{\it Emergence proposal} \cite{Harlow:2015lma,Grimm:2018ohb,Heidenreich:2018kpg} (see also \cite{Heidenreich:2017sim,Lee:2018urn,Ooguri:2018wrx,Marchesano:2022axe,Castellano:2022bvr,Castellano:2023qhp,Blumenhagen:2023tev,Hattab:2023moj}) which in its stronger version may be stated as follows:  
	
	{\it In a theory of Quantum Gravity all light particles in a perturbative regime have no kinetic terms in the UV.  The required kinetic terms appear
	as an IR effect due to loop corrections involving the sum over a tower of asymptotically massless states.}
 
A milder version of this proposal states that for any singularity at infinite distance in moduli space, there is an
infinite tower of states becoming massless which induces quantum corrections to the metrics matching the tree level behaviour.

	It is interesting to check whether our findings in the previous sections are consistent with this hypothesis or not. Specifically 
	let us consider the subsector of the theory in which two D6-branes,  D6$_\alpha$ and D6$_\beta$ branes intersect at small angles. There will be towers of gonion states	localised at their intersection. We would like to check whether the sum over one-loop diagrams involving the gonions are able to reproduce the form of the kinetic terms of the chiral massless states and the $U(1)$ gauge boson under which the tower is charged.

\begin{figure}[ht]
\centering
\begin{subfigure}{.5\textwidth}
\centering
% Aquí va el código de tu primera figura
\begin{tikzpicture}[very thick,q0/.style={->,blue,semithick,yshift=5pt,shorten >=5pt,shorten <=5pt}]

  % Loop
  \def\radius{1.5}
  \draw (0,0) circle (\radius);
  \node[above] (1) at (0,\radius) {$\psi_n$};
  \node[below] (2) at (0,-\radius) {$\psi_n$};
  \draw[q0] (140:0.75*\radius) arc (140:40:0.75*\radius) node[midway,below] {$q$};
   \draw[q0,yshift=-10pt] (320:0.75*\radius) arc (320:220:0.75*\radius) node[midway,above] {$q-p$};
   \draw (-3,0) node[left] {$V$};
   \draw (3.7,0) node[left] {$V$};

  % External lines
 % External lines with new style

% External lines with sinusoidal style
\draw[black, decorate, decoration={snake, amplitude=.1cm, segment length=5mm, post length=0mm}] (-2*\radius,0) -- (-\radius,0);
\draw[black, decorate, decoration={snake, amplitude=.1cm, segment length=5mm, post length=0mm}] (\radius,0) -- (2*\radius,0);

\filldraw
(-\radius,0) circle (2pt)
(\radius,0) circle (2pt);
  \draw[q0] (-2*\radius,0.1) -- (-\radius,0.1) node[midway,above] {$p$};
  \draw[q0] (\radius,0.1) -- (2*\radius,0.1) node[midway,above] {$p$};
\end{tikzpicture}
\vspace{0.95mm}
\end{subfigure}%
\begin{subfigure}{.5\textwidth}
\centering
% Aquí va el código de tu segunda figura
\begin{tikzpicture}[very thick,q0/.style={->,blue,semithick,yshift=5pt,shorten >=5pt,shorten <=5pt}]
% Loop con estilo de línea discontinua
\def\radius{1.5}
\draw[dashed] (0,0) circle (\radius);
\node[above] at (0,\radius) {$\tilde{\psi}_n$};
\node[below] at (0,-\radius) {$\tilde{\psi}_n$};
\draw[q0] (140:0.75*\radius) arc (140:40:0.75*\radius) node[midway,below] {$q$};
\draw[q0,yshift=-10pt] (320:0.75*\radius) arc (320:220:0.75*\radius) node[midway,above] {$q-p$};
\draw (-3,0) node[left] {$V$};
\draw (3.7,0) node[left] {$V$};

  % External lines
 % External lines with new style

% External lines with sinusoidal style
\draw[black, decorate, decoration={snake, amplitude=.1cm, segment length=5mm, post length=0mm}] (-2*\radius,0) -- (-\radius,0);
\draw[black, decorate, decoration={snake, amplitude=.1cm, segment length=5mm, post length=0mm}] (\radius,0) -- (2*\radius,0);

\filldraw
(-\radius,0) circle (2pt)
(\radius,0) circle (2pt);
  \draw[q0] (-2*\radius,0.1) -- (-\radius,0.1) node[midway,above] {$p$};
  \draw[q0] (\radius,0.1) -- (2*\radius,0.1) node[midway,above] {$p$};
\end{tikzpicture}
% Puedes añadir una leyenda individual si lo deseas
% \caption{Leyenda de la segunda figura}
\end{subfigure}
\caption{One-loop contributions to the gauge kinetic functions from a tower of gonions.}
\label{emergauge}
\end{figure}
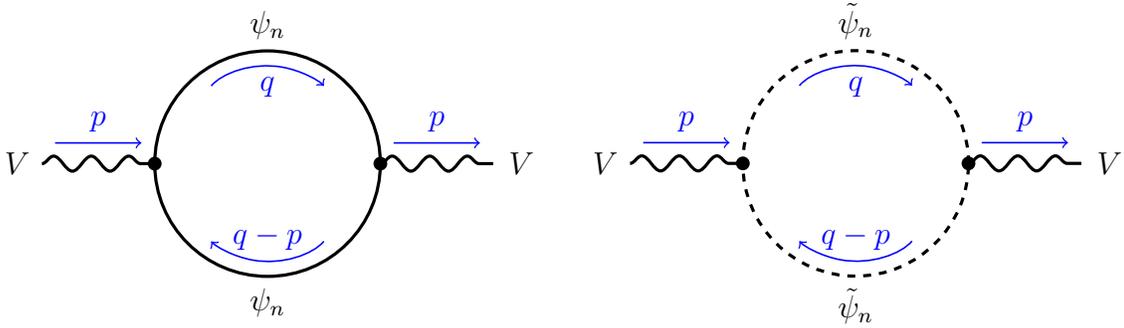

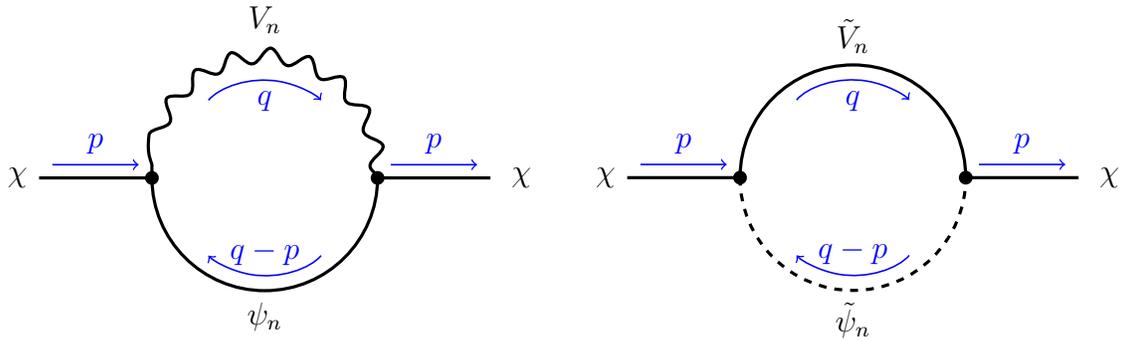
\begin{figure}[ht]
\centering
\begin{subfigure}{.5\textwidth}
\centering
% Aquí va el código de tu primera figura
\begin{tikzpicture}[very thick,q0/.style={->,blue,semithick,yshift=5pt,shorten >=5pt,shorten <=5pt}]
% Loop
\def\radius{1.5}

% Semicírculo superior con estilo snake
\draw[black, decorate, decoration={snake, amplitude=.1cm, segment length=5mm, post length=0mm}] (0:\radius) arc (0:180:\radius);
\node[above] at (0,1.18*\radius) {$V_n$};

% Semicírculo inferior normal
\draw (180:\radius) arc (180:360:\radius);
\node[below] at (0,-\radius) {$\psi_n$};
  \draw[q0] (140:0.75*\radius) arc (140:40:0.75*\radius) node[midway,below] {$q$};
   \draw[q0,yshift=-10pt] (320:0.75*\radius) arc (320:220:0.75*\radius) node[midway,above] {$q-p$};

% Semicírculo inferior normal

\draw (-3,0) node[left] {$\chi$};
\draw (3.7,0) node[left] {$\chi$};

% Líneas exteriores
\filldraw
(-2*\radius,0) -- (-\radius,0) circle (2pt)
(\radius,0) circle (2pt) -- (2*\radius,0);
\draw[q0] (-2*\radius,0) -- (-\radius,0) node[midway,above] {$p$};
\draw[q0] (\radius,0) -- (2*\radius,0) node[midway,above] {$p$};
\end{tikzpicture}
% Puedes añadir una leyenda individual si lo deseas
% \caption{Leyenda de la primera figura}
\vspace{0.95mm}
\end{subfigure}%
\begin{subfigure}{.5\textwidth}
\centering
% Aquí va el código de tu segunda figura
\begin{tikzpicture}[very thick,q0/.style={->,blue,semithick,yshift=5pt,shorten >=5pt,shorten <=5pt}]
% Loop
\def\radius{1.5}

% Semicírculo superior con estilo snake
% Semicírculo superior normal
\draw (0:\radius) arc (0:180:\radius);
\node[above] at (0,\radius) {$\tilde{V}_n$};

% Semicírculo inferior con estilo dashed
\draw[dashed] (180:\radius) arc (180:360:\radius);
\node[below] at (0,-\radius) {$\tilde{\psi}_n$};
  \draw[q0] (140:0.75*\radius) arc (140:40:0.75*\radius) node[midway,below] {$q$};
   \draw[q0,yshift=-10pt] (320:0.75*\radius) arc (320:220:0.75*\radius) node[midway,above] {$q-p$};

% Semicírculo inferior normal

\draw (-3,0) node[left] {$\chi$};
\draw (3.7,0) node[left] {$\chi$};

% Líneas exteriores
\filldraw
(-2*\radius,0) -- (-\radius,0) circle (2pt)
(\radius,0) circle (2pt) -- (2*\radius,0);
\draw[q0] (-2*\radius,0) -- (-\radius,0) node[midway,above] {$p$};
\draw[q0] (\radius,0) -- (2*\radius,0) node[midway,above] {$p$};
\end{tikzpicture}
% Puedes añadir una leyenda individual si lo deseas
% \caption{Leyenda de la segunda figura}
\end{subfigure}
\caption{One-loop contributions to a chiral fermion kinetic metric from a tower of gonions and
KK gauge tower.}
\label{emerfermion}
\end{figure}
	Let us start with $g_{\a\b}^{-2}$, the gauge kinetic function of the gauge group generated by  $Q_{\alpha \beta }\equiv Q_\alpha-Q_{\beta}$,
	which is the one felt by the towers at the intersection.  All the tower of gonions has the {\it same charge } which we denote by $q_{\alpha \beta}$.
	Gonions will include vector-like $\CN=1$ chiral multiplets which will contribute to the renormalisation of the gauge kinetic function
	at one loop. 	
	Loops like those in figure \ref{emergauge} contribute to the gauge kinetic function of the gauge group $Q_{\alpha \beta}$ 
	(see \cite{Castellano:2022bvr} for details of the computation)\footnote{ To match the results known in explicit string models it is crucial that
 all the states in the tower have the same charge. This is unlike e.g. $U(1)$ BPS towers in which charges grow with the level.}
	\beq
	\delta(g_{\a\b}^{-2})  \, \simeq \, q_{\a\b}^2 \sum_{k=1}^{N_g}   \log (\Lambda_{QG}^2/m_k^2) \, \simeq \, q_{\a \b }^2 \ N_g 
	= q_{\a\b}^2 \prod_{i=1}^{p_{\rm gon}/2} \left( \frac {M_s}{m_{\rm gon}^i}\right)^2  \, .
 \label{kineticgauge}
	\eeq
	In terms of gonion angles, i.e. $(m_{\rm gon}^i/M_s)^2\simeq \theta_i$,  one has 
	\beq
	\delta (g_{\a\b}^{-2})  \, \simeq \, q_{\a\b}^2 \ \prod_{i=1}^{p_{\rm gon}/2}  \frac {1}{\theta_i}  \, .
	\eeq
	So we learn that as the gonions become light, the gauge coupling becomes small.
	The idea of emergence implies that if $N_g=N$ and the tower saturates the species scale, the full
	gauge kinetic function will be given by
	\beq
	g_{\a\b}^{-2} \, \simeq \, q_{\a\b}^2 N \, \simeq \ q_{\a \b}^2 \left(\frac {M_{\rm P}}{M_s}\right)^2 \, =  \, q_{\a \b}^2 \ e^{-2\phi_4} \, ,
	\eeq
	where $e^{\phi_4}$ is the 4d dilaton. Note that this is independent of the tower structure and only rests on the equality $N_g=N$.

	Using the above expressions one can also write an expression in which $M_s$ does not appear explicitly
	\beq
	g_{\a\b}^{-2} \ \simeq \ \prod_{i=1}^{p_{\rm gon}/2} \left(\frac {M_{\rm P}}{m_{\rm gon}^i}\right)^{4/(2+p_{\rm gon})}\, .
	\eeq
 If the tower of gonions does not saturate the species scale, then there is a contribution 
 to the $g_{\a\b}^{-2}$ gauge kinetic function as in (\ref{kineticgauge}) but the above expression  does not apply.
 Note that $g_{\a\b}^{-2}$ is just  the contribution of the intersection $\alpha \beta$ to the gauge group with generator
 $Q_\a - Q_\b$. Additionally, there may be in general other contributions  to the gauge kinetic terms of 
 $U(1)_\a$ and $U(1)_\b$ from other intersections. Let us again consider the toy example (\ref{toymodel}).
 The contribution from the intersection of the $aa^*$ to the  gauge coupling $g_{a}$ will be
 \beq 
 \delta(g_a^{-2})  \, \simeq \, (2)^2N_g\, \simeq \, 4k^3 s \, .
 \eeq
Interestingly, this matches with the tree level value for the coupling constant  $g_a$ in (\ref{gacouplings}), as expected from emergence in the limit $s\sim  u^3\simeq \lambda$ in which the $u$-dependent pieces are 
subleading. In fact one can see that such subleading terms are induced by the contribution to $g_{aa}^{-2}$ from the 
intersections $ab$ and $ac$:
\beq
  \delta(g_a^{-2})\, \simeq \, \left(\frac {M_s}{m_{{\rm gon},2}}\right)^2 \, \simeq \, ku  \, , \qquad  
  \delta(g_a^{-2})\, \simeq \, \left(\frac {M_s}{m_{{\rm gon},3}}\right)^2 \, \simeq \, ku \, . 
 \eeq

Let us consider now the metric of the chiral fields at the D6-brane intersection $\a \b$. In this case in the loop circulates a tower of gonions,  with the same charge as the massless field, along with a tower of KK states living in the worldvolume of the intersecting branes. The trilinear couplings involving the massless chiral field, a gonion and a KK state were described in  e.g. \cite{Berasaluce-Gonzalez:2012abm,Hamada:2012wj}. The multiplicities  of both towers are the same  and the one loop diagrams as in figure \ref{emerfermion} have analogous overall behaviour than the  previous computation. One thus has again
\beq
         K_{\alpha \beta } \, \simeq    \,  \prod_{i=1}^{p_{\rm gon}/2} \left( \frac {M_s}{m_{\rm gon}^i}\right)^2\, .
         \eeq
         Again, if the tower of gonions saturates the spacies scale one gets
                  \beq
           K_{\alpha \beta }  \, \simeq \, \prod_{i=1}^{p_{\rm gon}/2}  \left(\frac {M_{\rm P}}{m_{\rm gon}^i}\right)^{4/(2+p_{\rm gon})}\, .
           \label{metricaMp}
           \eeq
          Note that this result matches with that in eq.(\ref{metricgon}), assuming the gonion tower saturates 
        the species scale.  Note also that in that saturation case one has
        \beq
         K_{\alpha \beta } \, \simeq \, g_{\a\b}^{-2} \, \simeq \, N_g\, \simeq N\, \simeq \,  \left(\frac{M_{\rm P}}{M_s}\right)^2 \, 
         =\, e^{-2\phi_4} \, ,
         \eeq
        which relates the metric at the singular intersection and the gauge kinetic function of the
        gauge boson coupling to $Q_\alpha -Q_\b$. This means that when the metric at the intersection becomes singular, these gauge couplings go to zero. This is in agreement with what we have seen in previous sections.
          
Emergence ideas may also give information about the kinetic term of the moduli involved in the problem. Thus e.g., consider the asymptotic EFT string limits. There the masses of the KK states behave like
\beq 
m_{\rm KK}(\lambda)  \, \simeq \, \frac {1}{\lambda^{w/2}} \, .
\eeq
One can then compute the metric  of the $\lambda$ field by computing the one-loop corrections involving the full tower of KK states up to the species scale. This is by now a well known computation which yields for the metric (see \cite{Grimm:2018ohb,Heidenreich:2018kpg,Castellano:2022bvr}):
                  \beq
                  K_\lambda \, \simeq \, \frac {1}{m_{\rm KK}^2} \left( \partial_\lambda m_{\rm KK}\right)^2\, \simeq \, \frac {1}{\lambda^2} \, ,
                  \eeq
                  and hence a leading term for the $\lambda$-dependent K\"ahler potential
                  \beq 
                  K_Q(\lambda) \, \sim \, - \log \lambda \ ,
                  \eeq
                 as expected.

 \subsection{Singular Yukawa couplings and emergence}

 In \cite{Castellano:2022bvr,Castellano:2023qhp} it was already stated, using emergence arguments, that the (non-vanishing) Yukawa couplings  of 3-particles
 with associated towers of states with characteristic scales $m_a$ takes the qualitative form
 \beq
 Y_{ijk}  \simeq  \left(\frac {m_i}{M_s}\right)^{\gamma_i}\left(\frac {m_j}{M_s}\right)^{\gamma_j}\left(\frac {m_k}{M_s}\right)^{\gamma_k} \, ,
 \eeq
 with the three factor arising from wavefunction renormalisation of the three legs, and $\gamma_{a,b,c}$ being 
 positive rational numbers depending on the structure of the towers. In the present paper we confirm that expected structure 
 but within the realm of specific string models, which also feature specific charged towers: the gonion particles. We also work within  ${\cal N }=1$ supergravity (not only rigid SUSY as in \cite{Castellano:2023qhp}). Thus in section \ref{s:Ylimit} we obtained 
 the expression
\beq
Y_{ijk}  \simeq  e^{\phi_4/2} 
\left(\frac {m^i_{\rm gon}}{M_s}\right)^{1/2}\left(\frac {m^j_{\rm gon}}{M_s}\right)^{1/2}\left(\frac {m^k_{\rm gon}}{M_s}\right)^{1/2} \, ,
\eeq
where $m^i_{\rm gon}$ is the lightest of the gonion towers at the intersection corresponding to the particle $i$, and the
same for $j,k$. In the interesting case in which only one of the intersections (say that associated to particle $a$) 
is singular, there are no towers associated to the other two particles in the coupling and one has
\beq
Y_{ijk} \, \simeq \, e^{\phi_4/2} 
\left(\frac {m^i_{\rm gon}}{M_s}\right)^{1/2} \, .
\label{Yg1}
\eeq
If in addition the gonion towers associated to particle $i$ saturate the species scale so that $N_g=N$
one can write 
\beq
Y_{ijk} \, \simeq \, e^{\phi_4/2} N^{-1/2}\, \simeq \, e^{\phi_4} \, \simeq \, g_{\a \b} \, ,
\label{Yg2}
\eeq
where $g_{\a \b}$ is the gauge coupling of the gauge boson coupling to $Q_\alpha-Q_\b$. 
An example of this setup is the toy model with the limit \eqref{quasiEFTIIa} where $s,u^{(2)}\sim \lambda$ and $u^{(1)}$, $u^{(3)}$ are bounded.
This yields a spectrum  of the EFT string II-a) type, as in figure \ref{fig:scalesEFT}.
In this case we have $p=p_{\rm gon}=2$ and a tower of gonions in the
intersection $ab$ in the second complex plane.
Since $p=2$ one has $M_s^2=m_{\rm KK}M_{\rm P}\sim M_{\rm P}^2/u$, so that $N=u$. On the other hand the number of gonions is
$N_g=(M_s/m_{{\rm gon},2})^2=u$, so the  gonion tower saturates. So one concludes that the Yukawa coupling 
should go like $Y\sim g_a\sim 1/\lambda^{1/2}\simeq 1/u^{1/2}$, which agrees with what is found in the previous chapters.

\subsection{Chiral towers and the (sub)Lattice/Tower WGC}

 As we have seen, the towers of gonions at an intersection of two branes $a,b$ all have the same bifundamental 
 charge. They are 
 in fact charged  under the linear combination of $U(1)$'s proportional to $Q_a-Q_b$, but not the 
 orthogonal one. This $U(1)$ gets a St\"uckelberg mass, but this mass is in general much lighter 
 than the gonion masses, so that there is typically a large range of scales at which we have a tower of
 states, with the same charge, coupling to an approximately massless $U(1)$.  As we mentioned in the introduction, this violates 
 the (sub)Lattice and Tower  WGC, since at least at the particle level, there is no trace of a tower containing a
full spectrum of $U(1)$ charges. In fact this is not totally surprising since examples of violations have appeared already in  the literature in non-chiral setups  \cite{Alim:2021vhs,Cota:2022yjw,Cota:2022maf,Cota:2023uir}. The inspiration to put forward those extensions came from demanding  the consistency of the WGC under dimensional reduction \cite{Heidenreich:2015nta,Heidenreich:2016aqi,Andriolo:2018lvp,Montero:2016tif}. 
In particular, starting with an extremal particle under a $U(1)$ and compactifying on a circle, an extra $U(1)_{\rm KK}$ gauge boson appears. Studying the 
convex hull for  these two $U(1)$'s one can easily see that the WGC is violated. This problem can be
cured if there is a full tower of states with all the charges, as in those mentioned extensions. However, in our case, a different remedy is at play. 

\begin{figure}[ht]
\centering
\begin{tikzpicture}
% Eje horizontal más largo
\draw[-,thick] (0,0) -- (3.5,0);
\draw[-,black, very thick,dotted] (3.5,0) -- (4.5,0);
\draw[->,thick] (4.5,0) -- (8,0) node[right] {$Z_{U(1)}$};
% Cuarto de circunferencia de radio 2
% Cuarto de circunferencia de radio 2 pintado de gris oscuro
\filldraw[fill=gray!60] (0,0) -- (2.5,0) arc (0:90:2.5) -- cycle;
% Eje vertical más corto
\draw[->,thick] (0,0) -- (0,4) node[above] {$Z_{KK}$};

\node at (6.7,-0.4) {$g_{U(1)}^{-1}$};
\draw[.,dotted, line width=0.4mm,blue!95] (0,2.5) -- (3.2,2.44);
\draw[.,dotted, line width=0.4mm,blue!35] (3.2,2.44) -- (3.53,2.425);
\draw[.,dotted, line width=0.4mm,blue!35] (4.5,2.36) -- (4.69,2.30);
\draw[dotted, line width=0.4mm, blue!95] (4.7,2.3) .. controls (6.3,1.6) and (6.5,0.8) .. (6.7,0);
% Etiquetas de los ejes
\node at (2.5,-0.3) {$1$};
\node at (-0.3,2.5) {$1$};
\end{tikzpicture}
\caption{Blue line corresponds to the convex hull of the charge-to-mass ratio $\vec{Z}=\vec{Q}/m^{\rm KK}_{\rm gon}$, where $m^{\rm KK}_{\rm gon}$ are the gonion Kaluza-Klein replicas. If the radius of the circle compactification satisfies $R\gtrsim M_{\rm P}^{-1}$, the convex hull always contains the extremal region (grey area, see \cite[eq.(83)]{Heidenreich:2015nta}). Here we stay in a perturbative regime $g_{U(1)} \ll 1$.}
 \label{fig: convexhull}
\end{figure}
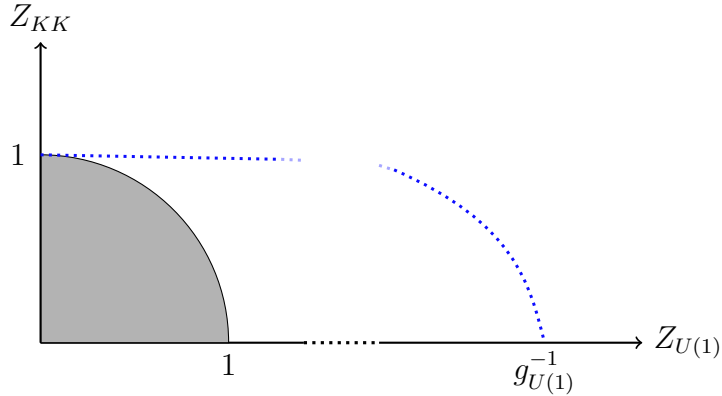

Indeed, the results of section \ref{s:torito} (see limits \eqref{quasiEFTIIa} and \eqref{guaynonEFT}) and the discussion in the last subsection (namely eqs.\eqref{Yg1} and \eqref{Yg2}) show that in some instances the charged  gonion is not extremal, but has a further power of $g$ suppression, with $m_{\rm gon}\simeq g_{U(1)}^2 M_{\rm P}$ and $g_{U(1)} \simeq e^{-\phi_4}$. Then one can see that a compactification down to three dimensions does not lead to a violation of the
of the Convex Hull WGC (see figure \ref{fig: convexhull}). Naively, this seems like a novel way of avoiding WGC inconsistencies, and it would be interesting to explore possible connections with other exceptions to the (sub)Lattice/Tower WGC
discussed in the literature \cite{Cota:2022yjw,Cota:2023uir}. 

As a final comment, let us stress an important difference between the $U(1)$'s we are referring to and those present in non-chiral theories. In our setup, the $U(1)$ coupling to the massless chiral fields oftentimes leads to anomalies, which are  cancelled by a 4d Green--Schwarz mechanism, as it is well-known to happen for this class of theories. As a result these $U(1)$'s are not strictly massless, but in general get a St\"uckelberg mass.

%%%%%%%%%%%%%%%%%%%
%%%%%%%%%%%%%%%%%%%
	
\section{Final comments and outlook}
\label{s:conclu}

In recent years much has been learned about the behaviour of  EFT theories from QG and string theory  at infinite
distance limits in moduli space.  However most of the setups explored  so far display $\geq 8$ supercharges. 
In the present paper we have concentrated  instead on  the case of  4d ${\cal N}=1$ chiral theories, which are more
relevant for possible phenomenological applications. In fact, independently of any possible phenomenological application,
this class of theories present novel features like chiral fermions and  Yukawa couplings, which we have found give rise to 
new,  not yet well explored phenomena.
We have used as a laboratory type IIA CY orientifolds with chiral charged multiplets localised at D6-brane intersections.
We believe though that,  thanks to the different dualities to type IIB, F-theory and heterotic strings, the qualitative results
are expected to apply in more general settings. 

 One first question we have asked is whether, in analogy to gauge couplings, the  limit $Y\to 0$ is at infinite distance. We have argued that this is indeed the case when moving in complex structure moduli space, where it is clear that this limit can be attained, unveiling a rich casuistic of limits with novel physical features.  Indeed, in addition to the expected KK and string towers, we have found that in that limit a slightly more exotic type of towers appear. They correspond to towers of {\em gonions}: charged chiral states in massive vector-like representations of the D6-brane gauge group,  all having the same bifundamental charge as the zero modes and also living at the D6-brane intersections. Although in type IIA these states come from open string oscillators stretched from one intersecting brane to another, the spectrum is not Hagedorn-like but rather has the form $m_{gon}^2=\theta_{ab} M_s^2$. In fact one can write the Yukawa couplings in these limits 
in terms of the gonion masses, as in \eqref{yukigon}, showing explicitly how a vanishing Yukawa coupling implies in these limits
the presence of a light tower of charged gonions.  

We find that, in limits parametrised by a single complex structure field one gets  a general behaviour of the form $Y_*\simeq 1/u^r$,
with $r>0$ a rational number, which take values $r=1/4,1/3,1/2,3/4,1$ in most of the examples explored. Interestingly, this type of behaviour is consistent with recent results \cite{Butbaia:2024tje,Constantin:2024yxh}
in heterotic string theory, and it would be interesting to explore possible connections. 
One also finds that, as the gonion towers become light, some of the kinetic terms  of chiral fermions involved in the Yukawa coupling diverge, which is at the root of the vanishing of the Yukawa, once one goes to canonical kinetic terms. At the same time, some D6-brane gauge group factors coupling to both chiral matter and their gonions  go also to zero as the Yukawa does, like $Y_*\sim g_*^{2r}$. This also explains why the  limit $Y_*\rightarrow 0$ must be at infinite distance, since such a limit would imply the presence of global  symmetries.

The towers of charged gonions are quite interesting theoretically, since they appear to provide examples in which a (sub)Lattice/Tower WGC conjecture does not hold. In particular, the  gonion tower states have all the same charge as the massless chiral fermions and there is no trace of particles with multiple charges. Still, this does no lead to any contradiction upon dimensional reduction since gonion masses are not extremal particles and have a further $g_*$ suppression factor in their mass.  A different direction that we study is if our findings concerning the kinetic terms of chiral fields and gauge bosons as well as the Yukawa couplings are consistent with the {\it Emergence} hypothesis. We find that indeed that is the case, and the particular form of the gonion towers is crucial to 
reproduce what is known from explicit examples. It would be interesting to make a more thorough analysis in a variety of 
explicit models to make further tests of this hypothesis, as this class of ${\cal N}=1$ 4d models offers very rich possibilities.

We think we have only scratched the surface of 
 ${\cal N}=1$  4d models in what concerns infinite distance properties and Swampland constraints,  and we 
believe that they may yet give us many surprises. Furthermore, the comparison of our findings with alternative computations 
of Yukawa couplings in other type of compactifications \cite{Butbaia:2024tje,Constantin:2024yxh} would be very interesting.
Last but not least, the limit of small Yukawas may be of much phenomenological relevance since, as we said, most of the
Yukawa couplings of the SM are small. We will present results in this direction, particularly in the context of neutrino masses, in a companion paper \cite{Casas:2024clw}.

\bigskip

\bigskip

\bigskip

\centerline{\bf \large Acknowledgments}

\vspace*{.5cm}

We would like to thank Alberto Castellano, José Luis Hernando and Ignacio Ruiz for discussions.  This work is supported through the grants CEX2020-001007-S and PID2021-123017NB-I00, funded by MCIN/AEI/10.13039/501100011033 and by ERDF A way of making Europe. G.F.C. is supported by the grant PRE2021-097279 funded by MCIN/AEI/ 10.13039/501100011033 and by ESF+. 

%%%%%%%%%%%%%%%%%%%
%%%%%%%%%%%%%%%%%%%

\appendix

\section{Gonion spectrum for D6-branes at angles}
\label{ap:spectrum}

In this appendix we describe the mass spectrum of open string states between two intersecting D6-branes, following \cite{Aldazabal:2000dg,MarchesanoBuznego:2003axu}. For examples of such spectra in models with realistic gauge sectors see for instance \cite{Ibanez:2001nd}.

Two static D6-branes can intersect transversely in at most three angles. At such intersection, one can model the open-string spectrum by considering the flat-space setup of \cite{Berkooz:1996km,Arfaei:1996rg}, and then applying the bosonisation language that is standard from orbifold twisted sectors \cite{Green:2012oqa,Bailin:1994qt}. In this language the intersection angles are encoded in the four-dimensional twist vector $v_{\th} = (\th_{\a\b}^1,\th_{\a\b}^2,\th_{\a\b}^3,0)$ of the $\a\b$ sector, where $-1 < \th_{\a\b}^r \leq 1$ measures the three intersection angles in units of $\pi$ and the last entry is identified with the complex dimension filled by the D6-branes besides the light-cone directions. Oscillation states are described by the sum $\chi = r + v_{\th}$, with $r \in \Z^4$ in the Neveu--Schwarz and $r \in (\Z + \oh)^4$ in the Ramond sector, and where the GSO projection is implemented via $\sum_i r^i =$ odd. Finally, if one of the angles vanishes then the D6-branes are parallel in one dimension and can be separated by a distance $L$. In terms of these quantities the mass of an open string state in the $\a\b$ sector is given by
\beq
\a' m_{\a\b}^2 = \frac{L^2}{4\pi\a^\prime} + N_{bos}(\th_{\a\b}^r) + \frac{(r + v_\th)^2}{2} -\oh + E_{\a\b},
\label{mass2}
\eeq
where $E_{\a\b}$ is the vacuum energy
\beq
E_{\a\b} = \sum_{r=1}^3 \oh |\th_{\a\b}^r| \left(1 - |\th_{\a\b}^r |\right)\, .
\label{vacuumenergy}
\eeq
Finally,  $N_{bos}(\th_{\a\b}^r)$ stands for the contribution coming from worldsheet bosonic oscillations on top of either of the above R or NS states. The creator operators that implement twisted oscillations are of the form $\a_{-|\th^r|}$ or $\a_{|\th^r|-1}$ and increase the mass of the state by $|\th_{\a\b}^r|/\a'$ and $(1 - |\th_{\a\b}^r|)/\a'$, respectively.

Given these formulae it is easy to construct the spectrum at each D6-brane intersection and see its explicit dependence on the intersection angles. Notice that the fourth entry of the vector $r + v_\th$, which will be either an integer (NS sector) or a half integer (R sector), provides the 4d Lorentz quantum number of the open string state. Let us for instance consider an intersection such that $\th_{\a\b}^1, \th_{\a\b}^2 > 0$, $\th_{\a\b}^3 < 0$. and $|\th_{\a\b}^r| \ll 1$. Then it is easy to check that we will have a unique massless state coming from the R sector:
\beq
\chi = r + v_\vt= ( \th_{\a\b}^1-\oh, \th_{\a\b}^2-\oh, \th_{\a\b}^3+\oh, -\oh)\, ,
\label{fermion}
\eeq
which, in terms of four-dimensional physics, corresponds to a Weyl fermion of negative chirality. On the other hand, the lightest excitations in the NS sector are given by:
\beq
\begin{array}{ccc}
{\bf State} & & {\bf Mass}^2  \\
\chi = (\th_{\a\b}^1-1, \th_{\a\b}^2, \th_{\a\b}^3, 0)  & & \a' m_{\a\b}^2 = \oh \left(-|\th^1_{\a\b}| + |\th_{\a\b}^2| + |\th_{\a\b}^3| \right) \\
\chi = (\th_{\a\b}^1, \th_{\a\b}^2-1, \th_{\a\b}^3, 0)  & & \a' m_{\a\b}^2 = \oh \left(|\th^1_{\a\b}| - |\th_{\a\b}^2| + |\th_{\a\b}^3| \right) \\
\chi = (\th_{\a\b}^1, \th_{\a\b}^2, \th_{\a\b}^3+1, 0) & & \a' m_{\a\b}^2 = \oh \left(|\th^1_{\a\b}| + |\th_{\a\b}^2| - |\th_{\a\b}^3| \right) \\
\chi = (\th_{\a\b}^1+1, \th_{\a\b}^2, \th_{\a\b}^3, 0)  & & \a' m_{\a\b}^2 = \oh \left(3|\th^1_{\a\b}| + |\th_{\a\b}^2| + |\th_{\a\b}^3| \right) \\
\chi = (\th_{\a\b}^1, \th_{\a\b}^2+1, \th_{\a\b}^3, 0)  & & \a' m_{\a\b}^2 = \oh \left(|\th^1_{\a\b}| +3 |\th_{\a\b}^2| + |\th_{\a\b}^3| \right) \\
\chi = (\th_{\a\b}^1, \th_{\a\b}^2, \th_{\a\b}^3-1, 0) & & \a' m_{\a\b}^2 = \oh \left(|\th^1_{\a\b}| + |\th_{\a\b}^2| +3 |\th_{\a\b}^3| \right) 
\end{array}
\label{scalars}
\eeq
These states will all correspond to $4d$ scalars, as its associate helicity shows. To each of these states we can apply the bosonic creation operators
\be
(\a_{-|\th_{\a\b}^1|})^{k_1} \, (\a_{-|\th_{\a\b}^2|})^{k_2}\, (\a_{-|\th_{\a\b}^2|})^{k_3} \, ,
\label{oscil}
\ee
from where the spectrum \eqref{towergon} is obtained. Finally, the lightest $W$-boson state is 
\beq
\begin{array}{ccc}
\chi = (\th_{\a\b}^1, \th_{\a\b}^2, \th_{\a\b}^3, \pm 1) & & \a' m_{\a\b}^2 = \oh \left(|\th^1_{\a\b}| + |\th_{\a\b}^2| + |\th_{\a\b}^3| \right) 
\end{array}
\label{Wvector}
\eeq
and so from applying \eqref{oscil} we recover the tower of states \eqref{towerWgon}. As we reach the string scale, further states will appear that come from either applying bosonic oscillators of the form $\a_{|\th^r|-1}$ or considering more general vectors $r$. The latter incorporates 4d helicities of higher spin, and so at this scale we recover the higher-spin, Hagedorn-like spectrum of a critical string. 

%%%%%%%%%%%%%%%%%%%
%%%%%%%%%%%%%%%%%%%

\section{Mirror duals of type I magnetised Calabi--Yau compactifications}
\label{ap:mirror}

In this appendix we translate the model building setup of \cite{Blumenhagen:2005pm,Blumenhagen:2005zg}, which features type I compactifications on smooth Calabi--Yau with $U(n)$ bundles, to the type IIA setting that we use along the main text. This framework is particularly useful for the construction of STU-like models in section \ref{s:Ylimit}.

Let us consider type I string theory compactified on a smooth Calabi--Yau $Y$, with a basis of Nef divisors $\ell_s^{-2} \om_i$ and triple intersection numbers $\kappa_{ijk} = \ell_s^{-6} \int_Y \om_i \wedge \om _j \wedge \om_k$. The K\"ahler potential for the K\"ahler sector reads
\be
K_Q = - \log s - \log \frac{1}{6} \kappa_{ijk} u^i u^j u^k\, ,
\ee
with
\be
s  = e^{-\phi} V_Y \, , \qquad 
u^i  =  e^{-\phi} t^i\, ,
\ee
$J = t^i \om_i$ the string frame K\"ahler form of $Y$ and $V_Y = \frac{1}{6} \kappa_{ijk} t^it^jt^k$ its volume. A single D9-brane $\a$ with a worldvolume flux $f_\a^i \om_i$ can be represented as
\be
[\Pi_\a] =  \oh [A] + \oh f_\a^i [D_i] - \left( \oh \kappa_{ijk} f_\a^j f_\a^k + \frac{1}{24} c_{2\, i} \right)[C^i] + \left(\frac{1}{6}  \kappa_{ijk} f_\a^i f_\a^j f_\a^k + \frac{1}{24} c_{2\, i} f_\a^i  \right)[B]\, ,
\ee
with $[D_i] = 2\ell_s^{-2}  \om_i$, and $[C^i] = \ell_s^{-4} \tilde{\om}^i$ a four-forms basis such that $\ell_s^{-6} \int_Y \om_i\wedge \tilde{\om}^j = \delta_i^j$. In addition $[B]= d{\rm vol}_Y/V_Y$ is a normalised volume form and $[A] = 2$.  Finally, $c_2(Y) = c_{2\, i} [C^i]$ is the decomposition of the second Chern class of $Y$. The orientifold image of D9-brane is obtained by performing the sing flip $f_\a^i \to - f_\a^i$. The chiral index between two D9-branes reads 
\be
[\Pi_\a] \cdot [\Pi_\b] = \frac{1}{6}  \kappa_{ijk} (f_\b^i - f_\a^i) (f_\b^j - f_\a^j) (f_\b^k - f_\a^k) + \frac{1}{12} c_{2\, i} (f_\b^i - f_\a^i)\, ,
\ee
where we have used the intersection rules $[A]\cdot[B]= 2$, $[C^i] \cdot [D_j] = 2\delta^i_j$. This choice of normalisation reflects that $[A]$ is Poincar\'e dual to a pair of Euclidean D5-branes wrapping $Y$ and with gauge group USp$(2)$, while $[C^i]$ are instead duals to $O(1)$ instantons made up from Euclidean D1-branes wrapping two-cycles of $Y$. Correspondingly, the class $[C^i]$ also represents space-time filling D5-branes wrapped on such two-cycles. Finally, the orientifold plane content amounts to the charge $[\Pi_{\rm O}] = -16 [A] + 2c_{2\, i} [C^i]$. 

From here one can derive the basic model-building rules used in \cite{Blumenhagen:2005pm,Blumenhagen:2005zg}, and one can see that the gauge couplings and FI-terms of these references map under mirror symmetry to the ones used in the main text. The mirror map to type IIA translates the classes $[A]$ and $[C^i]$ into odd three-forms, while $[B]$ and $[D_i]$ are mapped to even three-forms of the mirror Calabi--Yau $X$.

%%%%%%%%%%%%%%%%%%%
%%%%%%%%%%%%%%%%%%%

\section{Selection rules for Yukawas}
\label{ap:Hmom}

In this appendix we describe the selection rules for the Yukawa couplings. In doing so, we will get a rich picture of the constraints this imposes in terms of angle selection and gonion mass dependencies, which are important for the results of the main text.

Yukawa selection rules have been studied extensively in the context of heterotic orbifold theory \cite{Dixon:1986qv, Hamidi:1986vh, Cvetic:1987qx, Font:1988mm,Kobayashi:1990fx}. They are derived from the structure of the vertex operators and the orbifold symmetries. In particular, twisted vertex operators in the right-moving sector of the theory carry the so-called H-momentum, that is, the momentum along the bosonised fermions. Such momentum conservation is responsible for providing non-trivial selection rules that indeed forbid many couplings for both twisted and untwisted fields. 

This rationale is also applicable to the intersecting branes picture. Indeed, by dimensional reduction from 10d theory, one should be able to analogously translate the above H-momentum conservation condition into constraints on the twisted vector $r$ defined in \eqref{fermion}. These constraints read
\be
r^f_i + r^f_j + r^b_k = \left(0,0,0,\pm1\right) \label{eq: srulesint}
\ee
or equivalently,
\be
\sum_i r_i^b = \left(1,1,1,0\right),\label{eq: srulesb}
\ee
where $r^f$ and $r^b$ denote the fermionic and bosonic twisted vectors, respectively. One way to motivate these rules is from the viewpoint of a Higgssed 4d $\CN=4$ gauge theory. Indeed, let us consider a stack of three coincident D6-branes in flat space, with a $U(3)$ gauge theory in their worldvolume, and with an $\CN=4$ superpotential $W = W_{ijk} \tr \{ \Phi^i [\Phi^j, \Phi^k]\}$ for its adjoint fields. We now  give a linear vev to the off-diagonal entries of its adjoint fields such that it implements the gauge breaking $U(3)\rightarrow U(1)_a\times U(1)_b\times U(1)_c$.\footnote{Geometrically, we are describing the intersection of three D6-branes. See e.g. \cite{Marchesano:2010bs} for a similar analysis in the context of intersecting D7-branes.} From the initial $\CN=4$ superpotential one obtains a Yukawa coupling of the form $Y_{ijk}\sim Y\varepsilon_{ijk}\psi^i\psi^j\phi^k$, in agreement with the selection rules given by \eqref{eq: srulesb}. In a mirror picture of magnetised D9-branes, the same kind of rules would arise from the dimensional reduction procedure described in \cite{Cremades:2004wa}. For further selection rules on type II Yukawas see e.g. \cite{Higaki:2005ie,Abe:2009vi,Marchesano:2013ega}.

It should be emphasised that the above equations place strong constraints on the allowed Yukawa coupling, which are directly linked to the patterns and findings observed along this paper. To illustrate this with an example, let us consider three stacks of D6-branes with the following intersecting angles
\bea
\theta_{ab} &=& \left(\theta_{ab}^1, \theta_{ab}^2, \theta_{ab}^3, 0\right),\\
\theta_{bc} &=& \left(\theta_{bc}^1, \theta_{bc}^2, \theta_{bc}^3, 0\right),\\
\theta_{ca} &=& \left(\theta_{ca}^1, \theta_{ca}^2, \theta_{ca}^3, 0\right), 
\eea
and impose the supersymmetry condition $\sum_{i}\theta_{\a\b}^i =0$.\footnote{In general, supersymmetry requires $\sum_{i}\theta_{\a\b}^i =0$ mod $2\pi$. We will not consider $\sum_i \theta_{\a\b}^i\neq 0$ for the present example. }  Moreover, by the definition of $\theta_{\a\b}^i$, the sum $\theta_{ab}^i+\theta_{bc}^i+\theta_{ca}^i $ also cancels. This greatly restricts the number of possible configurations of $\theta_{\a\b}^i$. The parameter space is given by four variables, namely given $\{\theta^1_{ab},\theta^2_{ab},\theta^1_{bc},\theta^2_{bc}\}$ the full configuration is determined. Let us in particular take the following configuration
\bea
\theta_{ab} &=& \left(|\theta_{ab}^1|, |\theta_{ab}^2|, -|\theta_{ab}^3|, 0\right),\\
\theta_{bc} &=& \left(|\theta_{bc}^1|, -|\theta_{bc}^2|, |\theta_{bc}^3|, 0\right),\\
\theta_{ca} &=& \left(-|\theta_{ca}^1|, |\theta_{ca}^2|,| \theta_{ca}^3|, 0\right), 
\eea
and assume that none of the angles vanish. The lightest boson in each intersection is given by the following twisted vectors
\be
r_{ab}^b =\left(0,0,1,0\right),\quad r_{bc}^b = \left(0,1,0,0\right),\quad r_{ca}^b = \left(1,0,0,0\right), \ \label{eq: rbosons}
\ee
or equivalently, in terms of their supersymmetric fermionic partners,
\be
r_{ab}^f =\frac{1}{2}\left(-,-,+,-\right),\quad r_{bc}^f = \frac{1}{2}\left(-,+,-,-\right),\quad r_{ca}^f = \frac{1}{2}\left(+,-,-,-\right).
\ee
As can be seen, the selection rules for the lightest fields in each intersection are fulfilled. Notice that this will not hold for all Yukawas, and in the model of section \ref{s:torito} some Yukawas will be forbidden at tree level. In that case, Yukawa couplings arise between zero modes and heavier modes.

Applying equations \eqref{metricgen} and \eqref{eq: rbosons}, the K\"ahler metrics entering the Yukawa are
\bea
K_{ab}&=& \frac{e^{K/2-\phi_4}}{\sqrt{2\pi}}  \left( \frac{\G(|\theta^1_{ab}|)\,\G(|\theta^2_{ab}|)\,\G(1-|\theta^3_{ab}|)}{ \Gamma(1-|\theta^1_{ab}|)\,\Gamma(1-|\theta^2_{ab}|)\,\Gamma(|\theta^3_{ab}|)}\right)^{1/2}\\
K_{bc}&=& \frac{e^{K/2-\phi_4}}{\sqrt{2\pi}}  \left( \frac{\G(|\theta^1_{bc}|)\,\G(1-|\theta^2_{bc}|)\,\G(|\theta^3_{bc}|)}{ \Gamma(1-|\theta^1_{bc}|)\,\Gamma(|\theta^2_{bc}|)\,\Gamma(1-|\theta^3_{bc}|)}\right)^{1/2}\\
K_{ca}&=& \frac{e^{K/2-\phi_4}}{\sqrt{2\pi}}  \left( \frac{\G(1-|\theta^1_{ca}|)\,\G(|\theta^2_{ca}|)\,\G(|\theta^3_{ca}|)}{ \Gamma(|\theta^1_{ca}|)\,\Gamma(1-|\theta^2_{ca}|)\,\Gamma(1-|\theta^3_{ca}|)}\right)^{1/2}
\eea
which, at small angles (or diluted fluxes from the mirror IIB perspective), reduces to
\be
K_{\a\b}\simeq\left[ \frac{e^{K-2\phi_4}|\th_{\a\b}^i|}{2\pi |\th_{\a\b}^j||\th_{\a\b}^k|}\right]^{1/2}\label{eq: kabapp}
\ee
as it is well known in the literature \cite{Cremades:2004wa,DiVecchia:2008tm}. Finally, if one rewrites  \eqref{eq: kabapp} in terms of gonion masses
\be
K_{\a\b} \simeq e^{K/2}\frac{M_{\rm P}}{m_{\rm gon}}  \, ,
\ee
the Yukawa coupling takes the form already presented in \eqref{YWGCcpx}.

\bibliographystyle{JHEP2015}
\bibliography{bibliography}

\end{document}